\documentclass[]{aastex63}

\newcommand{\Ntargets}{338}        \newcommand{\NtargetsHERE}{292}    \newcommand{\Nsample}{229}         \newcommand{\Norbits}{54}          \newcommand{\NorbitsNEW}{28}       
\newcommand{\NorbitsUPDATED}{26}
\newcommand{\NorbitsNEWsoaronly}{20} 

\newcommand{\Nobservations}{1697}   \newcommand{\NobservationsNEW}{1066}        \newcommand{\NresolvedNEW}{900}     \newcommand{\Nresolvedpairs}{212}

\newcommand{\Porb}{$P_\mathrm{orb}$}
\newcommand{\Msun}{M$_\odot$}
\newcommand{\Gaia}{Gaia}  

\received{10 October 2025}
\revised{16 January 2026}
\accepted{16 January 2026}

\shorttitle{M Dwarf Orbits with SOAR}
\shortauthors{Vrijmoet et al.}

\begin{document}

\title{The Solar Neighborhood LIV: \Norbits{} Orbits of M Dwarf Multiples within 30 Parsecs with Speckle Interferometry at SOAR}

\correspondingauthor{Eliot Halley Vrijmoet}
\email{evrijmoet@smith.edu}

\author[0000-0002-1864-6120,gname='Eliot Halley',sname=Vrijmoet]{Eliot Halley Vrijmoet}
\affiliation{Five College Astronomy Department, Smith College, Northampton, MA 01063, USA}
\affiliation{RECONS Institute, Chambersburg, PA 17201, USA}

\author[0000-0002-2084-0782,gname=Andrei,sname=Tokovinin]{Andrei Tokovinin}
\affiliation{Cerro Tololo Inter-American Observatory | NSF's NOIRLab, 
Casilla 603, La Serena, Chile}

\author[0000-0002-9061-2865,gname=Todd,sname=Henry]{Todd J.\ Henry}
\affiliation{RECONS Institute, Chambersburg, PA 17201, USA}

\author[0000-0001-6031-9513,gname=Jennifer,sname=Winters]{Jennifer G.\ Winters}
\affiliation{RECONS Institute, Chambersburg, PA 17201, USA}
\affiliation{Bridgewater State University, 131 Summer St, Bridgewater, MA 02324, USA}

\author[0000-0003-0193-2187,gname=Wei-Chun,sname=Jao]{Wei-Chun Jao}
\affiliation{RECONS Institute, Chambersburg, PA 17201, USA}
\affiliation{Department of Physics and Astronomy, Georgia State University, Atlanta, GA 30303, USA}

\author[0000-0003-2159-1463,gname=Elliott,sname=Horch]{Elliott Horch}
\affiliation{RECONS Institute, Chambersburg, PA 17201, USA}
\affiliation{Department of Physics, Southern Connecticut State University, 501 Crescent Street, New Haven, CT 06515, USA}

\begin{abstract}

We present \NobservationsNEW{} speckle measurements of M dwarf multiples observed over 2021--2024, all taken with HRCam on the Southern Astrophysical Research 4.1~m telescope. Among these, \NresolvedNEW{} observations resolve companions in \Nresolvedpairs{} pairs, with separations spanning 17~milliarcseconds to 3$\farcs$4 and brightness differences ranging from 0 to 4.9 magnitudes in the $I$ filter. We have characterized the orbits of \Norbits{} of these companions, spanning periods of 0.67--30~yr, by combining our data with literature astrometry, radial velocities, and, in four cases, \textit{Hipparcos}-\Gaia{} accelerations. Among the orbits presented here are \NorbitsNEW{} that are the first-ever such characterizations for their systems, and \NorbitsUPDATED{} that revise previously-published orbits, thus providing a significant update to the observed dynamics of M dwarfs in the solar neighborhood. From these orbits, we provide new and updated dynamical total masses for these systems, precise to 0.7--7\% in nearly all cases. 
Future mass derivations for components in these systems will contribute to efforts in refining the mass-luminosity relation for the smallest stars, and will enhance investigations of age, magnetism, and metallicity effects on luminosities at a given mass.

\end{abstract}

\keywords{Astrometric binary stars (79), Astrometry (80), Binary
stars (154), M stars (985), Speckle interferometry (1552), Low mass stars (2050)}

\section{Introduction} 
\label{sec:intro}

Orbits of binary and multiple stars contain valuable information on their origins and masses. A large ensemble of known orbits can test models of formation and dynamical evolution of stellar systems, and comparing ensembles of different stellar types can reveal their differing histories. For example, very low-mass binary systems with primary stars $\lesssim$0.1 $M_\odot$ were shown by \citet{Dup17} to have less eccentric orbits at long periods than the systems with primaries $\sim$1 $M_\odot$ in \citet{Rag10}. When sufficiently accurate masses are available for a set of stars in a given mass regime, effects of age, magnetism, and metallicity can be discerned.

In this paper we focus on the ubiquitous M dwarfs, the stars that account for three of every four members of the solar neighborhood \citep{Hen24}. Assembling a large number of orbits for these low mass stars has historically been challenging due to their intrinsic faintness. Early systematic surveys complemented the historic visual discoveries of binary M dwarfs and revealed candidates, along with a modest number of orbits, using infrared speckle imaging \citep{Hen90,Hen-thesis,Lei97}, while more recent efforts used adaptive optics at near-infrared wavelengths \citep{War15,Man19}. Large efforts have been undertaken in the past decade at optical wavelengths that offer improved resolution limits for a given telescope aperture compared to observations made in the near-infrared. Notable recent strides in taking snapshots of M dwarf multiples that may be used to map orbits include \citet{Mas18} and \citet{Cla22} in the northern hemisphere, the targeted all-sky speckle survey of M dwarfs by \citet{Win-sample}, the AstraLux monitoring program \citep[e.g.,][]{Cal20}, as well as the {\it TESS} objects of interest follow-up program \citep{How21} that includes some M dwarfs. For stars within 100 pc, such surveys are particularly sensitive to separations comparable to our Solar System, thereby mapping real estate that may be occupied by stars rather than the exoplanets sought after by many researchers.

Adding to these efforts, we are carrying out a southern hemisphere speckle interferometry campaign to map the orbits of M dwarf multiples using the SOuthern Astrophysical Research (SOAR) telescope at Cerro Tololo Inter-American Observatory (CTIO). Notably, our survey reaches M dwarfs out to 25 pc, whereas most of the earlier efforts typically had horizons of 15 pc or less.  Our first results \citep{Vri22} covered observations from 2019--2020 and presented data for 211 detected companions, 97 of which were newly discovered through this program. Updating that work, this current paper covers observations over and focuses on 2021--2024  the \Norbits{} orbits we are now able to characterize via these new data. In many cases the orbit calculations have been made possible by combining the SOAR results with data from the other efforts discussed above. The sample of M dwarfs within 25 pc observed in our program is described in $\S$\ref{sec:sample}. The observations and data reduction processes are reviewed in $\S$\ref{sec:observations}, and the results are presented in $\S$\ref{sec:results}. A brief discussion of the results is given in $\S$\ref{sec:discussion}, followed by a summary of this work in $\S$\ref{sec:conclusion}. Overall, these orbits represent an 11\% increase to the current 257 orbits of M dwarfs within 25 pc in the Sixth Catalog of Orbits of Visual Binary Stars \citep[as of April 2025;][]{ORB6}. We also revise $\sim$9\% of that catalog's existing orbits.

\section{Sample} 
\label{sec:sample}

The selection of stars for this program was discussed in detail in \citet{Vri22}, so here we only briefly summarize that process. Because the program aims to map the orbits of nearby M dwarf multiples, our targets are all known binaries or higher-order systems, or stars that exhibit some evidence of potential multiplicity. Wide pairs ($\gtrsim 2\farcs5$) have been excluded in order to focus on orbits shorter than 30 yr. Membership is also restricted to systems within 25 pc, observable by SOAR (Decl.\ $\lesssim$ $+$20$^\circ$), and with photometry indicating the primary to be an M dwarf ($V-K_s > 3.70$ and $M_V > 9.02$). For comparison, the blue and red cutoffs in $V-K_s$ used to develop this sample correspond to $M_G$ = 8.23 and 17.79, respectively, which are consistent with the magnitude range of $M_G$ = 8.1--17.8 for red dwarfs outlined in \cite{Hen24}.

Three sources for targets were used: the RECONS\footnote{\url{http://www.recons.org}} long-term astrometry program at the CTIO/SMARTS 0.9~m \citep{Jao05,Hen18}, the Sixth Catalog of Visual Binary Orbits \citep{ORB6} that is consistently updated as publications are added to the literature, and suspect systems (candidate multiples) selected from Gaia DR2 \citep{Vri20}. The original full target list included 338 M dwarfs, but to focus on mapping orbits with likely periods less than 30 yr, in March 2022 we pruned the sample to retain only the \Nsample{} pairs that had been successfully resolved at SOAR. Of those resolved pairs, \Nresolvedpairs{} are presented here; the remainder have not yet been re-observed, mainly because their motions thus far have been slow.

\begin{figure}[h!] \centering
\includegraphics[width=0.7\textwidth]{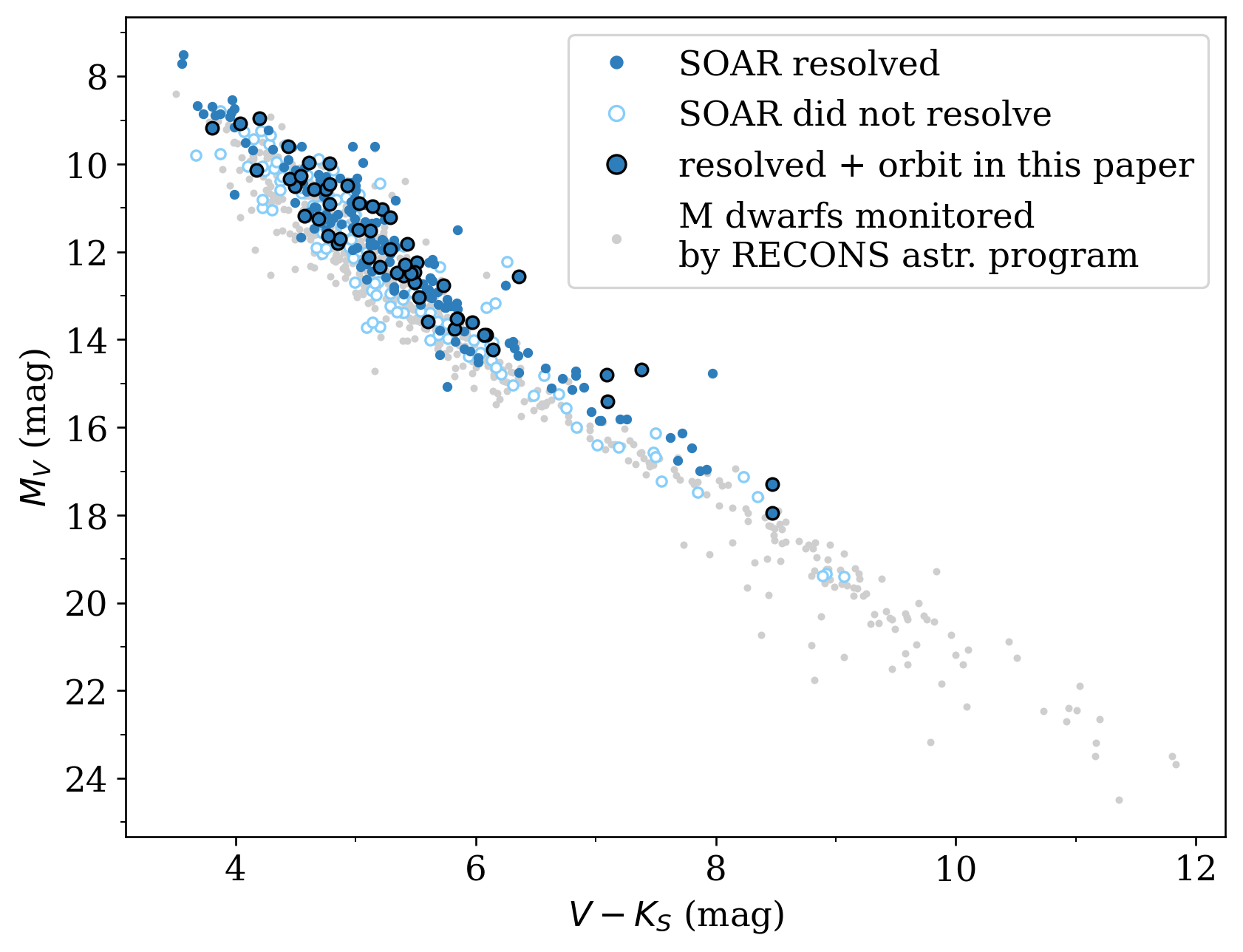}

\caption{Color-magnitude diagram illustrating the distribution of targets in our sample with respect to the M dwarf main sequence. Each light or dark blue point is a pair that has been resolved at SOAR (filled dark blue circles) or has been observed but not resolved (open light blue circles); their photometry measurements are listed in \citet[][Table 1]{Vri22}. The \Norbits{}~points circled in black are pairs for which an orbit is presented in this paper. To illustrate the rest of the low-mass main sequence, these points are overlaid on M and L dwarfs within 25 pc (light grey points) from the RECONS astrometry program \citep{Hen18}.
\label{fig:hrd}}
\end{figure}

The full list of 338 M dwarf targets in the original sample is given in Table~1 of \citet{Vri22}. The pruned, continuing target list is constructed from those marked as resolved in that Table~1 (i.e., ``Y'' in their column~14), plus those that were resolved after the publication of that work. Those new resolutions are indicated in this current publication with one or two asterisks in column~2 of our Table~\ref{tab:results} (discussed in further detail later in $\S$\ref{sec:observations} and $\S$\ref{sec:results}). For consistency, the star system names used throughout this paper match those in the rest of the \textit{Solar Neighborhood} series and the internal RECONS catalog. Full identifiers (coordinates, names used in SIMBAD, etc.) are given in the Appendix. M dwarfs within 25~pc are illustrated on the observational Hertzsprung-Russell diagram in Figure~\ref{fig:hrd}, in which resolved pairs are marked with filled blue circles, unresolved with open blue circles, and pairs with orbits presented here with black outlines. Resolved pairs are not overluminous in all cases, but are concentrated to $M_V \lesssim 18$~mag, reflecting the magnitude limit of HRCam at SOAR.

\section{Observations and Data Reduction with HRCam} 
\label{sec:observations}

Optical speckle observations for this program were taken at the 4.1~m SOAR telescope with the high-resolution camera (HRCam), using time awarded to the first two authors on this paper, supplemented with data secured during a few engineering nights. The instrument, observing procedure, and data reduction are described in \citet{Tok10} and \citet{Tok18_10yr}. All observing programs are executed in a common pool, optimizing priorities, telescope slews, and cadence under variable conditions that strongly influence performance for faint stars. Efficiency is increased because orbit phases can be timed among all observing nights and because all programs share common astrometric calibration of HRCam. Details of calibration using binaries with well-known orbits are published in a series of papers \citep[the most recent are][]{Tok22,Mas23,Tok24}.

This SOAR M dwarf speckle program was started in 2019 July and has continued through 2025. The new data included in this publication cover the SOAR observations from January 2021 through January 2025. Fast-moving M dwarf pairs are prioritized, allowing for a better coverage of their orbits. On a single visit, typically two data cubes of 400 images each are taken; occasionally, additional cubes are taken, for example, under poor seeing or for faint targets. Each frame in the cube has a short exposure time of 25 ms (50 ms for fainter stars), is $200 \times 200$ pixels in size, and covers a field $3\farcs15$ square; for pairs known to be separated by more than $\sim$$1\farcs4$, a wider $400 \times 400$ pixel field is used. Nearly all images are taken in the broad $I$ filter (790/130~nm bandwidth), 
which is very similar to a Cousins $I$ filter, with the transmission curve given in \citet{Tok18_10yr}, Figure~2. Additionally, a handful of measures presented here are in the Str\"{o}mgren $y$ filter.

A speckle power spectrum and autocorrelation function were computed for each data cube, and the existence of a companion in those data was inferred from the fringes in the power spectrum and/or secondary peaks in the autocorrelation function. Companion detections in two sequential data cubes safeguard against artefacts (e.g., cosmic-ray hits), while the two binary-star measurements are checked for consistency and averaged. The three characteristics of a stellar pair --- separation, position angle, and magnitude difference --- are determined by fitting a model to the high-frequency portion of the power spectrum. A known single reference star was used for comparison in the cases of close and/or high-contrast pairs to account for the intrinsic shape of the power spectrum. A similar fitting procedure is used for processing triple stars. Approximate detection limits (resolution and contrast) are determined as well for each observation. Typically, companions fainter by up to $\sim$3 mag in the $I$ filter can be detected at 0\farcs15 and $\sim$5 mag fainter at 1\farcs0.

The results are given in Table~\ref{tab:results}, and described in $\S$\ref{sec:results} below. Additional identifiers for these pairs (coordinates, alternate names) are given in the Appendix. Several flags regarding the observations are included in column 11 of Table~\ref{tab:results}. A ``q" flag is given when the true quadrant of the companion has been determined from the shift-and-add image that breaks the 180$^\circ$ ambiguity of the autocorrelation function for the position angle, $\theta$, of the companion at a given epoch. A ``p'' flag indicates that the $\Delta m$ for that observation was determined from the average image of a wide, classically resolved pair for which the PSFs are separated (produced from the same data cube) to avoid bias caused by speckle anisoplanatism. A ``:'' flag marks measurements from noisy data or below the diffraction limit that should be regarded as less certain. Finally, the few observations taken in the Stromgren $y$ filter are noted by ``$y$'' in the flags column.

\section{Results} 
\label{sec:results}

\subsection{SOAR Measurements for M Dwarf Systems}
\label{sec:soarobservations}

In \citet{Vri22} we published 830 optical speckle observations made primarily between 2019 July and 2020 December\footnote{A few observations were made in 2018, before the formal start of this program.} in the initial stages of this program. During that time window, a total of 333 M dwarfs were observed, of which 211 were found to have resolved companions. Many of those resolved systems are continuing to be observed and are among the systems with orbits reported here. All points used to calculate the orbits presented in this paper are given in Table~\ref{tab:results}. Among these are \Nobservations{} observations made from 2021 January through 2025 January, roughly half of which are listed for the first time in Table~\ref{tab:results} with the others reported previously in the yearly SOAR speckle results papers \citep[][and references therein]{Tok24}. Table~\ref{tab:results} also includes all imaging data from the literature that were used in orbit fits ($\S$\ref{sec:orbits}). Several orbit fits also incorporated radial velocity data gathered from the literature, as shown in Table~\ref{tab:RVdata}, and the fits to those RV data are presented in Figure~\ref{fig:RVorbits}.

\begin{figure}[h!] \centering
\includegraphics[width=0.6\textwidth]{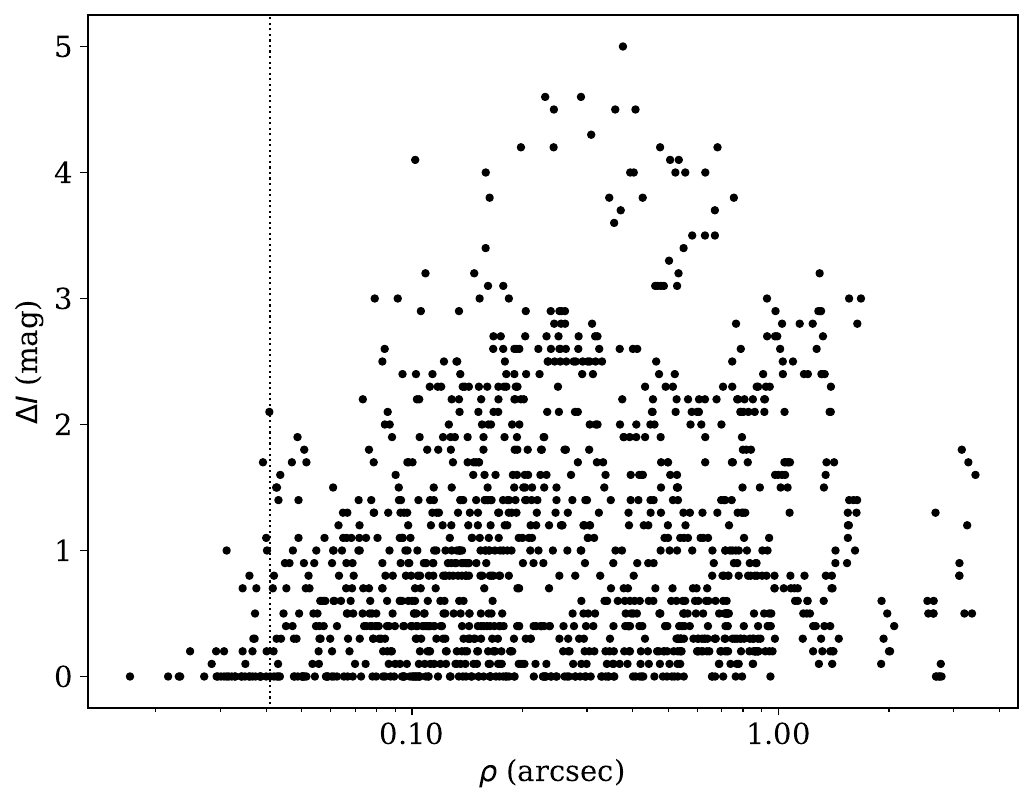}

\caption{Separation $\rho$ vs.\ magnitude difference $\Delta I$ for M dwarf pairs resolved at SOAR, including all of our program observations through 2019--2024 \citep[this work and][]{Vri22} but excluding those for which the data were noisy (``:'' in Table~\ref{tab:results}). The formal diffraction limit of SOAR in the $I$ filter is 41~mas and is marked by the vertical dotted line.  Measurements at smaller separations and small $\Delta I$, achieved by modeling asymmetric power spectra, are less accurate but still useful for orbit monitoring.
\label{fig:seps}}
\end{figure}

Now that this program is complete, \Ntargets{} M dwarf systems have been observed, of which \Nsample{} were resolved as pairs. This small increase since the initial program paper \citep{Vri22} reflects an effort to finish sweeping through the original target list, with emphasis shifted to continued monitoring of resolved systems to determine orbits. The SOAR data presented here describe the subset of \NtargetsHERE{} systems observed since the December 2020 cutoff of the previous paper; with an average of 1.9 observations per target, these objects were covered by \NobservationsNEW{} total observations. Over the full program's \Nobservations{} observations, separations of the resolved companions ranged from 17~mas to 4.9$\arcsec$ with brightness differences up to 5.0 mag in the $I$ band. Figure~\ref{fig:seps} illustrates the separations and magnitude differences of each observation for resolved pairs. The majority of measurements fall within 1$\arcsec$ and $\Delta I \lesssim 3.5$~mag because of the initial sample selection, which targeted systems with likely \Porb $\leq$ 30~yr. 
The full set of observations now in-hand allows us to determine \Norbits{} orbits of M dwarfs, including \NorbitsNEW{} new (first-time) and \NorbitsUPDATED{} updated orbits.

\subsection{High-Resolution Measurements}
\label{sec:results_noorbit}

High-resolution imaging results for our M dwarf targets are listed in Table~\ref{tab:results}. These results include:
\begin{itemize}
    \item all SOAR observations from January 2021 through January 2025, both resolved and unresolved systems
\item all literature astronometry that we used to fit orbits for \Norbits{} systems.
\end{itemize}

Each pair in Table~\ref{tab:results} is identified by its WDS code in column~1; additional names and full coordinates can be found in Table~\ref{tab:lookup}, and observed properties are in Table 1 of \cite{Vri22}. Each pair is a binary ``AB,'' except those with explicit component labels such as ``BC'' or ``Aa,Ab'' that are subsets of triples or higher-order multiples. The names are followed by the reference for its first resolution in column 2, with an asterisk (*) if we report the first resolution here for a system already known to be binary through other methods, or a double asterisk (**) for a first resolution of a pair not previously known as multiple. Systems that we reported as newly resolved in \citet{Vri22} reference that paper. Columns 3 and 4 list the Julian date of each observation (in decimal years) and whether the companion was resolved (Y) or not (N). For resolved pairs, columns 5--7 give the companion's separation $\rho$ in arcseconds, position angle $\theta$ in degrees (in the eastern direction, with respect to north at 0$^\circ$), and magnitude difference $\Delta m$ relative to the primary star. These position angles are in the equinox of their observed epoch (they are later converted to J2000 if used in orbit fitting; see $\S$\ref{sec:orbits}). 
The $\Delta m$ measurement is omitted for results from the literature, given the diversity of filters used by those other programs. If the target was unresolved, columns 8--10 outline the contrast curve with the minimum possible separation resolvable $\rho_\mathrm{min}$, the maximum magnitude difference that could be detected $0\farcs15$ from the primary, and the maximum magnitude difference $1\farcs0$ from the primary. The detection limits are approximate and likely conservative, as revealed by the parameters of resolved pairs in Figure~\ref{fig:seps}.  Column 11 gives the q, p, and : flags related to the SOAR observations, as outlined in $\S$3. For SOAR observations with no flags, position angles should be regarded as modulo 180$^\circ$, i.e., a system listed with position angle 20$^{\circ}$ may instead have angle 200$^{\circ}$. Pairs with orbits reported here (see column~13) have had that 180$^\circ$ ambiguity removed in order to generate reliable orbits. Column~12 notes the reference for the listed observation, and column~13 marks with a checkmark results that were used in an orbit fit (Table~\ref{tab:orbits} and Figures~\ref{fig:orbits}--\ref{fig:orbits5}). 
Note that overall, uncertainties on the SOAR measurements are typically $\sim$2~mas in separations and $\sim$0.2$^{\circ}$ in position angles.

\subsection{Radial Velocity Measurements}
\label{sec:RVs}

Among the \Norbits{}~pairs for which orbits are provided here, there are six that have astrometry data augmented by radial velocity (RV) measurements used to improve their orbits. Here we describe pertinent details of each system's RV data.

\textit{GJ~84~AB (WDS~02051$-$1737)} --- For this SB1 system, 10 RV points from 1998--2001 from HIRES on Keck and the Hamilton spectrograph at Lick Observatory were used, as described in \cite{Nid02}. Only a small portion of the 15-year orbit is covered with RVs.

\textit{GJ~2060~AB (WDS~07289$-$3015 AB)} --- We used two sources of RV points for this SB1, together comprising 21 RVs total. \citet{Rod18}, providing 11 points, measured velocities in 2014 and 2016 using HARPS on the ESO 3.6-m telescope at La Silla Observatory in Chile. \citet{Dur18} provided 10 points from FEROS at the ESO-MPG 2.2-m, also at La Silla. The combined RV data set spans 2005--2017, complementing the imaging data that mainly cover that same period (2005--2024, plus 1991 from \textit{Hipparcos}).

\textit{LHS~3117~AB (WDS~15474$-$1054)} --- This SB3 has 70 RV points spanning 2004--2020 from CARMENES on the 3.5-m at the Calar Alto Observatory \citep{Bar21}. We fit the system as an SB1, omitting the RVs from B and C, to accommodate limitations in our routines for fitting RVs and astrometry jointly. The wealth of data for component A still permit a remarkably precise solution, with an uncertainty of only $\sim$2~d in orbital period, as shown in $\S$\ref{sec:results}.

\textit{LHS~501~AC (WDS~20556$-$1402 Aa,Ab)} --- For this SB2 system, the 36 RV points of intense coverage from 2016--2018 originate from the CARMENES spectrograph, as described in \cite{Bar18}. This results in significant coverage of all phases of the orbit when the components' individual lines are separated. 

\textit{LTT~17066~AB (WDS~23585$+$0740)} --- For this SB3, we used 39 RV points spanning 2009--2021 from CARMENES as well as HARPS and FEROS at La Silla Observatory, as described in \cite{Bar21}. Because the system is resolved as SB3 in \citet{Bar21} but only two components are resolved in the astrometry, we used the weighted average RVs of B and C to complete our joint fit of RV and astrometry, thus treating it like an SB2. The 13-year orbit is well-covered with RVs, resulting in a high-quality fit in which the orbital period is determined with an error of only 5 days.

\textit{GJ~791.2~AB (WDS 20298$+$0941 AB)} --- For this SB2, we used 12 RV points from the Sandiford Cassegrain Echelle Spectrograph on the McDonald 2.1-m, as described in \cite{Ben16}. The RV epochs span six orbits from 1995--2005, allowing the orbital period to be derived with an error of only a few hours.

\startlongtable
\begin{deluxetable}{lllclrclccccc}
\tabletypesize{\scriptsize}
\tablecaption{
Compilation of new and previously reported SOAR speckle observations from the M dwarf multiples survey.
Additional imaging data used in orbit fits are also listed here, and points that were used to fit those orbits are marked in column~13.
(This table is available in its entirety in machine-readable form.)
\label{tab:results}}

\tablehead{
\colhead{WDS} & \colhead{First} & \colhead{Date obs.}  & \colhead{Resol.} & \colhead{$\rho$             } & \colhead{$\theta$} & \colhead{$\Delta m$} & \colhead{$\rho_\mathrm{min}$} & \colhead{$\Delta m$ ($0\farcs15$)} & \colhead{$\Delta m$ ($1\farcs0$)} & \colhead{Obs. } & \colhead{Data} & \colhead{Used in   } \\[-1em] 
\colhead{   } & \colhead{res. } & \colhead{(year)   }  & \colhead{(Y/N) } & \colhead{($^{\prime\prime}$)} & \colhead{(deg)   } & \colhead{(mag)     } & \colhead{($^{\prime\prime}$)} & \colhead{(mag)                   } & \colhead{(mag)                  } & \colhead{flags} & \colhead{ref.} & \colhead{orbit fit?} \\[-1em] 
\colhead{(1)} & \colhead{(2)  } & \colhead{(3)      }  & \colhead{(4)   } & \colhead{(5)                } & \colhead{(6)     } & \colhead{(7)       } & \colhead{(8)                } & \colhead{(9)                     } & \colhead{(10)                   } & \colhead{(11) } & \colhead{(12)} & \colhead{(13)      } \\[-2em] 
}
\startdata
00067$-$0706       & Jan14a & 2022.4419 & Y & 0.1019  & 204.8   & 0.7  &  ...   &  ...   &  ...  &  q   & Mas23      & ... \\ 
                   &        & 2022.6823 & Y & 0.1135  & 201.9   & 0.9  &  ...   &  ...   &  ...  &  :   & Mas23      & ... \\ 
                   &        & 2023.0063 & Y & 0.1349  & 198.1   & 1.0  &  ...   &  ...   &  ...  &  q   & Tok24      & ... \\ 
                   &        & 2024.7041 & Y & 0.2228  & 191.3   & 1.0  &  ...   &  ...   &  ...  &  :   & This paper & ... \\ 
00089$+$2050       & Beu04  & 2001.5980 & Y & 0.1110  & 169.9   & ...  &  ...   &  ...   &  ...  &  ... & Beu04      & $\checkmark$ \\ 
                   &        & 2012.0200 & Y & 0.1330  & 271.9   & ...  &  ...   &  ...   &  ...  &  ... & Jan14a     & $\checkmark$ \\ 
                   &        & 2014.5630 & Y & 0.1458  & 94.5    & ...  &  ...   &  ...   &  ...  &  ... & Hor15a     & $\checkmark$ \\ 
                   &        & 2014.5630 & Y & 0.1460  & 94.8    & ...  &  ...   &  ...   &  ...  &  ... & Hor15a     & $\checkmark$ \\ 
                   &        & 2019.5397 & Y & 0.1110  & 154.4   & ...  &  ...   &  ...   &  ...  &  ... & Tok20b     & $\checkmark$ \\ 
                   &        & 2019.8564 & Y & 0.1233  & 130.1   & ...  &  ...   &  ...   &  ...  &  ... & Tok20b     & $\checkmark$ \\ 
                   &        & 2020.8340 & Y & 0.1522  & 77.5    & ...  &  ...   &  ...   &  ...  &  ... & Tok21      & $\checkmark$ \\ 
                   &        & 2020.9241 & Y & 0.1514  & 74.2    & ...  &  ...   &  ...   &  ...  &  ... & Tok21      & $\checkmark$ \\ 
                   &        & 2021.5684 & Y & 0.1436  & 42.8    & 0.3  &  ...   &  ...   &  ...  &  ... & Tok22      & $\checkmark$ \\ 
                   &        & 2022.4420 & Y & 0.1243  & 349.3   & 0.2  &  ...   &  ...   &  ...  &  ... & Mas23      & $\checkmark$ \\ 
                   &        & 2023.5710 & Y & 0.1437  & 281.2   & 0.3  &  ...   &  ...   &  ...  &  q   & Tok24      & $\checkmark$ \\ 
                   &        & 2024.7017 & Y & 0.1249  & 218.1   & 0.4  &  ...   &  ...   &  ...  &  :   & This paper & $\checkmark$ \\ 
00482$-$0508       & none   & 2021.5657 & N & ...     &  ...    & ...  & 0.0415 & 2.5    &  5.0  &  ... & Tok22      & ... \\ 
03425$+$1232       &  **    & 2021.7516 & Y & 0.1276  & 66.6    & 1.6  &  ...   &  ...   &  ...  &  :   & This paper & ... \\ 
                   &        & 2022.7743 & Y & 0.1473  & 64.9    & 1.7  &  ...   &  ...   &  ...  &  q   & This paper & ... \\ 
                   &        & 2023.5714 & Y & 0.1418  & 64.6    & 1.9  &  ...   &  ...   &  ...  &  q   & This paper & ... \\ 
                   &        & 2024.7018 & Y & 0.0732  & 62.5    & 1.7  &  ...   &  ...   &  ...  &  :   & This paper & ... \\ 
03434$-$0934       &   *    & 2021.7542 & Y & 0.0729  & 125.9   & 2.6  &  ...   &  ...   &  ...  &  :   & This paper & ... \\ 
                   &        & 2022.7742 & N & ...     &  ...    & ...  & 0.0485 & 2.1    &  4.2  &  ... & This paper & ... \\ 
                   &        & 2023.8966 & Y & 0.1115  & 322.5   & 2.3  &  ...   &  ...   &  ...  &  :   & This paper & ... \\ 
                   &        & 2024.7018 & Y & 0.1068  & 282.8   & 2.4  &  ...   &  ...   &  ...  &  :   & This paper & ... \\ 
\enddata
 
\tablecomments{
The orbits fit from these imaging data are shown in Figures~\ref{fig:orbits}--\ref{fig:orbits5} and described in Table~\ref{tab:orbits}. All magnitude differences ($\Delta m$) are in the $I$ band, except where the $y$ band is noted in column 11. Literature observations use a wide variety of filters, thus their $\Delta m$ values are omitted here. Columns~1--11 mirror Table~2 of \citet{Vri22}, and abbreviated descriptions are given below. \\ \\
Column 1: For resolved systems not already listed in the WDS catalog \citep{WDS}, which uses epoch 2000.0 coordinates, here we list the WDS code we anticipate they will be assigned. \\ \\
Column 2: For systems whose first resolution is given by this work, a single asterisk (*) indicates a new resolution of a known but previously unresolved multiple, while a double asterisk (**) marks systems with newly discovered companions detected here for the first time. Targets with ``none'' in this column have never been resolved (either by this program or others). \\ \\
Columns 5--7: These parameters are given when a companion was detected (Y in column~4). Magnitude differences (column~7) are only noted for SOAR detections, and are in the $I$ band unless another filter is noted in column~11. \\ \\
Columns 8--10: These parameters are given when a companion was not detected, and describe the sensitivity of a given observation. \\ \\
Column 11: The flag ``q'' marks that the position angle indicates the true quadrant of the companion; the flag ``p'' notes that the observation's $\Delta m$ was determined from the average image; the flag ``:'' marks measurements from noisy data (i.e., more uncertain results); and ``$y$'' indicates an observation taken in the Str\:{o}mgren $y$ filter. These flags are described in more detail in $\S$\ref{sec:observations}. \\ \\
Column 12: The references corresponding to these codes are given in full in Table~\ref{tab:references}. \\ \\
Column 13: Observations marked here with checkmarks were used in an orbit fit; see Tables~\ref{tab:orbits} and ~\ref{tab:RVorbits} for orbital parameters and Figures~\ref{fig:orbits}--\ref{fig:orbits5} for orbit plots.
}
\end{deluxetable}

\startlongtable
\begin{deluxetable}{llllcc}
\tabletypesize{\scriptsize}
\tablecaption{
Radial velocities supplementing our SOAR imaging for six M dwarf multiples for which we fit orbits.
(This table is available in its entirety in machine-readable form.)
\label{tab:RVdata}}

\tablehead{
\colhead{WDS} & \colhead{Date obs.}  & \colhead{RV           } & \colhead{$\sigma$ RV  } & \colhead{Component}  & \colhead{Ref.} \\[-1em] 
\colhead{   } & \colhead{(year)   }  & \colhead{(km s$^{-1}$)} & \colhead{(km s$^{-1}$)} & \colhead{         }  & \colhead{    } \\[-1em] 
\colhead{(1)} & \colhead{(2)      }  & \colhead{(3)          } & \colhead{(4)          } & \colhead{(5)      }  & \colhead{(6) } \\[-2em] 
}
\startdata
02051$-$1737       & 1998.070  & 23.604     & $\pm$0.030  &  A  & Nid02  \\ 
                   & 1998.542  & 23.344     & $\pm$0.030  &  A  & Nid02  \\ 
                   & 1998.698  & 23.274     & $\pm$0.030  &  A  & Nid02  \\ 
                   & 1998.977  & 23.140     & $\pm$0.030  &  A  & Nid02  \\ 
                   & 1999.533  & 23.052     & $\pm$0.030  &  A  & Nid02  \\ 
                   & 1999.637  & 23.084     & $\pm$0.030  &  A  & Nid02  \\ 
                   & 1999.998  & 23.248     & $\pm$0.030  &  A  & Nid02  \\ 
                   & 2000.099  & 23.334     & $\pm$0.030  &  A  & Nid02  \\ 
                   & 2000.107  & 23.346     & $\pm$0.030  &  A  & Nid02  \\ 
                   & 2000.929  & 24.634     & $\pm$0.030  &  A  & Nid02  \\ 
07289$-$3015 AB    & 2005.138  & 29.930     & $\pm$0.100  &  A  & Dur18  \\ 
                   & 2005.143  & 30.090     & $\pm$0.100  &  A  & Dur18  \\ 
                   & 2007.183  & 28.310     & $\pm$0.100  &  A  & Dur18  \\ 
                   & 2010.902  & 27.740     & $\pm$0.110  &  A  & Dur18  \\ 
                   & 2012.673  & 28.080     & $\pm$0.100  &  A  & Dur18  \\ 
                   & 2014.319  & 28.990     & $\pm$0.010  &  A  & Rod18  \\ 
                   & 2014.883  & 28.910     & $\pm$0.120  &  A  & Dur18  \\ 
                   & 2015.096  & 28.740     & $\pm$0.120  &  A  & Dur18  \\ 
                   & 2015.391  & 28.900     & $\pm$0.130  &  A  & Dur18  \\ 
                   & 2016.762  & 28.340     & $\pm$0.020  &  A  & Rod18  \\ 
                   & 2016.762  & 28.270     & $\pm$0.020  &  A  & Rod18  \\ 
\enddata
 
\tablecomments{The orbits fit to these RVs are illustrated in Figure~\ref{fig:RVorbits} and described in Table~\ref{tab:RVorbits}. The reference abbreviations in column~6 are defined in Table~\ref{tab:references}.}
\end{deluxetable}

\startlongtable
\begin{deluxetable}{cc}
\tabletypesize{\scriptsize}
\tablecaption{Codes for the references cited in tables throughout this work.
\label{tab:references}}

\tablehead{
\colhead{Code} & \colhead{Ref. } \\[-1em] 
\colhead{(1) } & \colhead{(2)  } \\[-2em] 
}
\startdata
Bal06  & \citet{Bal06} \\
Bal13  & \citet{Bal13} \\
Ben00  & \citet{Ben00} \\ 
Ben16  & \citet{Ben16} \\
Ber10  & \citet{Ber10} \\
Beu04  & \citet{Beu04} \\
Bil06  & \citet{Bil06} \\ 
Bla87  & \citet{Bla87} \\
Bon09  & \citet{Bon09} \\
Bow15  & \citet{Bow15} \\  
Bow19  & \citet{Bow19} \\ 
Cal20  & \citet{Cal20} \\
Cal22  & \citet{Cal22} \\
Cha10  & \citet{Cha10} \\  
Cor17  & \citet{Cor17} \\
Dah76  & \citet{Dah76} \\
Dae07  & \citet{Dae07} \\
Del99  & \citet{Del99} \\ 
Dup10  & \citet{Dup10} \\
Dup16  & \citet{Dup16} \\
Dup17  & \citet{Dup17} \\
Dur18  & \citet{Dur18} \\
ElB18  & \citet{ElB18} \\
ESA97  & \citet{ESA97} \\
Fin37  & \citet{Fin37} \\
For05  & \citet{For05} \\ 
Fra98  & \citet{Fra98} \\
Gat19  & \citet{Gat19} \\ 
Gol04  & \citet{Gol04} \\
Hei87  & \citet{Hei87} \\
Hei90  & \citet{Hei90} \\
Hen94  & \citet{Hen94} \\ 
Hen04  & \citet{Hen04} \\ 
HIP97  & \citet{HIP97} \\
Hor10  & \citet{Hor10} \\
Hor11  & \citet{Hor11} \\
Hor12  & \citet{Hor12} \\
Hor15a & \citet{Hor15a} \\ 
Hor15b & \citet{Hor15b} \\
Hus00  & \citet{Hus00} \\
Jan12  & \citet{Jan12} \\
Jan14a & \citet{Jan14a} \\
Jan14b & \citet{Jan14b} \\
Jan17  & \citet{Jan17} \\
Jao14  & \citet{Jao14} \\ 
Jod13  & \citet{Jod13} \\ 
Kna19  & \citet{Kna19} \\
Kon10  & \citet{Kon10} \\
Kui34  & \citet{Kui34} \\
Kui43  & \citet{Kui43} \\
Law06  & \citet{Law06} \\  
Lei94  & \citet{Lei94} \\ 
Lip55  & \citet{Lip55} \\ 
Luy49  & \citet{Luy49} \\
Man19  & \citet{Man19} \\
Mar00  & \citet{Mar00} \\ 
Mar90  & \citet{Mar90} \\ 
Mas18  & \citet{Mas18} \\
Mas23  & \citet{Mas23} \\
Mon06  & \citet{Mon06} \\
Reu38  & \citet{Reu38} \\
Reu41  & \citet{Reu41} \\
Rie14  & \citet{Rie14} \\
Rod18  & \citet{Rod18} \\
Ros55  & \citet{Ros55} \\
Sch14  & \citet{Sch14} \\
Sha17  & \citet{Sha17} \\
Sie05  & \citet{Sie05} \\
Ste39  & \citet{Ste39} \\
Sto16  & \citet{Sto16} \\
Tok10  & \citet{Tok10} \\
Tok15  & \citet{Tok15} \\
Tok16a & \citet{Tok16a} \\
Tok17  & \citet{Tok17} \\
Tok18  & \citet{Tok18} \\
Tok19a & \citet{Tok19a} \\
Tok19b & \citet{Tok19b} \\
Tok20a & \citet{Tok20a} \\
Tok20b & \citet{Tok20b} \\
Tok21  & \citet{Tok21} \\
Tok22  & \citet{Tok22} \\
Tok24  & \citet{Tok24} \\
VBi60  & \citet{VBi60} \\
Vil84  & \citet{Vil84} \\ 
Vri22  & \citet{Vri22} \\ 
War15  & \citet{War15} \\
WDS    & \citet{WDS} \\
Win11  & \citet{Win11} \\
Win19  & \citet{Win19} \\
\enddata
 
\end{deluxetable}

\subsection{Orbits with High-Resolution Imaging Data Only}
\label{sec:orbits}

\begin{figure} \centering
\includegraphics[scale=0.4]{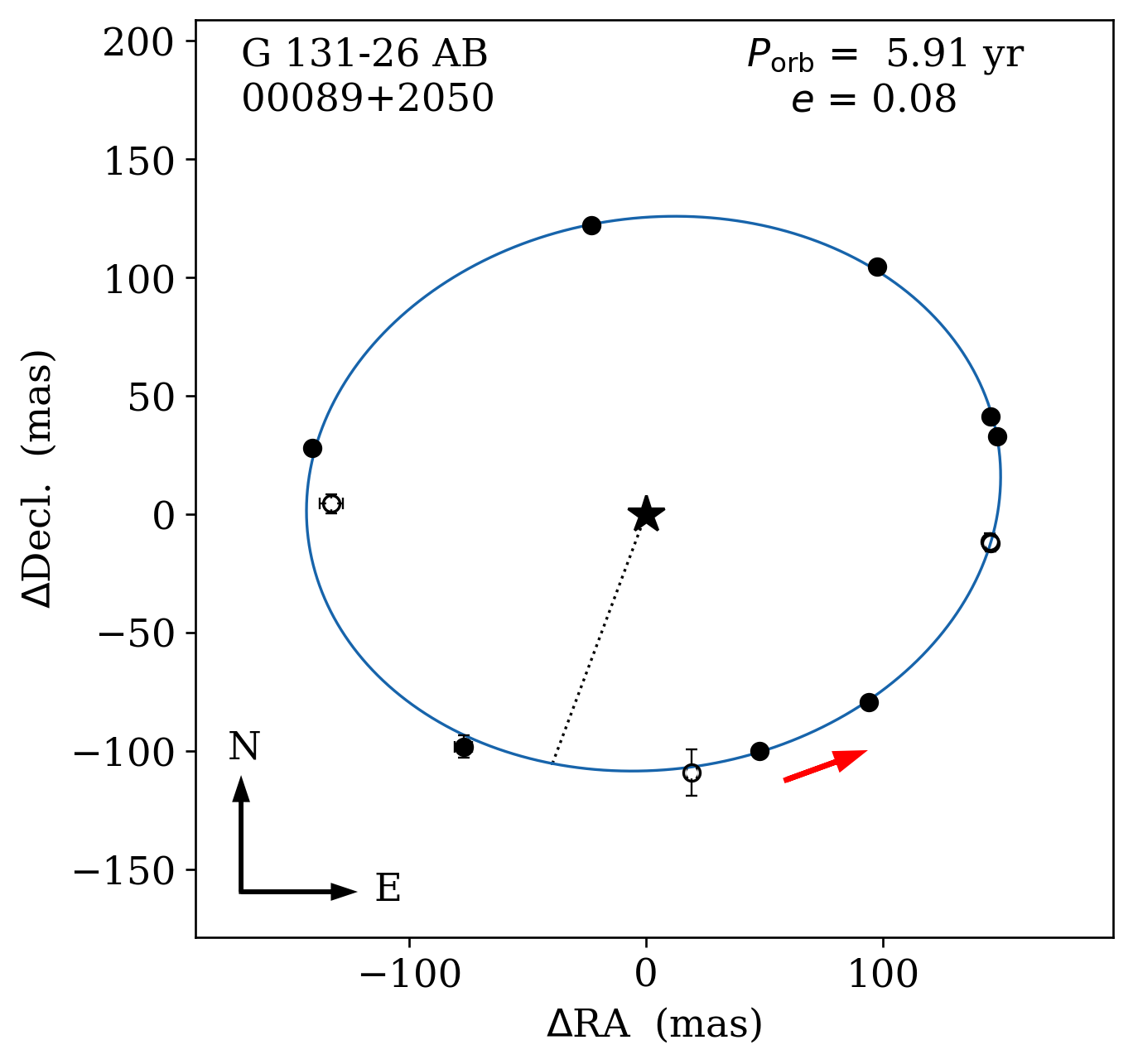}
\includegraphics[scale=0.4]{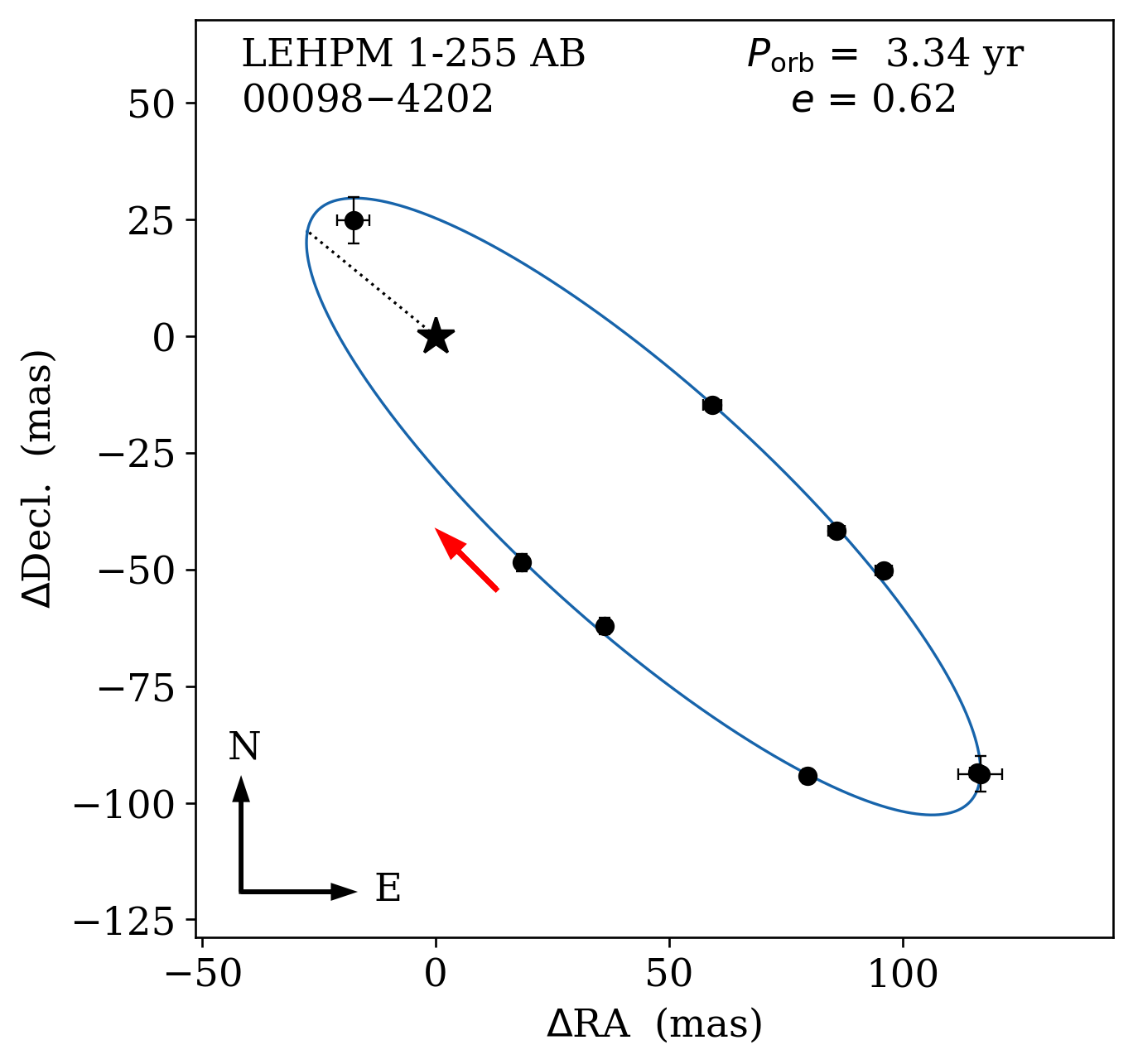}
\includegraphics[scale=0.4]{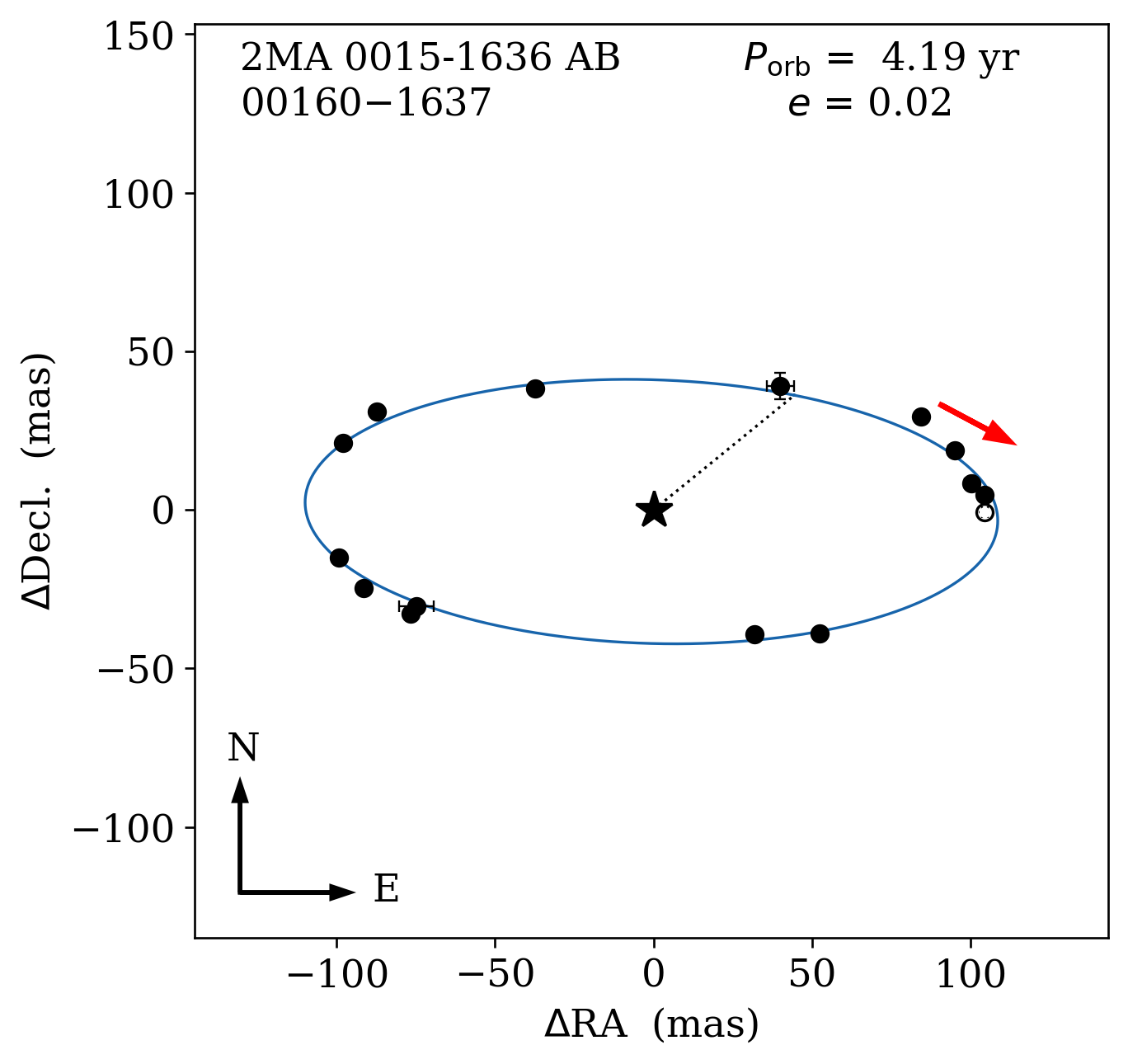}
\includegraphics[scale=0.4]{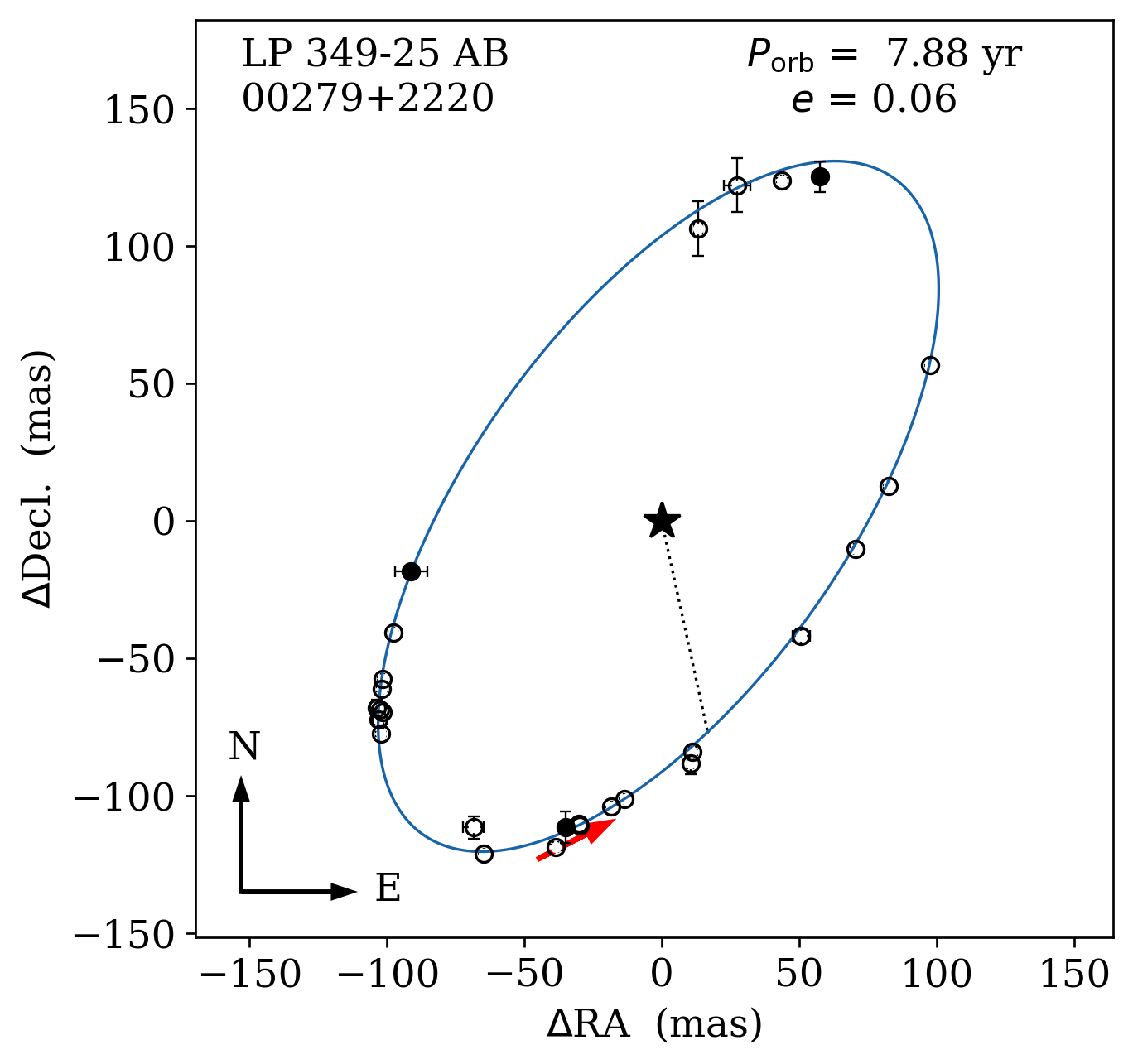}
\includegraphics[scale=0.4]{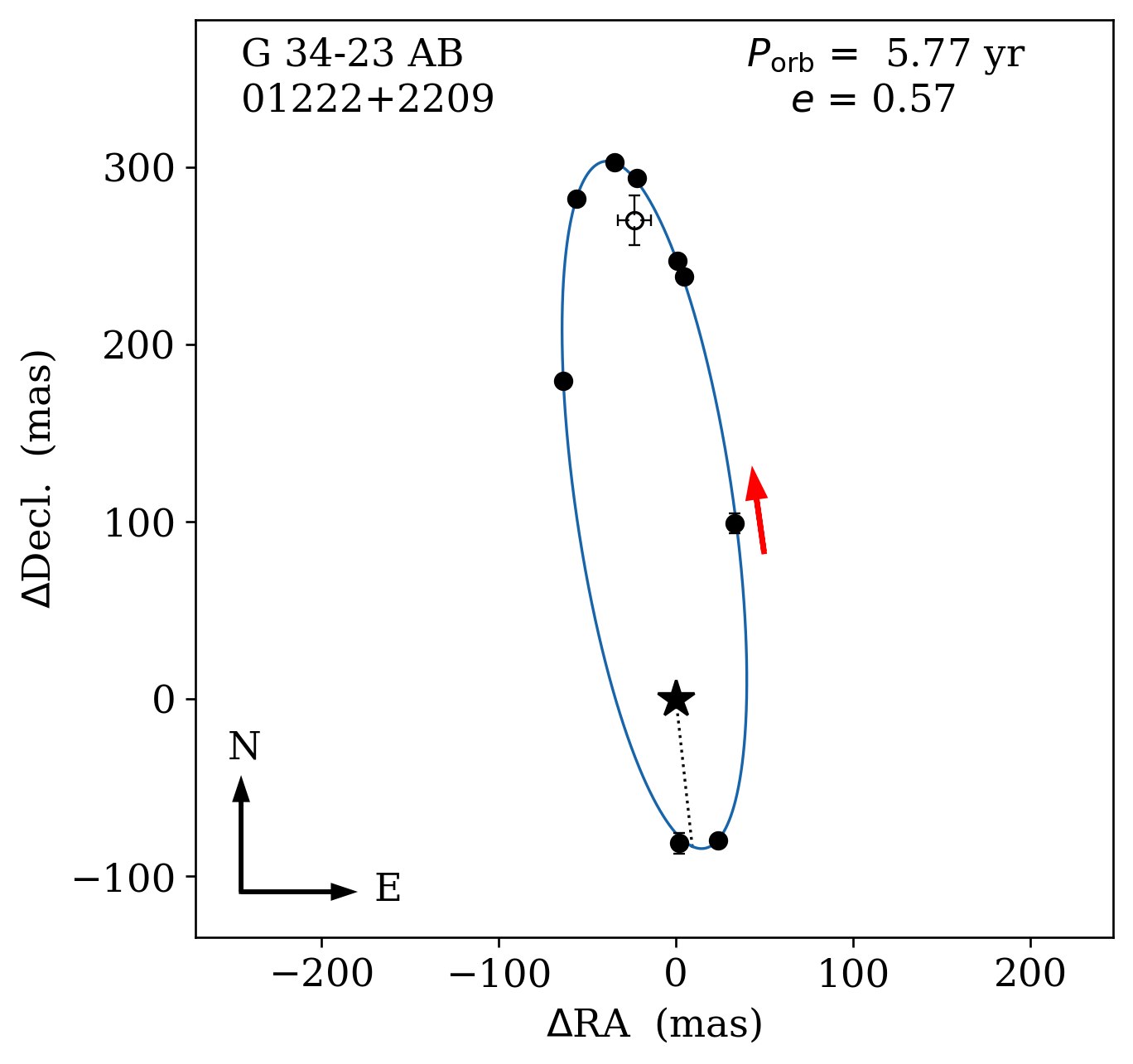}
\includegraphics[scale=0.4]{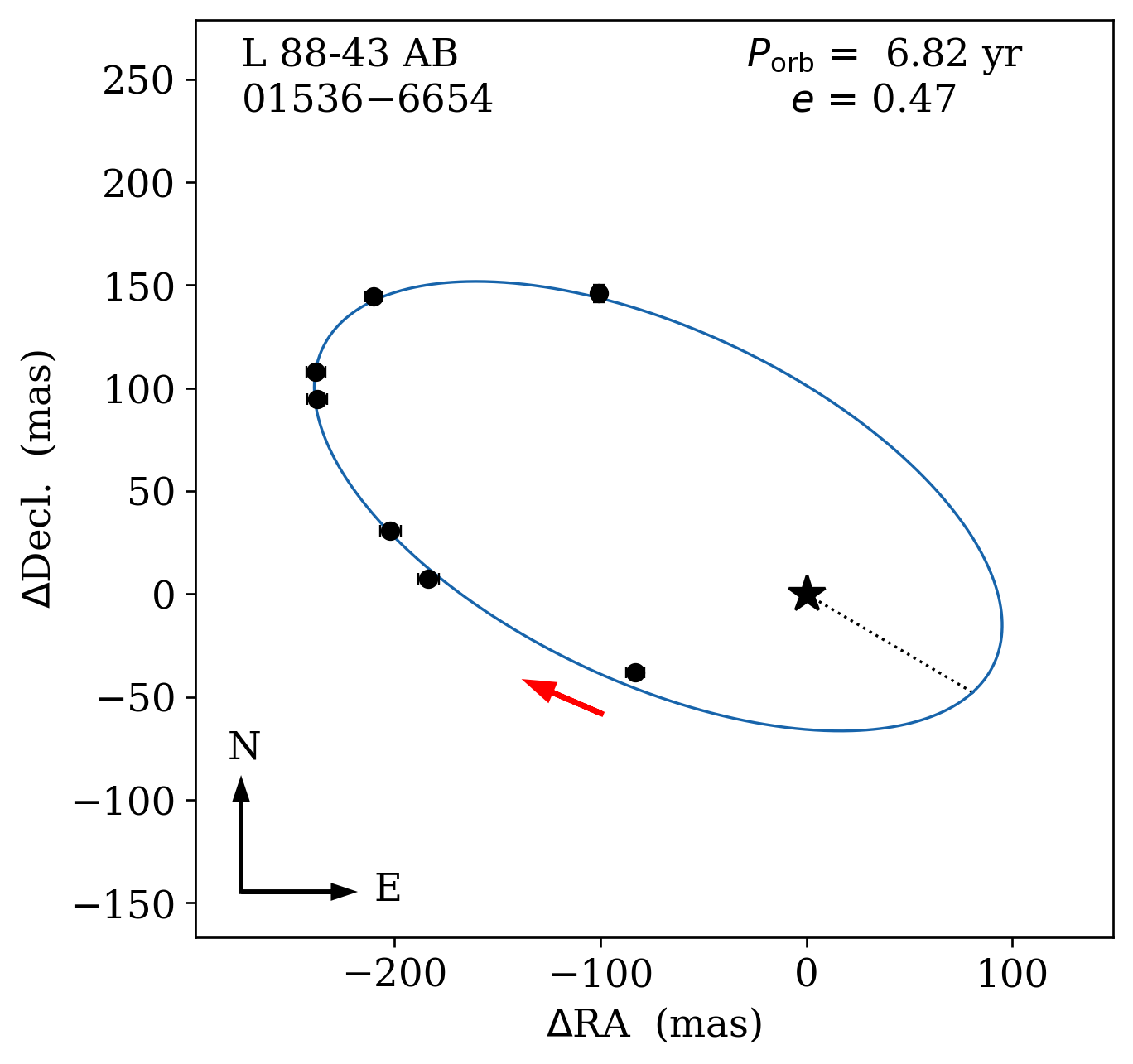}
\includegraphics[scale=0.4]{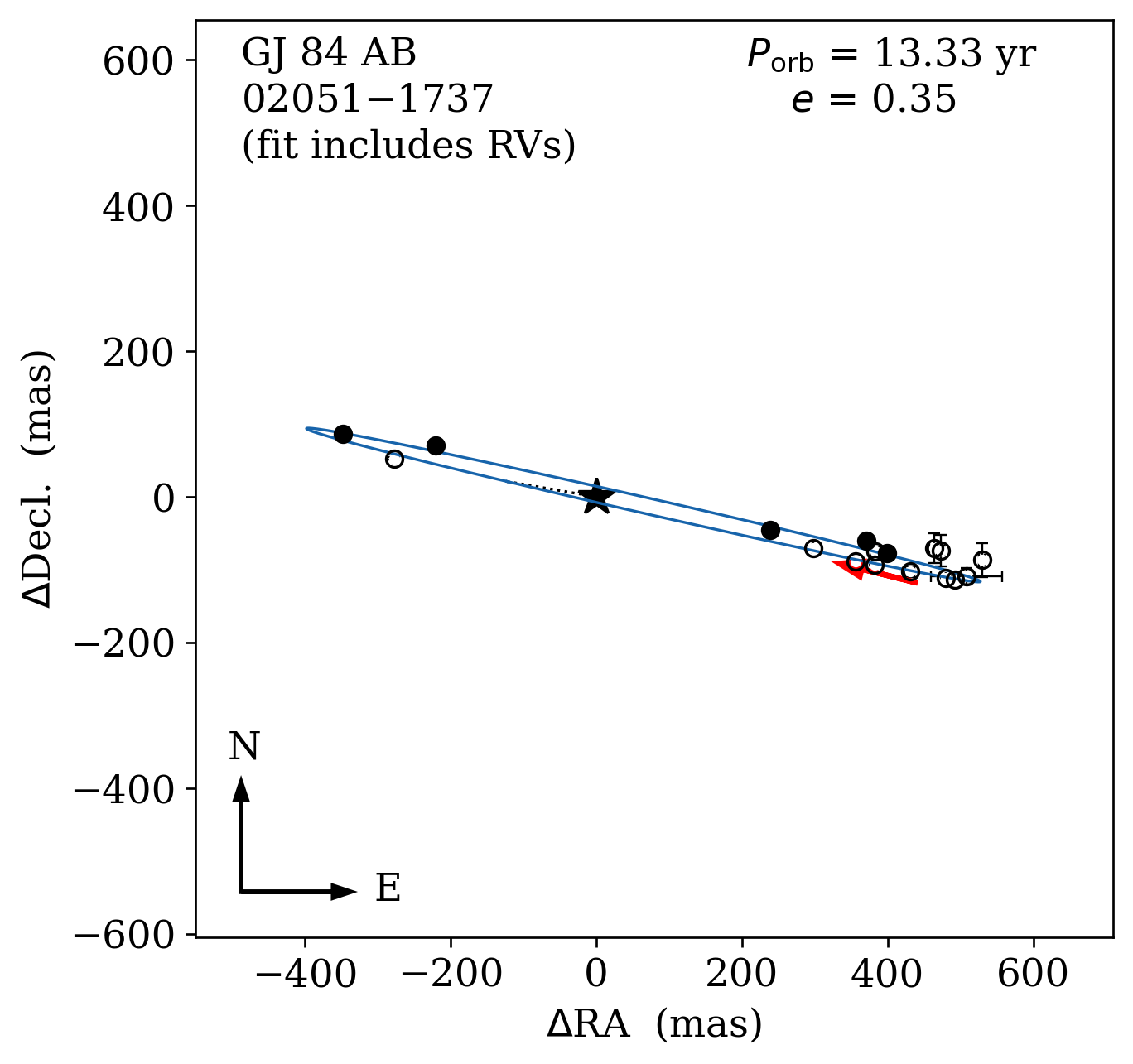}
\includegraphics[scale=0.4]{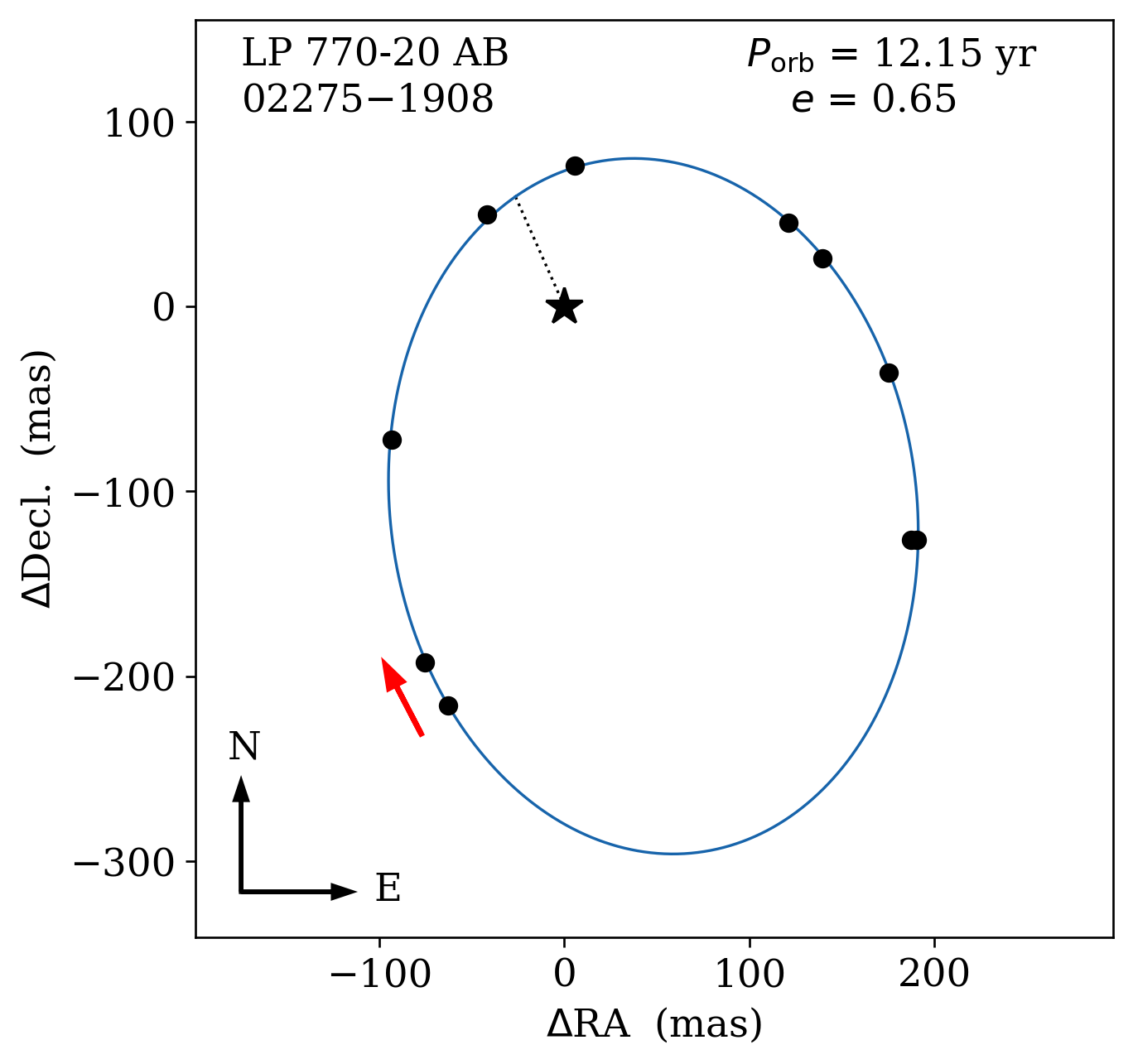}
\includegraphics[scale=0.4]{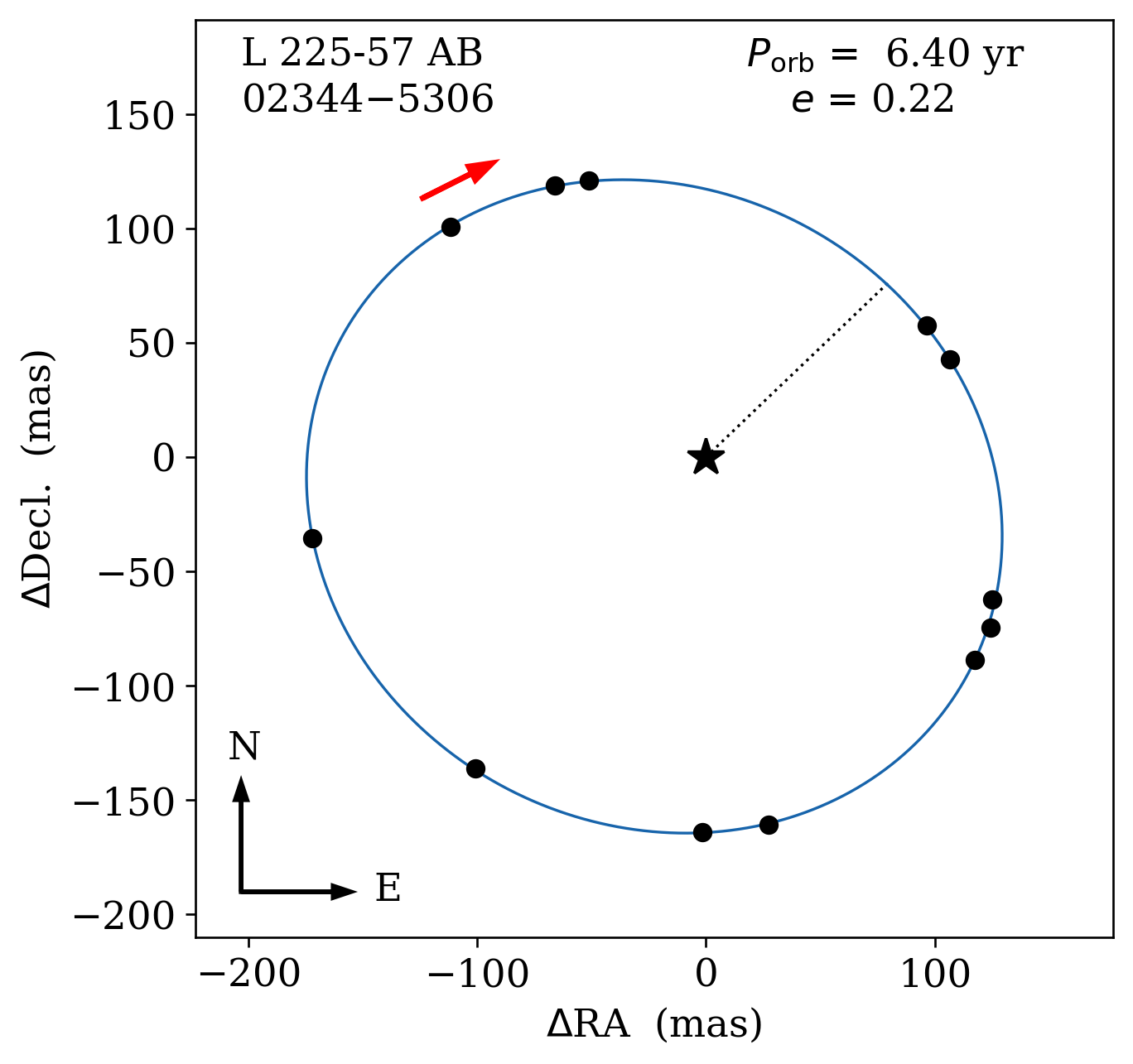}
\includegraphics[scale=0.4]{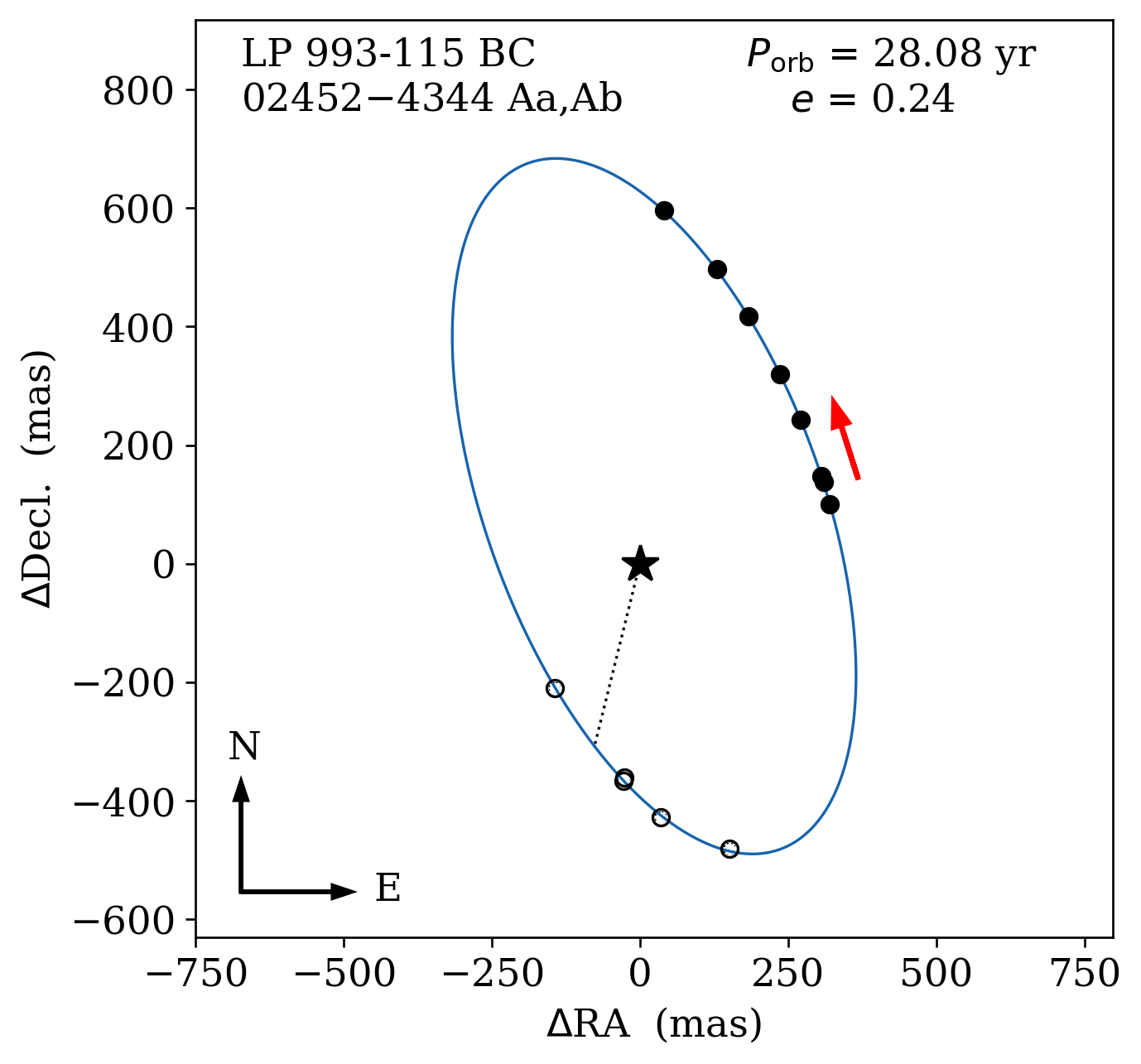}
\includegraphics[scale=0.4]{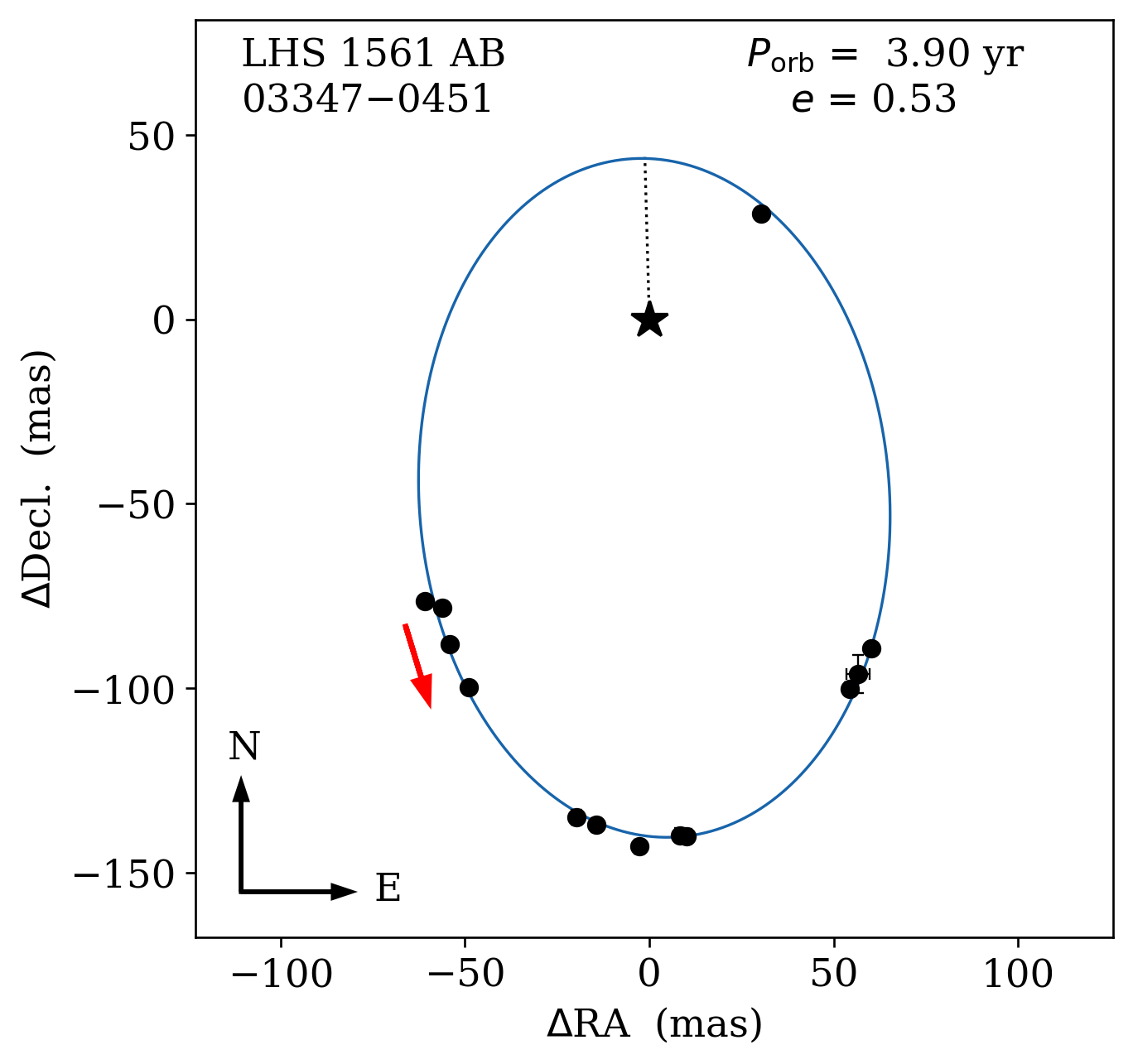}
\includegraphics[scale=0.4]{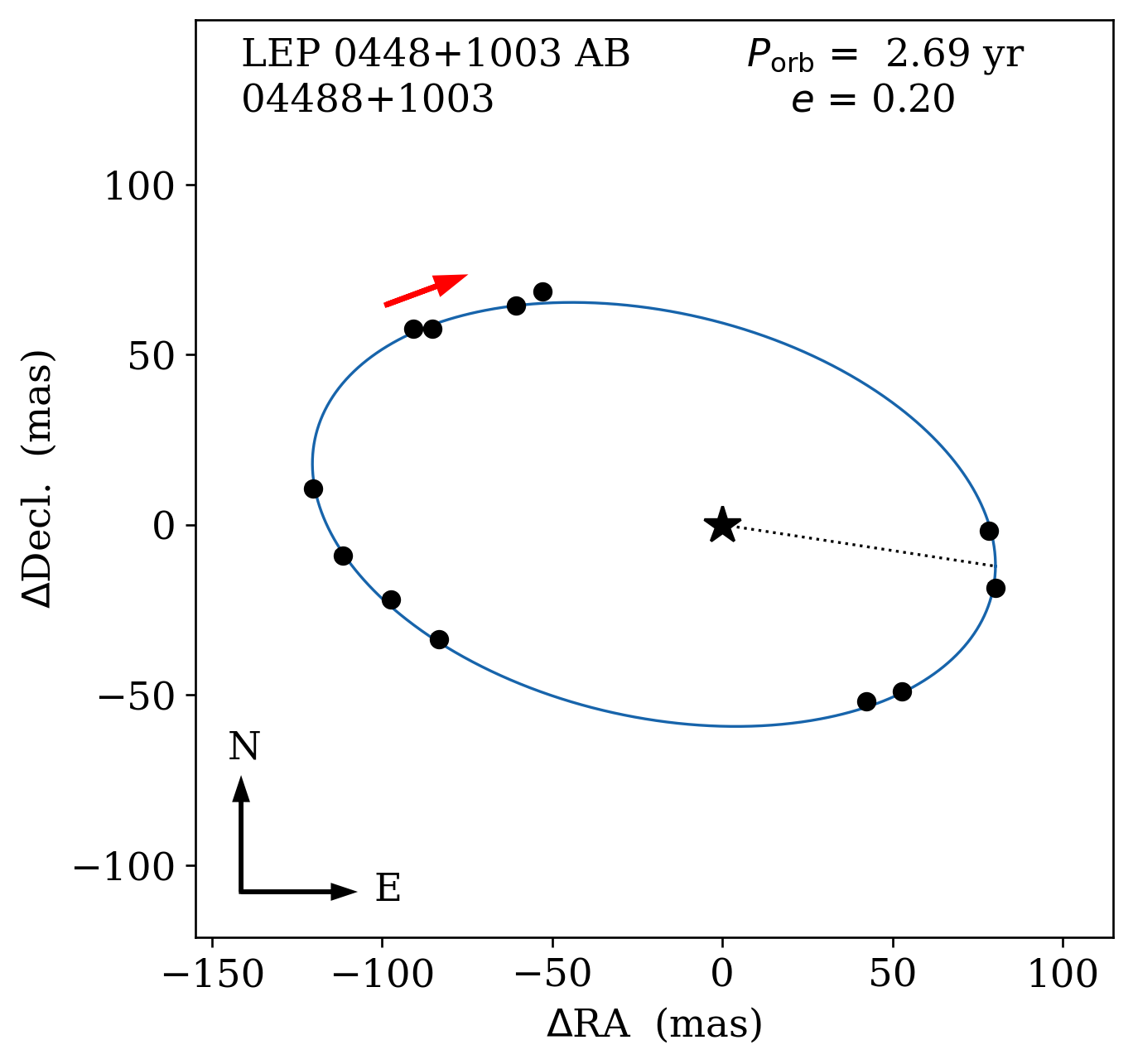}
\caption{
Orbits of M dwarf systems determined from the imaging data in Table~\ref{tab:results}. One system shown here (GJ~84~AB) also incorporated the RV data in Table~\ref{tab:RVdata}; its corresponding RV fit is shown in Figure~\ref{fig:RVorbits}. Each orbit's parameters are given in Table~\ref{tab:orbits} or Table~\ref{tab:RVorbits}, as fit by the code \texttt{ORBIT} ($\S$\ref{sec:orbits}). 
The points mark observations from SOAR (filled circles) and the literature (open circles), with each indicating the location of the secondary star with respect to the primary (marked by the filled black star). Uncertainties on each observation are plotted here as error bars, but are often smaller than these point sizes. In each plot, the dotted line shows the location of the periastron and the red arrow shows the direction of motion. 
\label{fig:orbits}}
\end{figure}

\begin{figure} \centering
\includegraphics[scale=0.4]{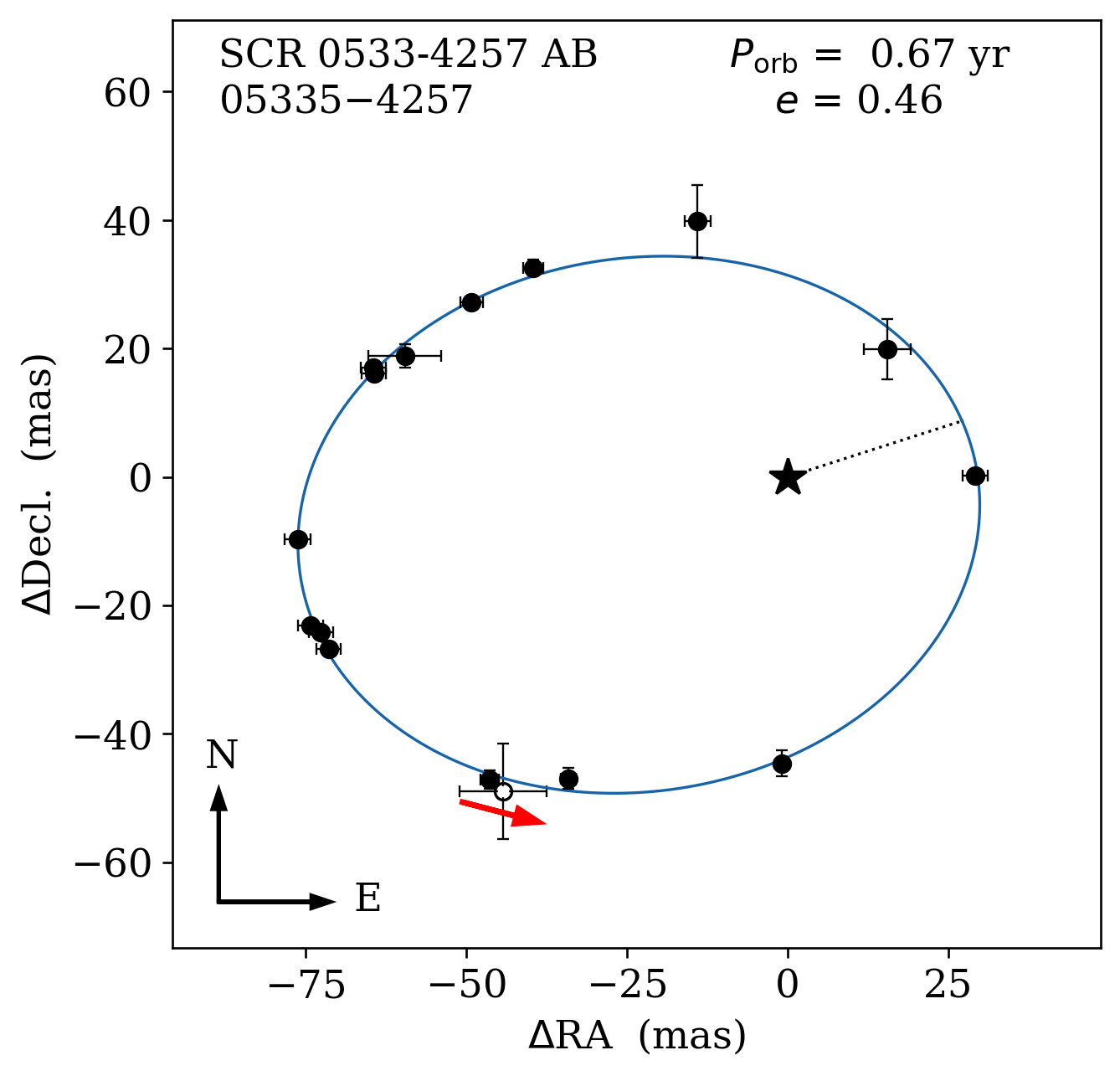}
\includegraphics[scale=0.4]{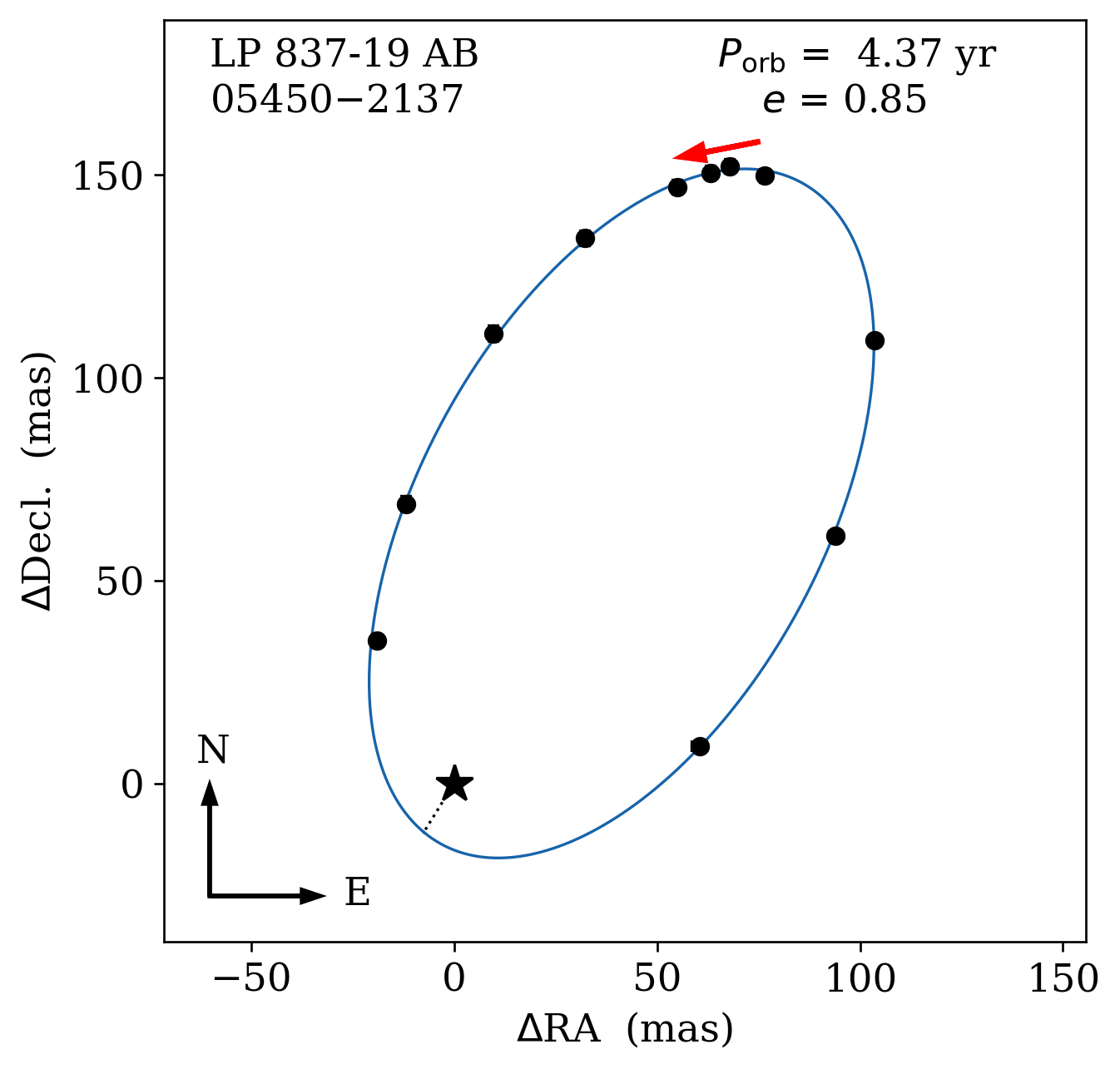}
\includegraphics[scale=0.4]{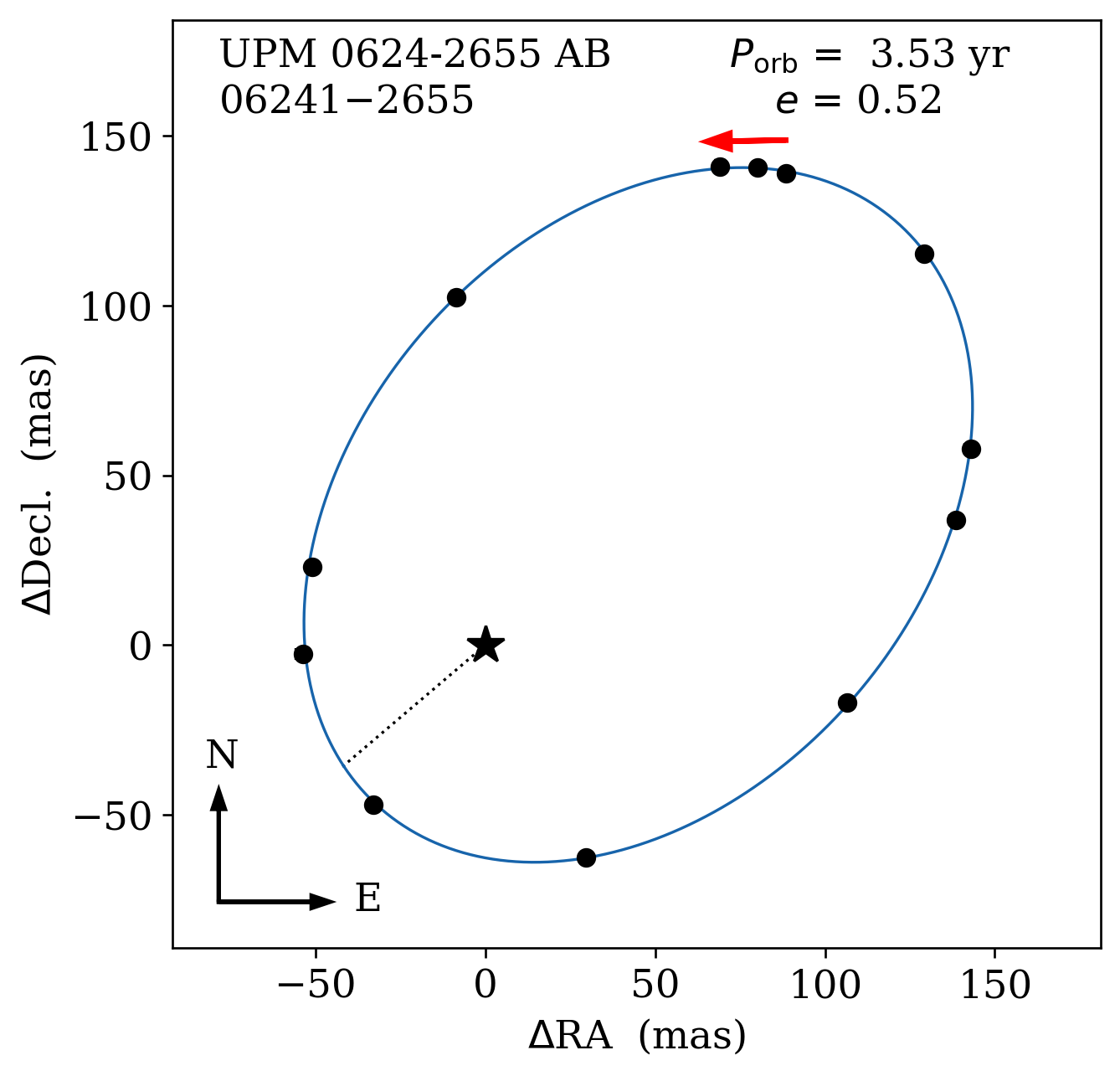}
\includegraphics[scale=0.4]{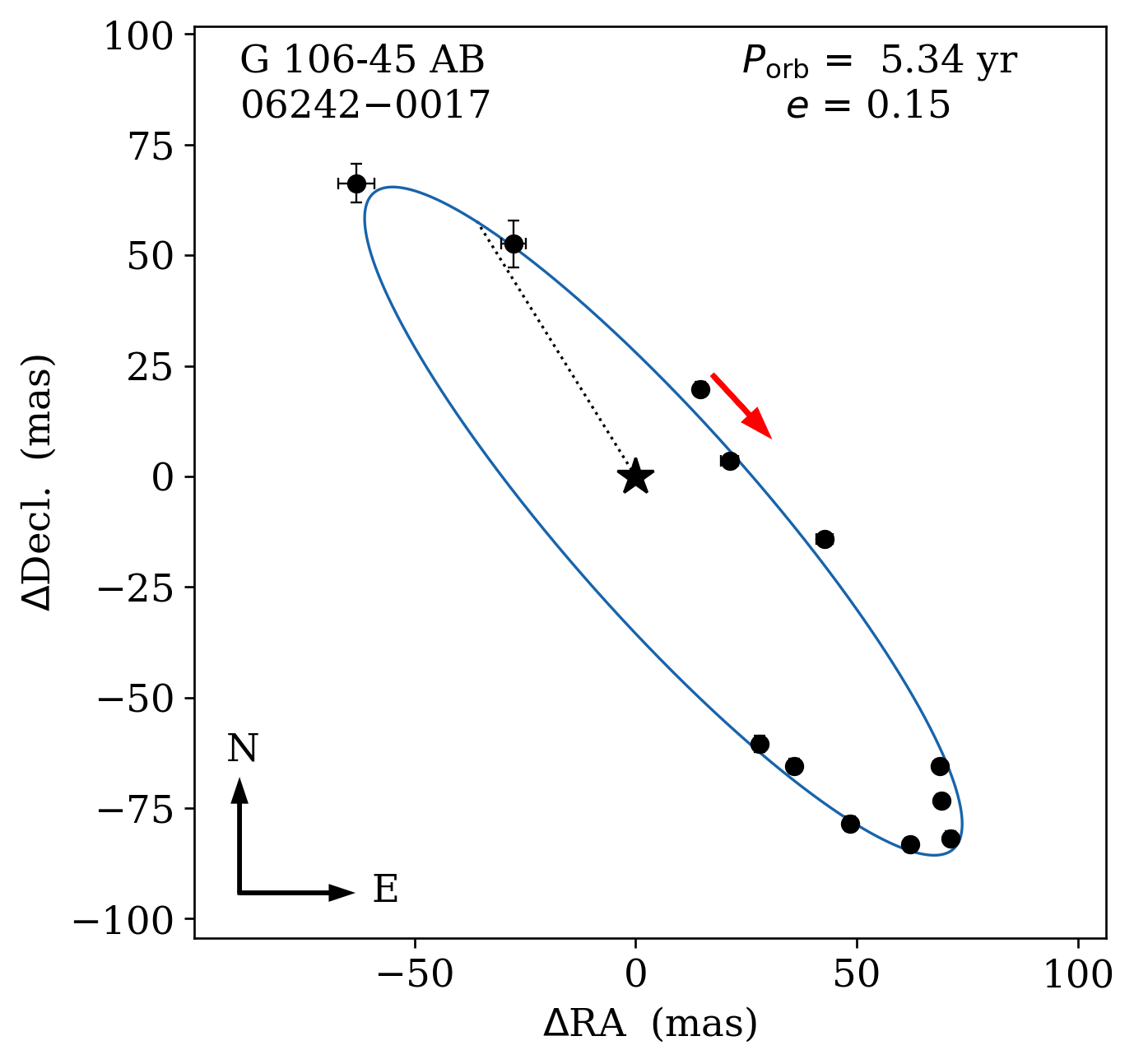}
\includegraphics[scale=0.4]{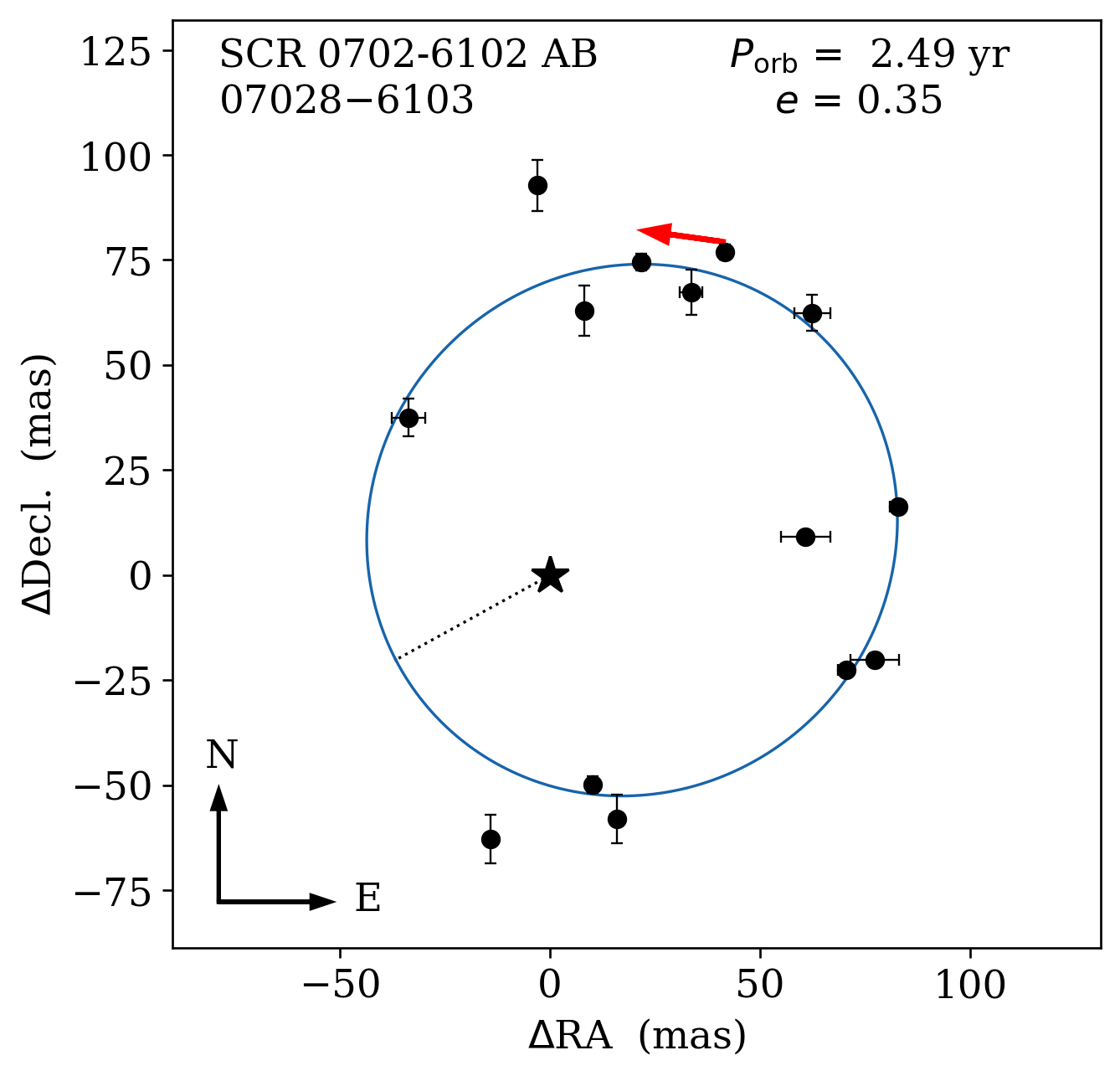}
\includegraphics[scale=0.4]{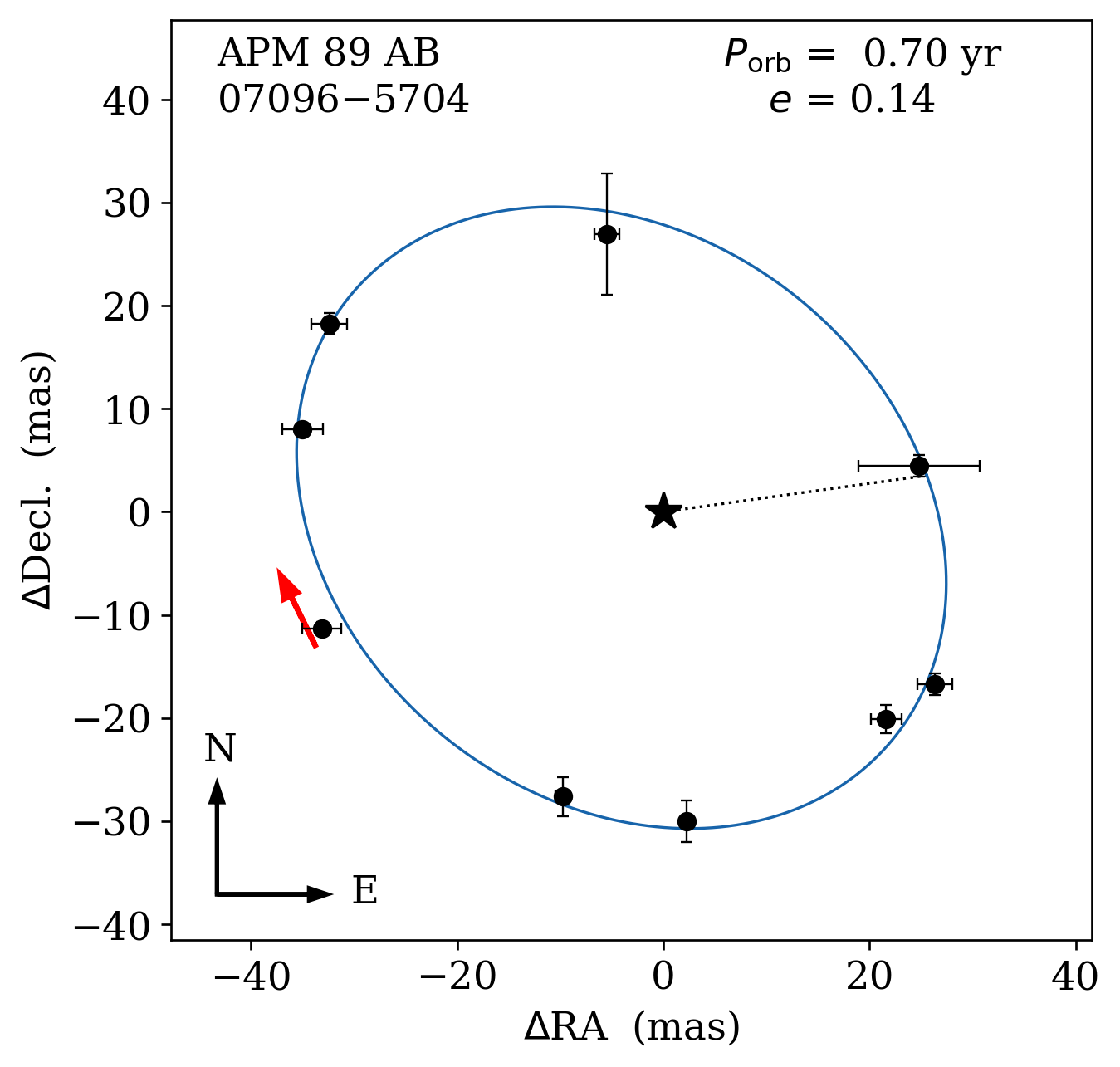}
\includegraphics[scale=0.4]{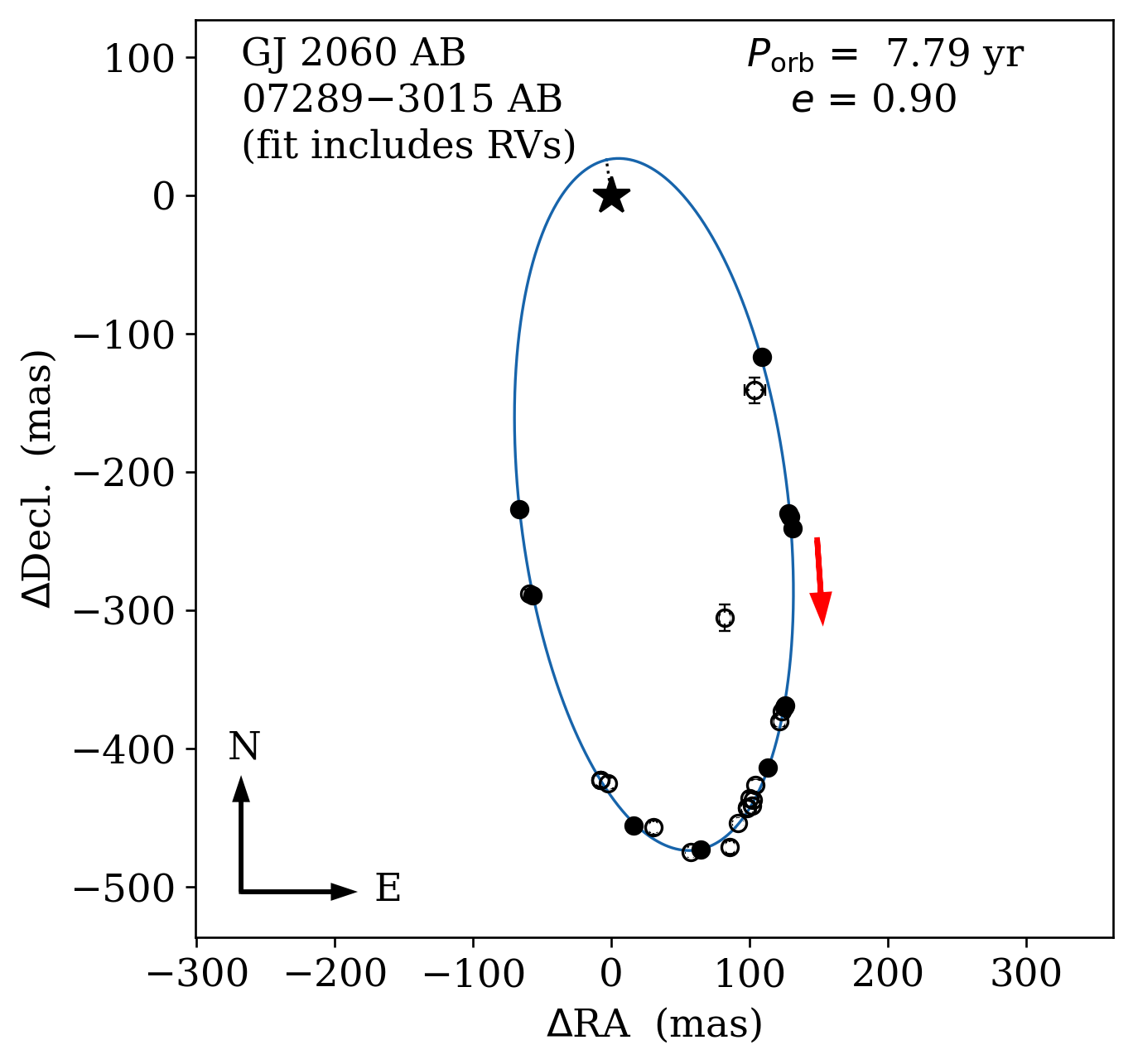}
\includegraphics[scale=0.4]{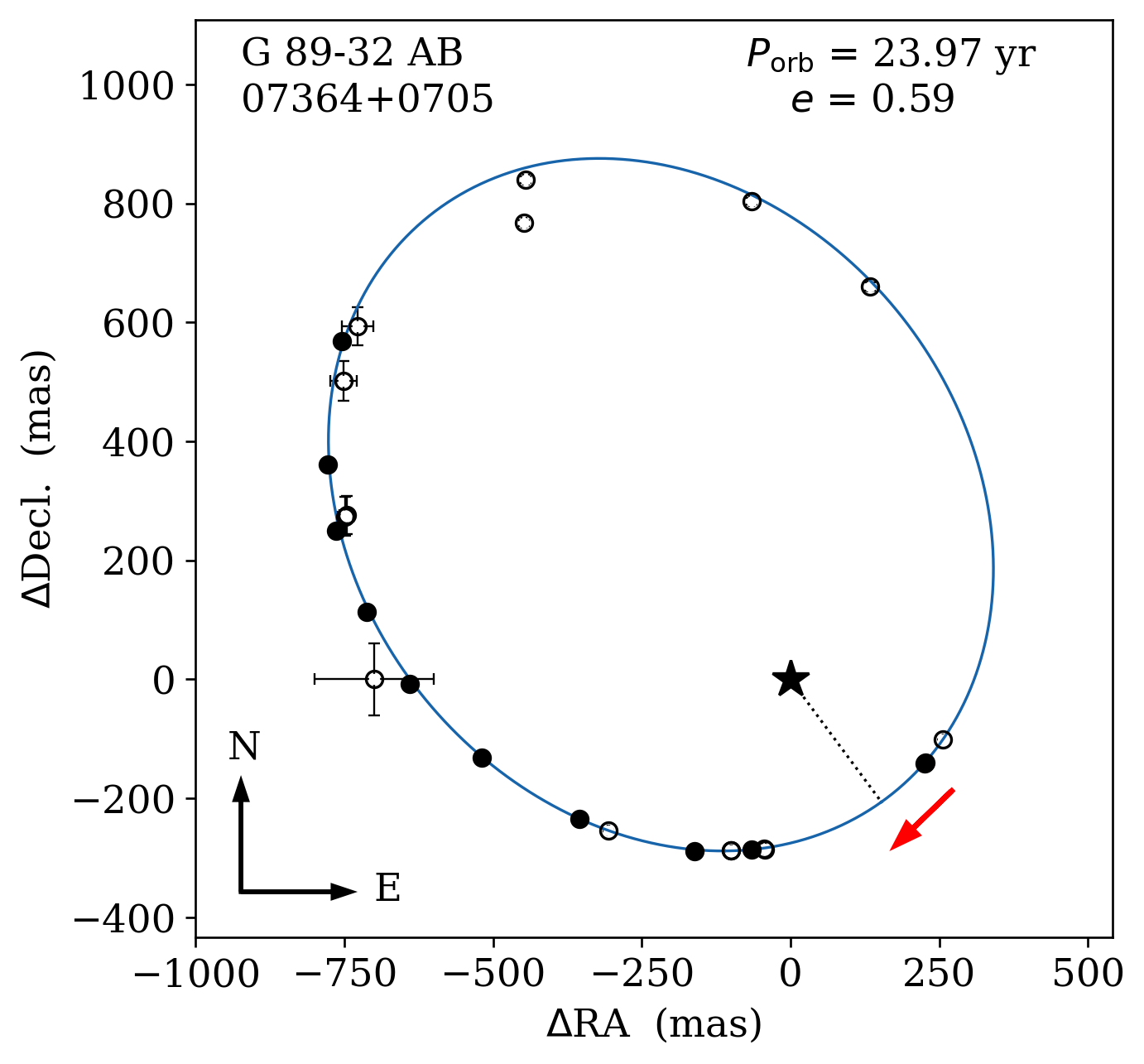}
\includegraphics[scale=0.4]{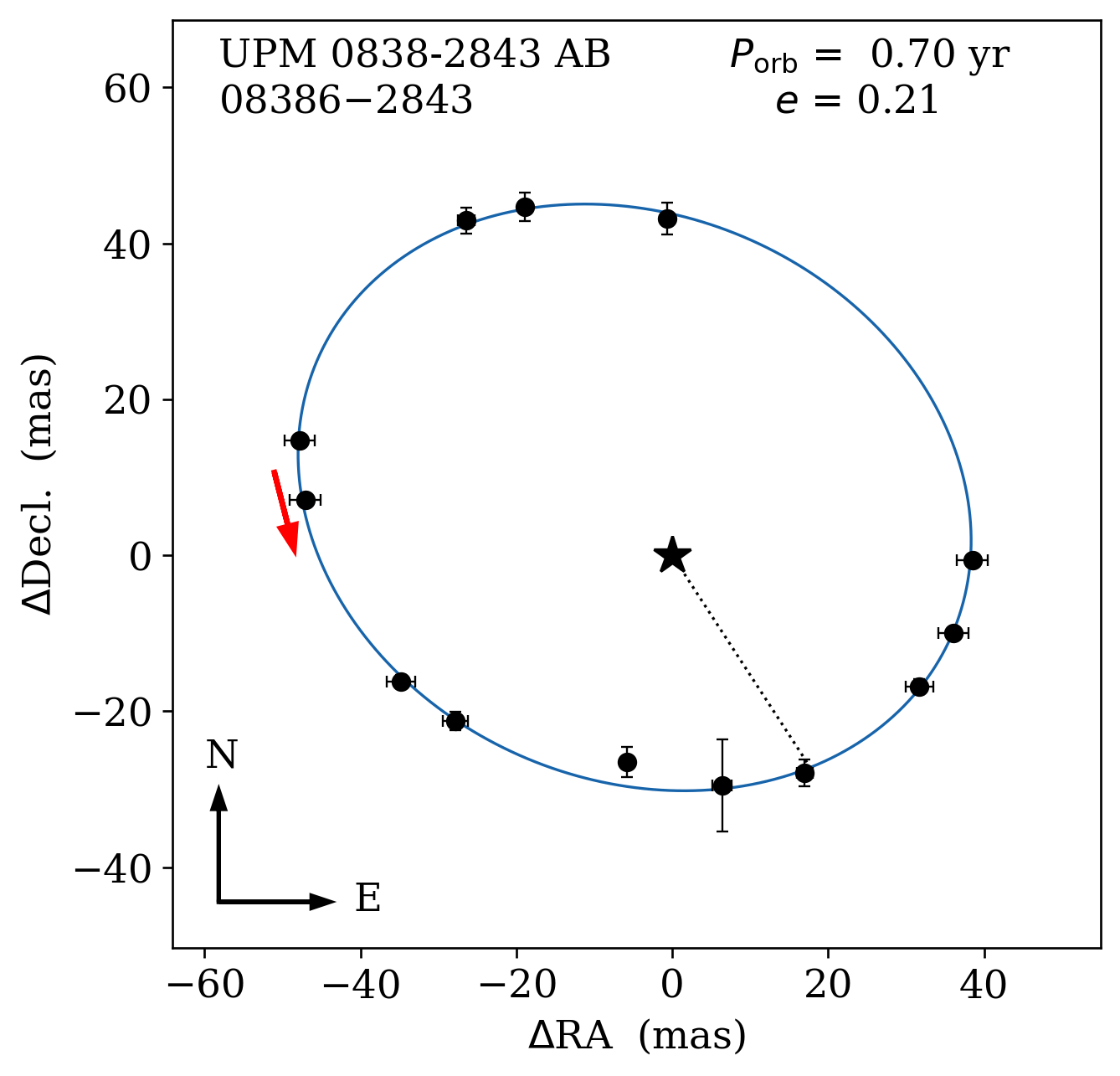}
\includegraphics[scale=0.4]{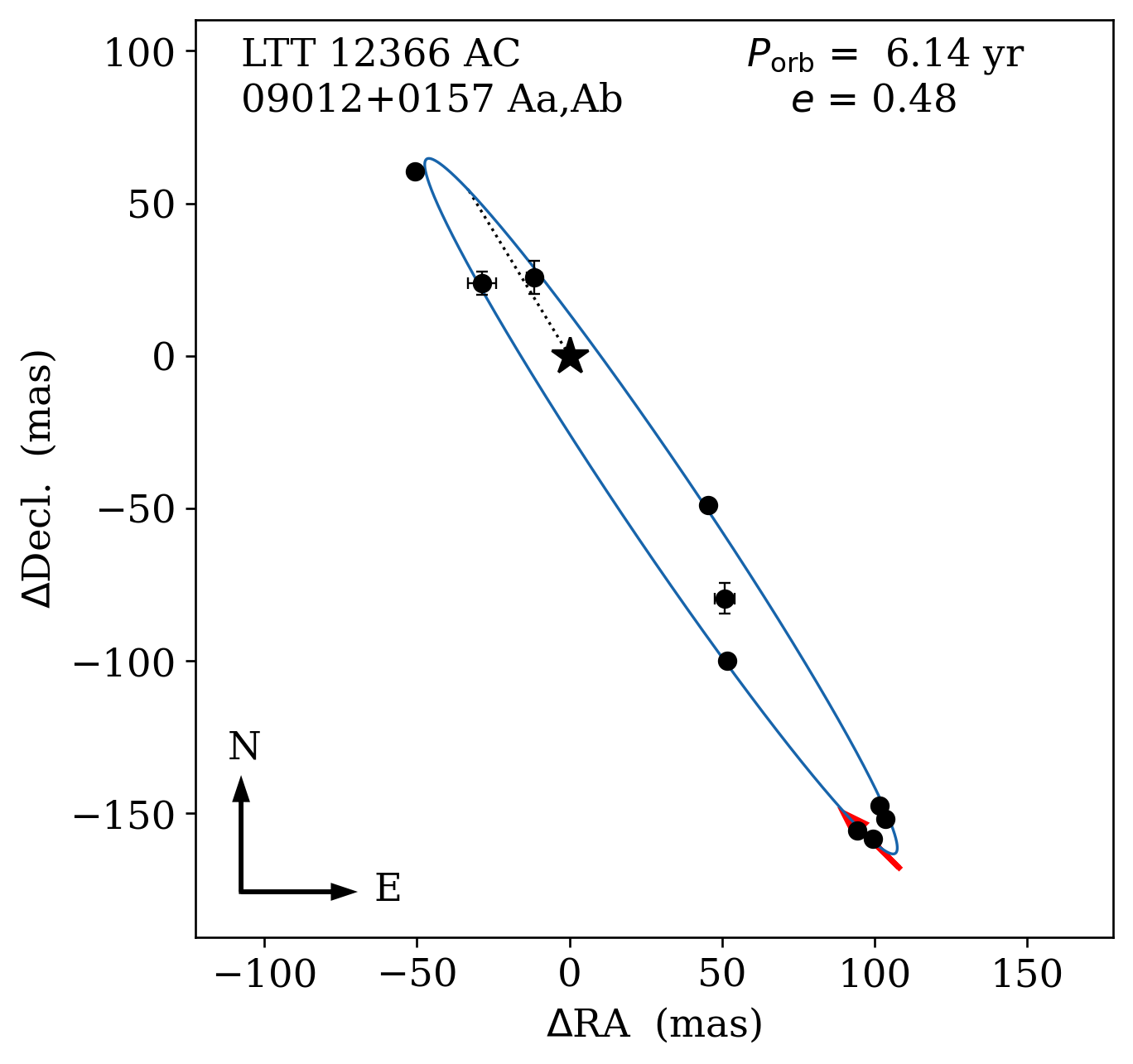}
\includegraphics[scale=0.4]{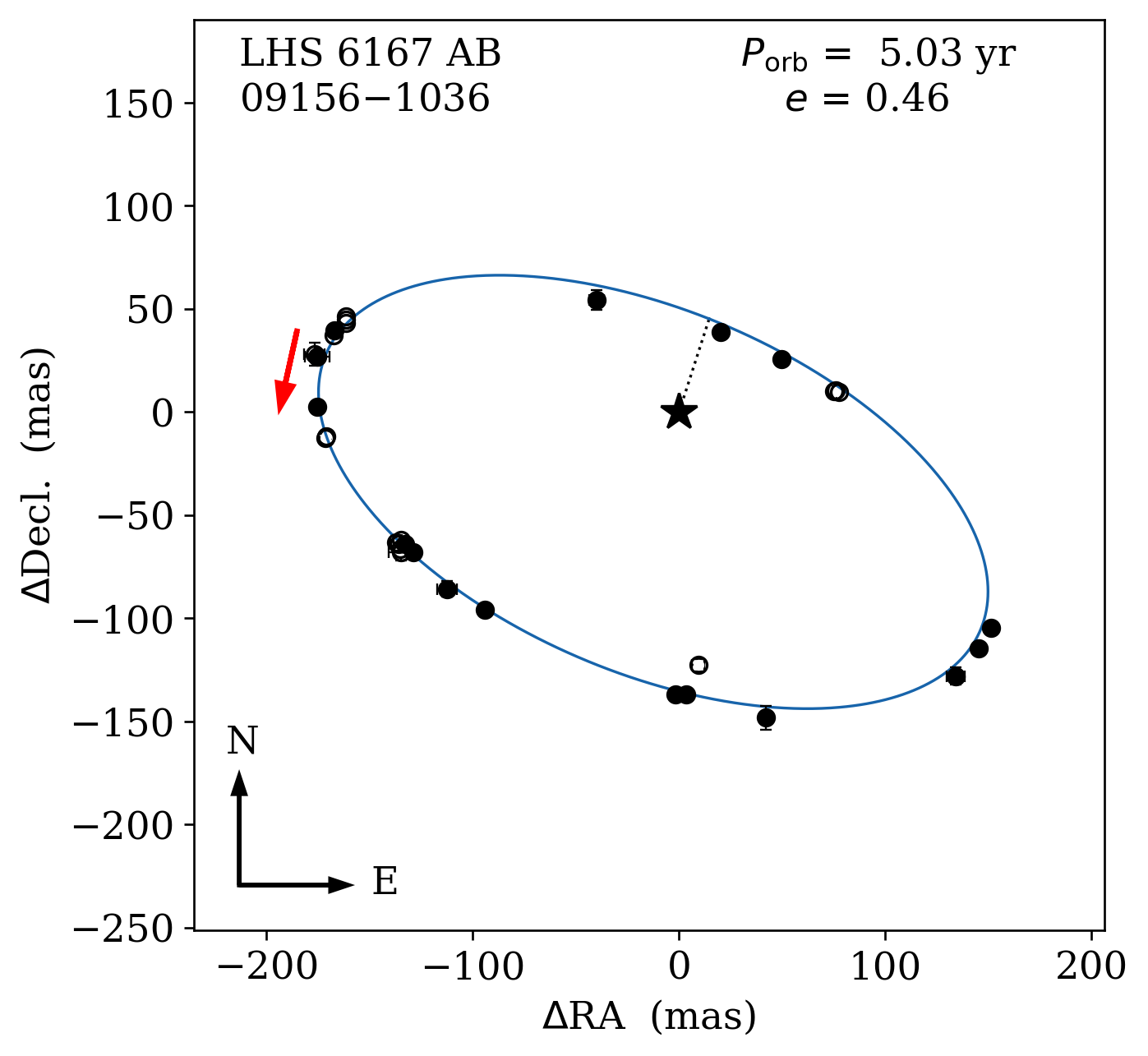}
\includegraphics[scale=0.4]{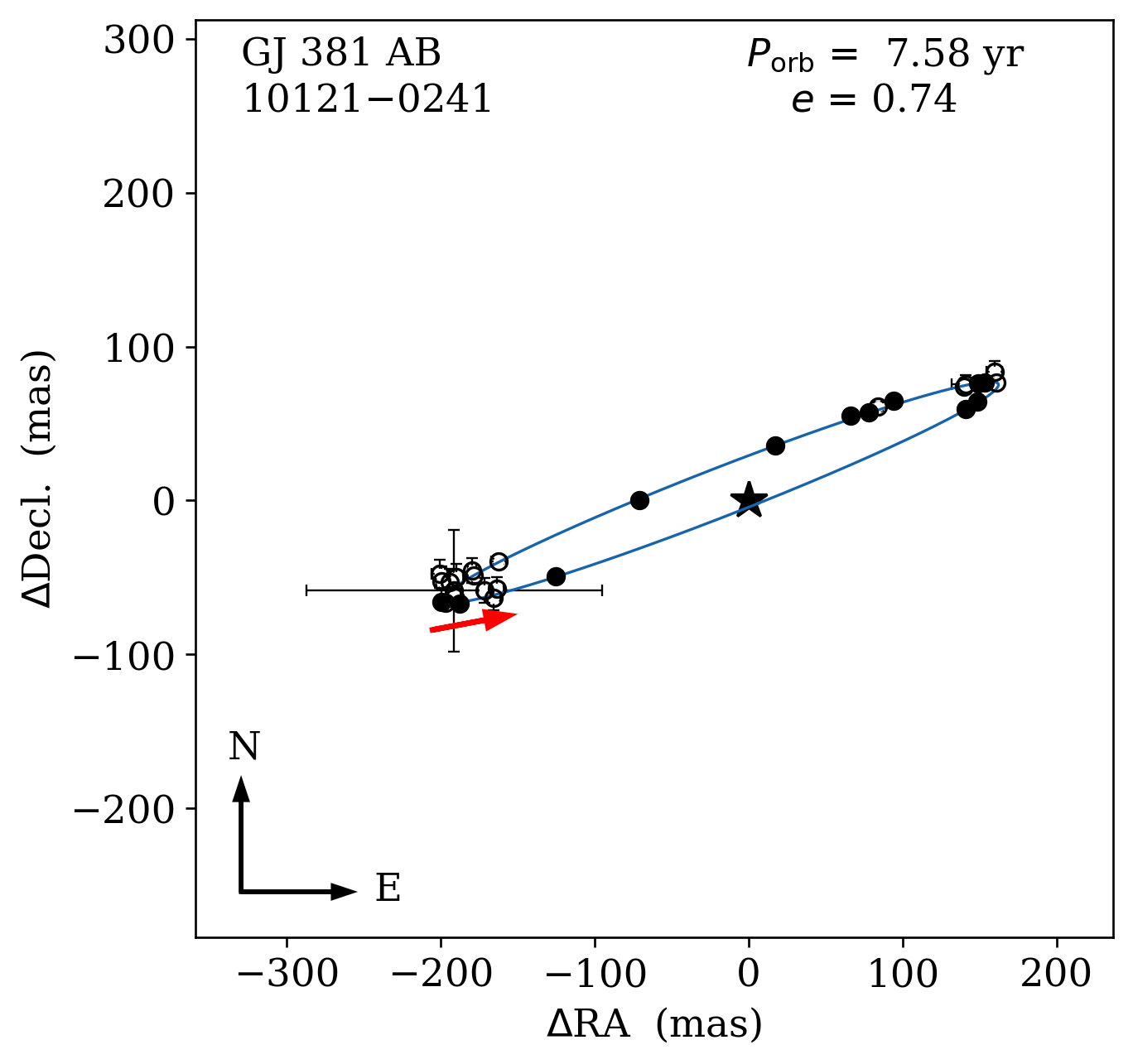}
\caption{Orbits of M dwarf systems determined from the data presented in Table~\ref{tab:results}. One system shown here (GJ~2060~AB) also incorporated the RV data in Table~\ref{tab:RVdata}; its corresponding RV fit is shown in Figure~\ref{fig:RVorbits}. 
For additional details, see the caption to Figure~\ref{fig:orbits}.
\label{fig:orbits2}}
\end{figure}

\begin{figure} \centering
\includegraphics[scale=0.4]{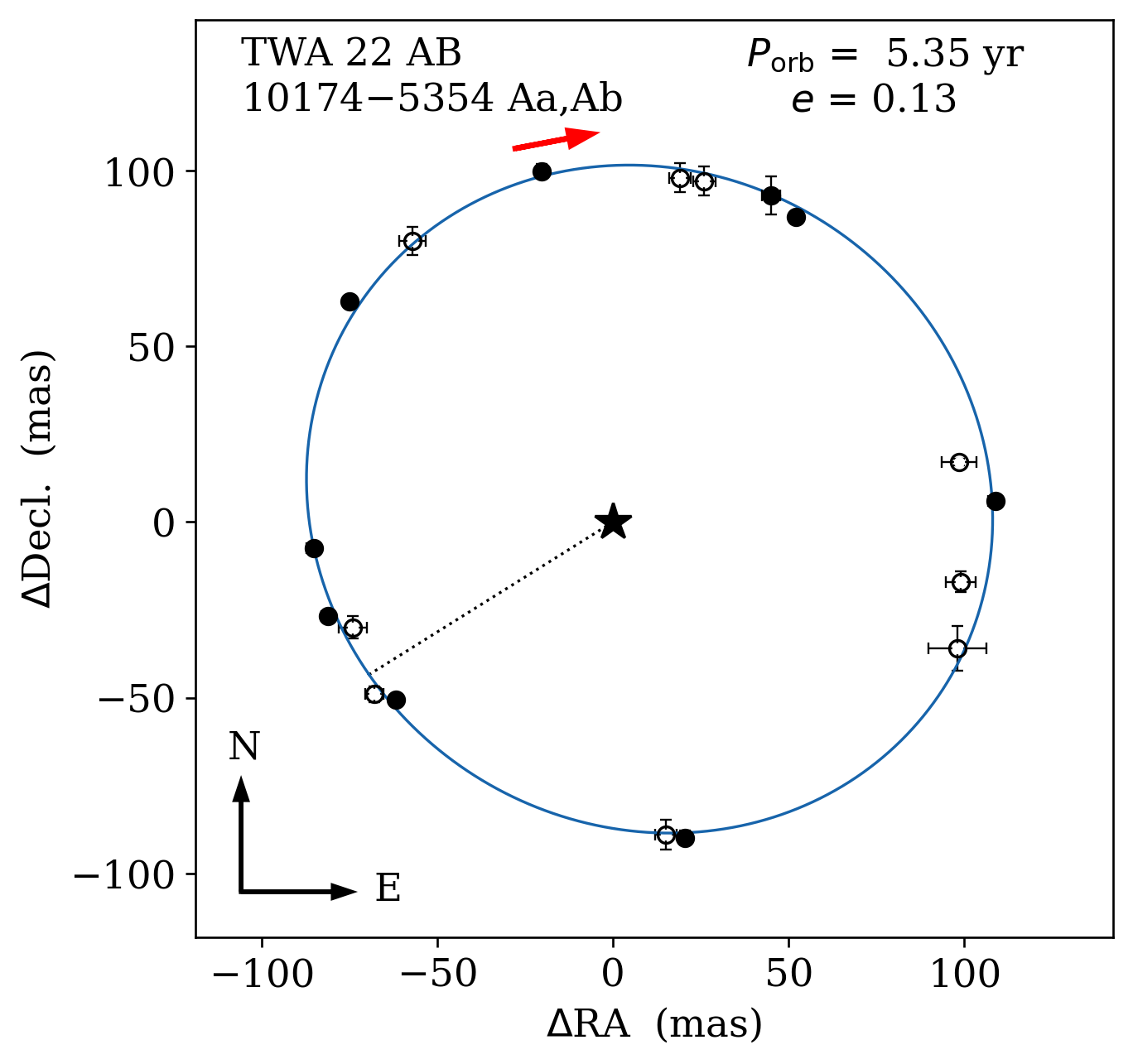}
\includegraphics[scale=0.4]{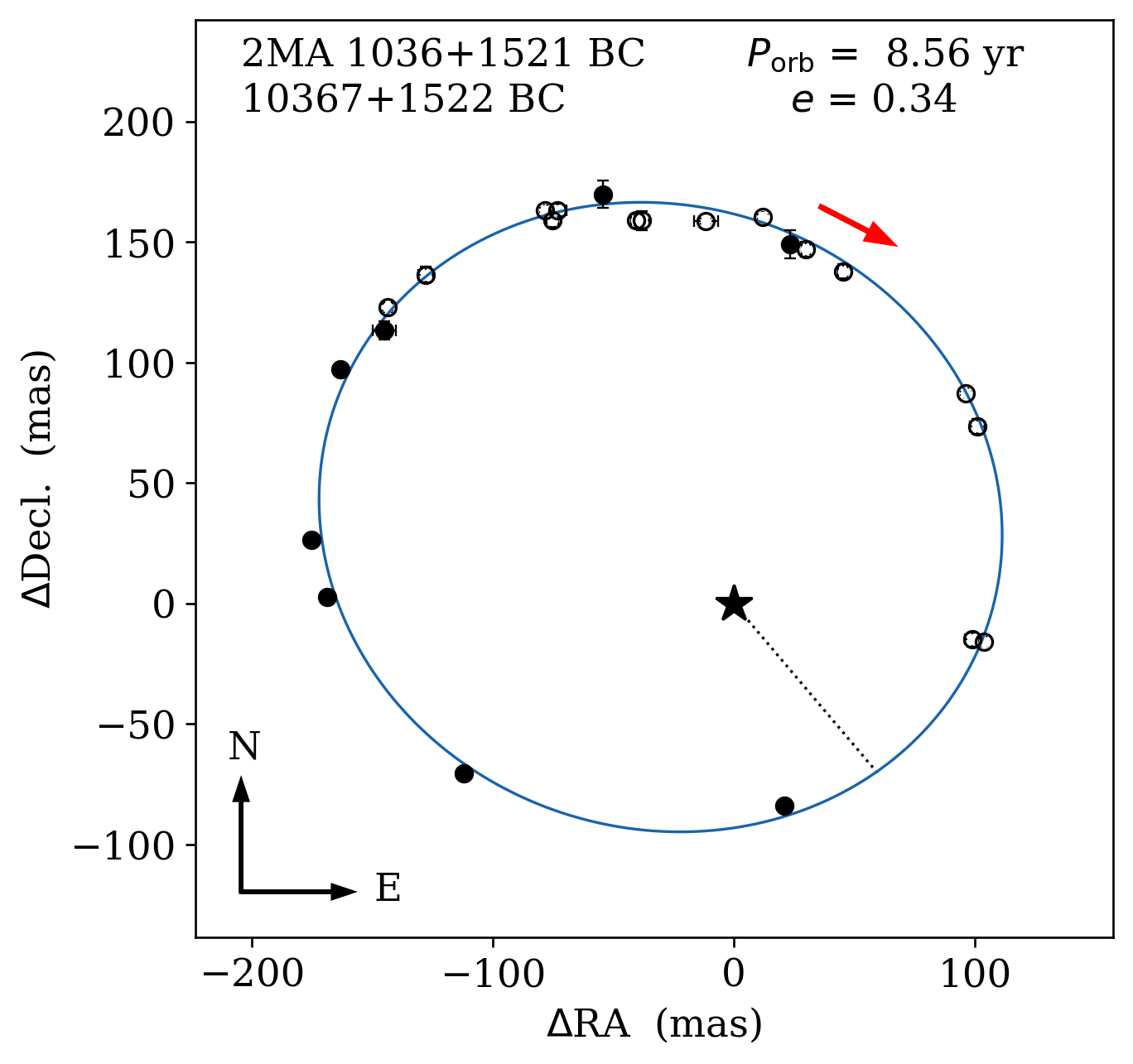}
\includegraphics[scale=0.4]{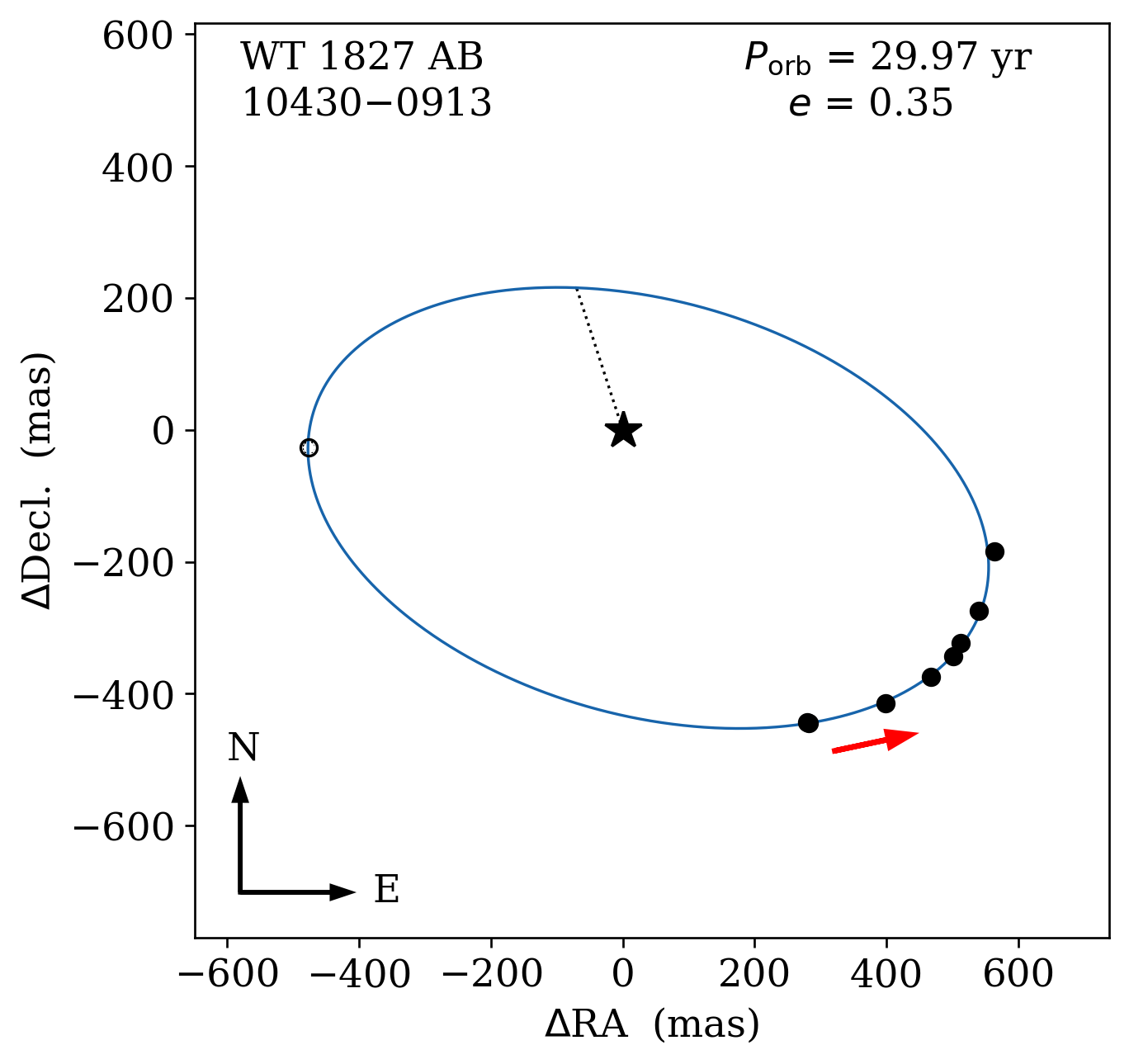}
\includegraphics[scale=0.4]{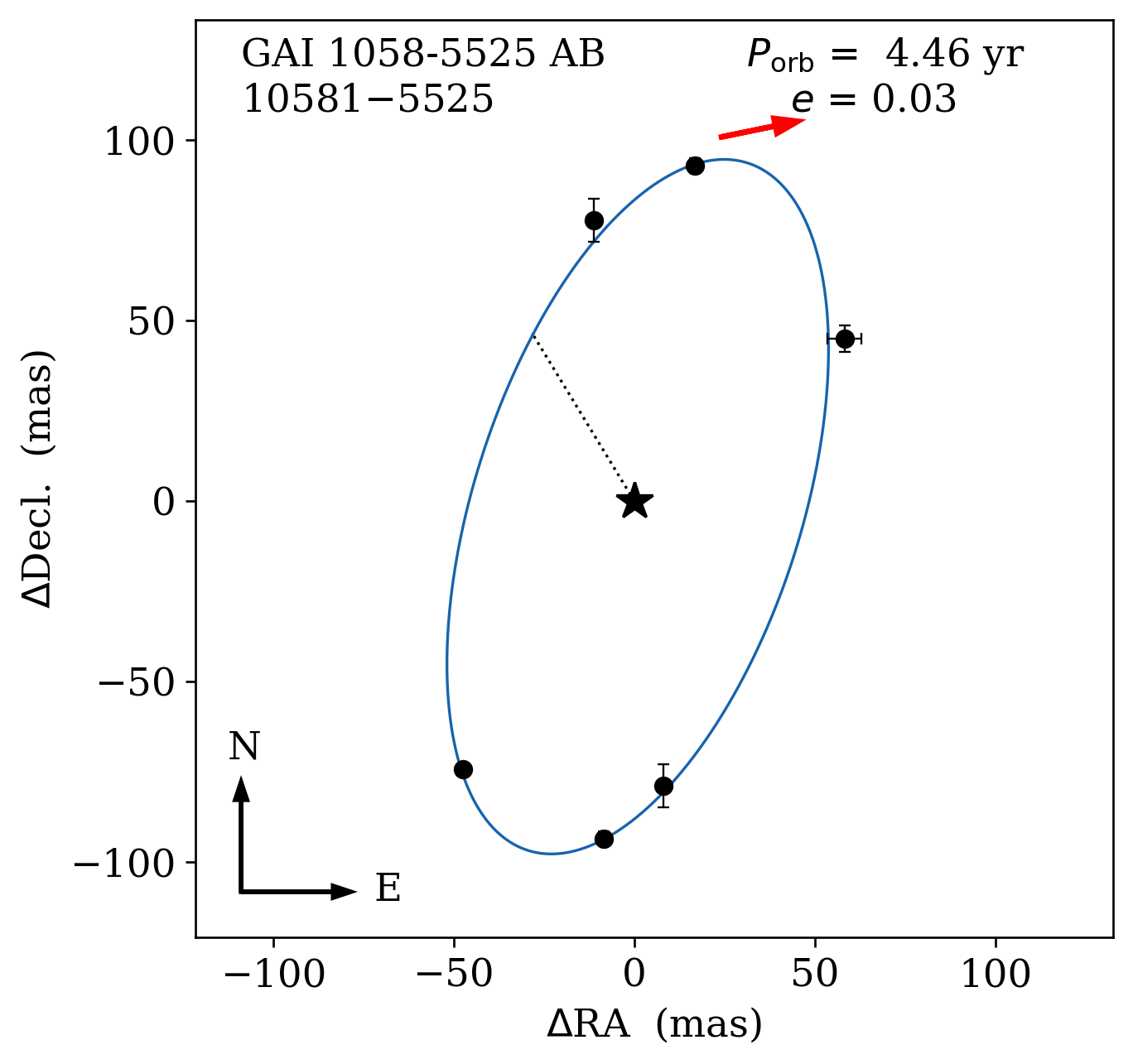}
\includegraphics[scale=0.4]{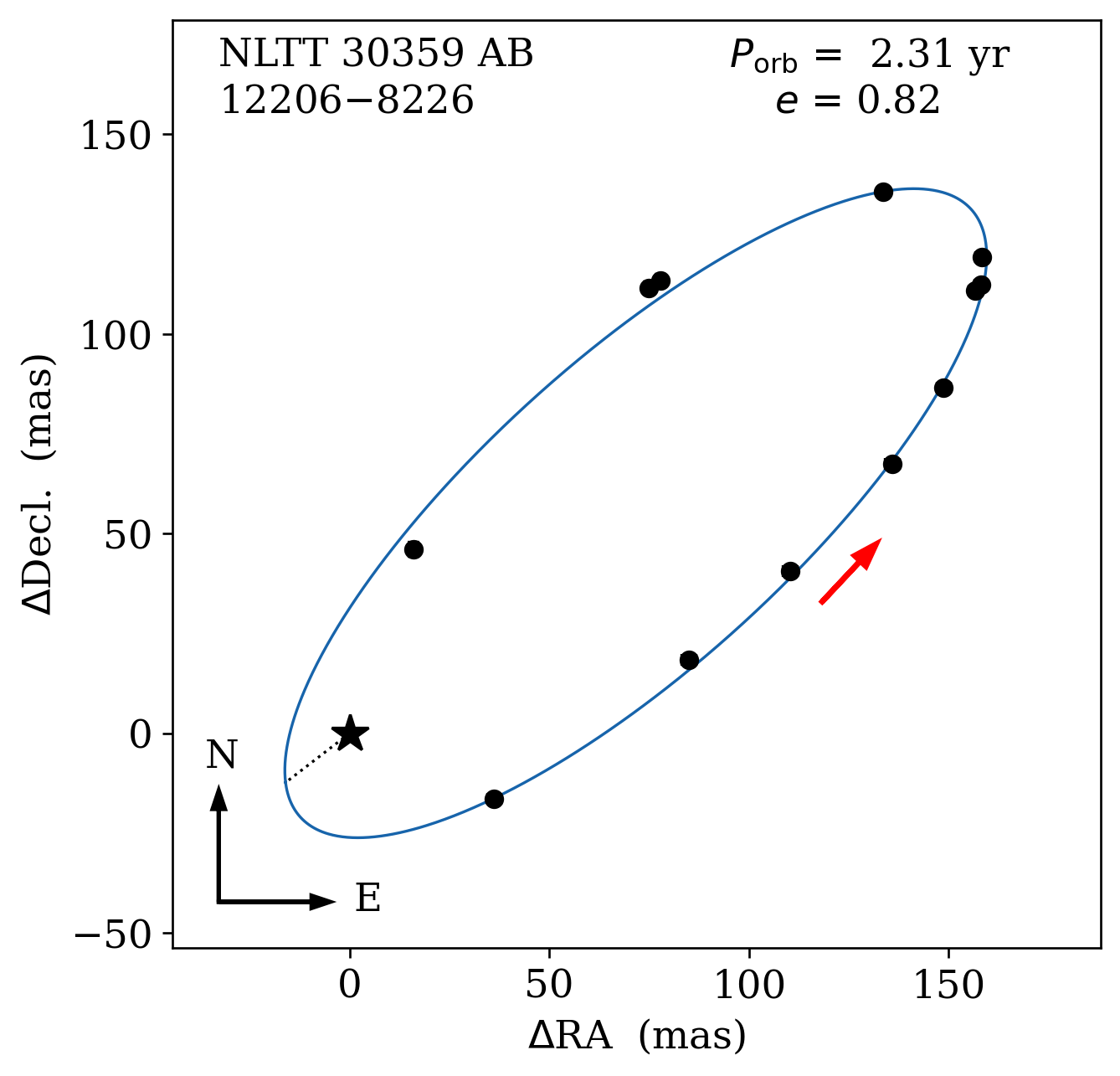}
\includegraphics[scale=0.4]{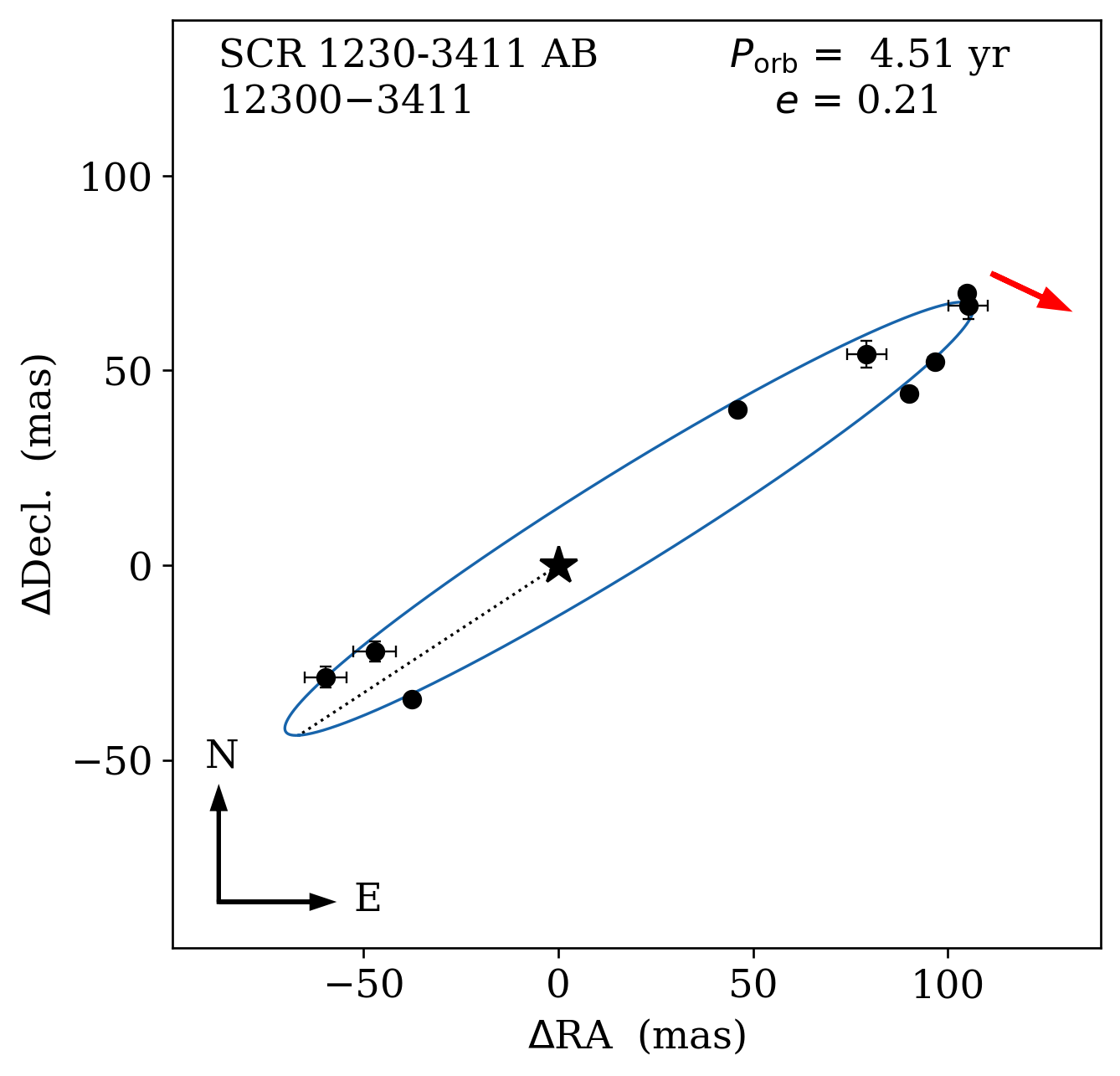}
\includegraphics[scale=0.4]{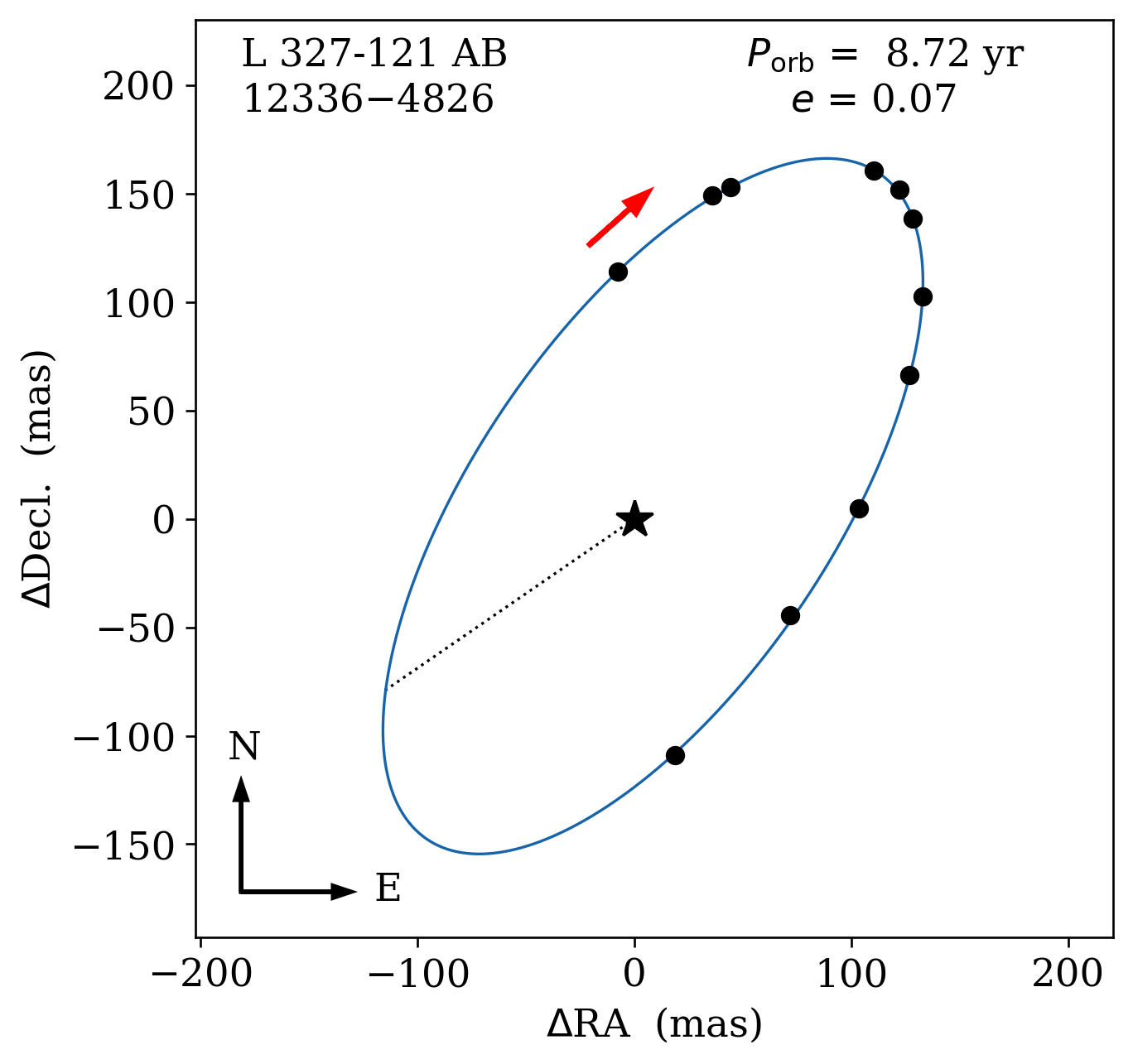}
\includegraphics[scale=0.4]{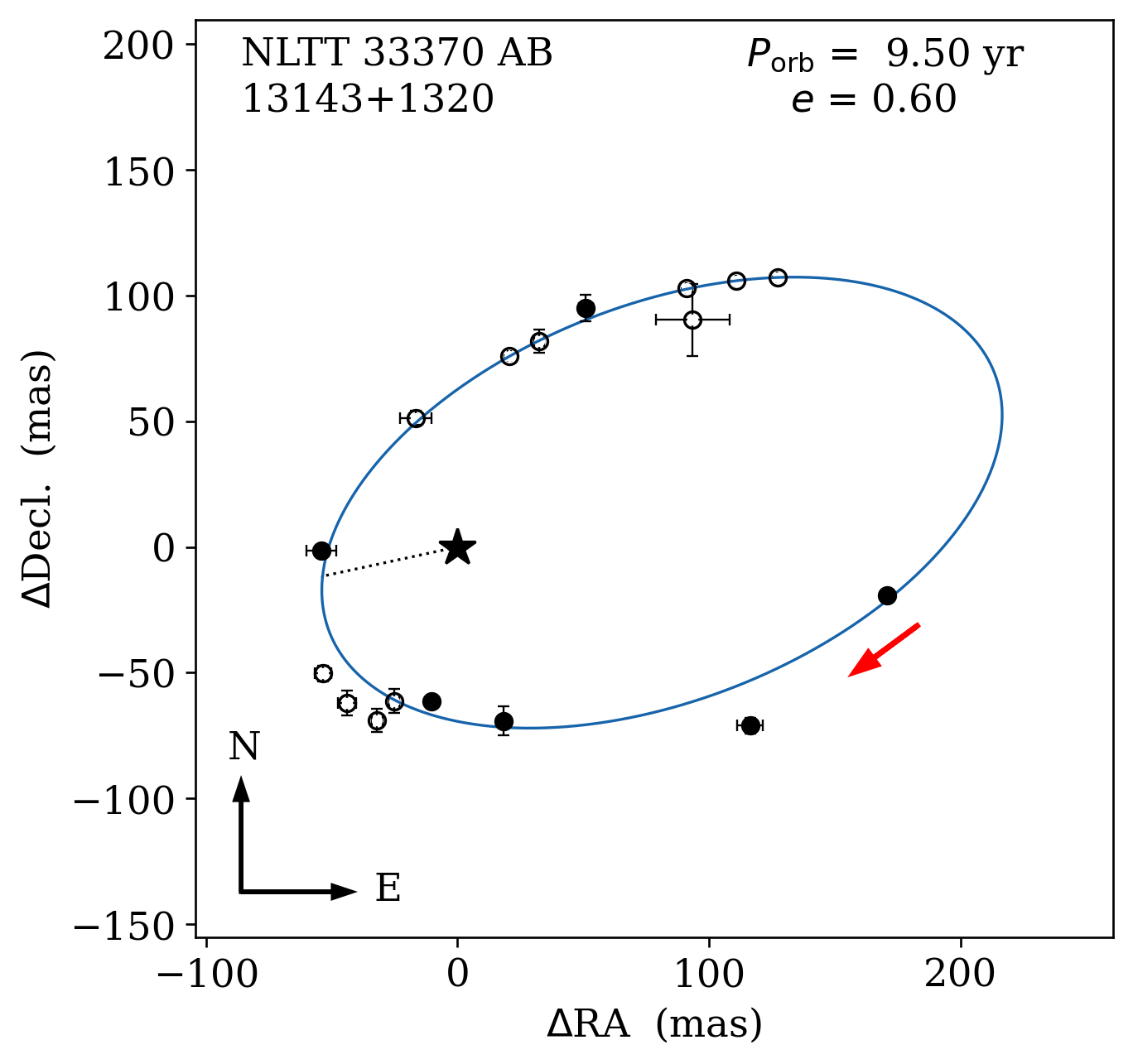}
\includegraphics[scale=0.4]{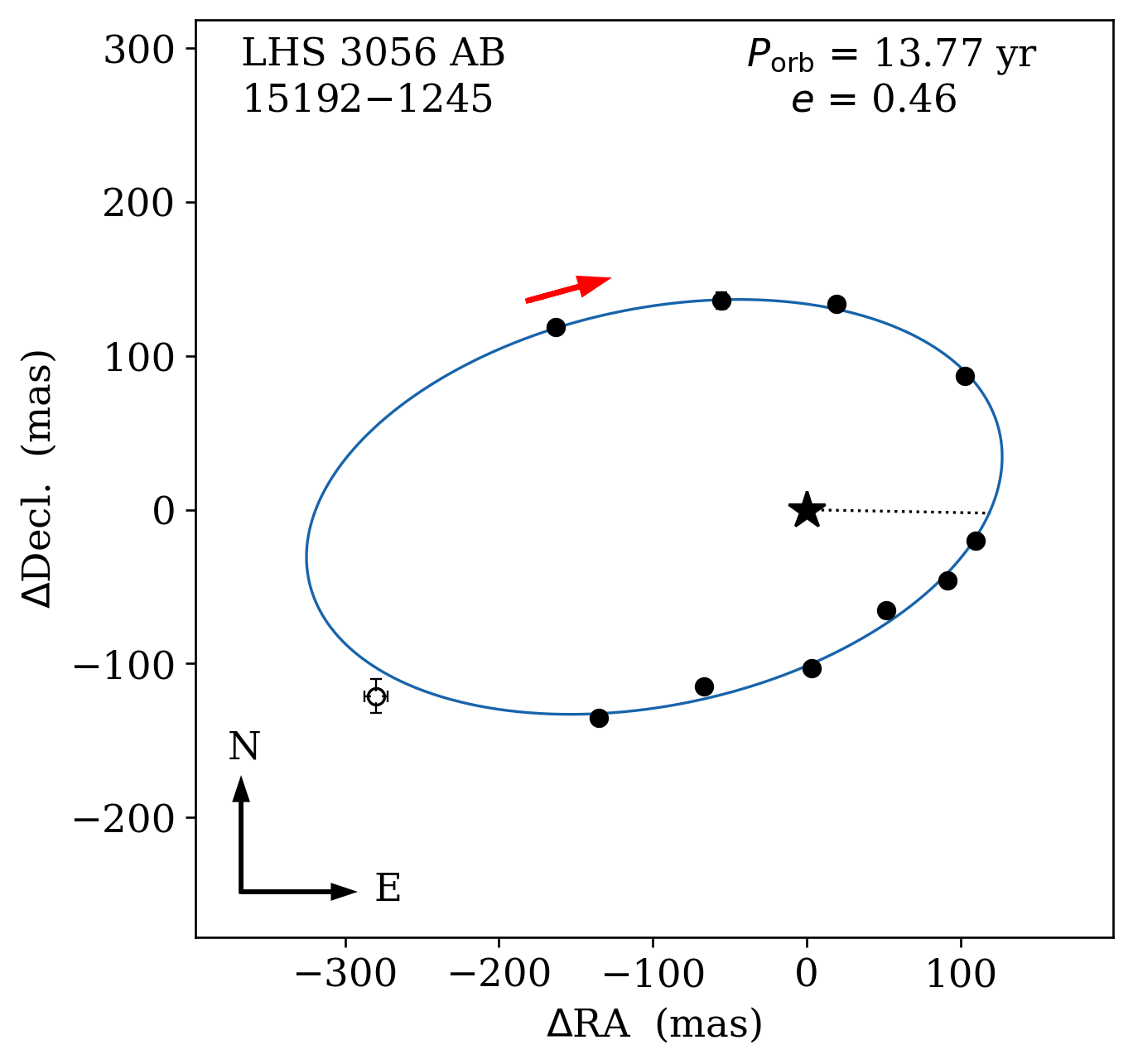}
\includegraphics[scale=0.4]{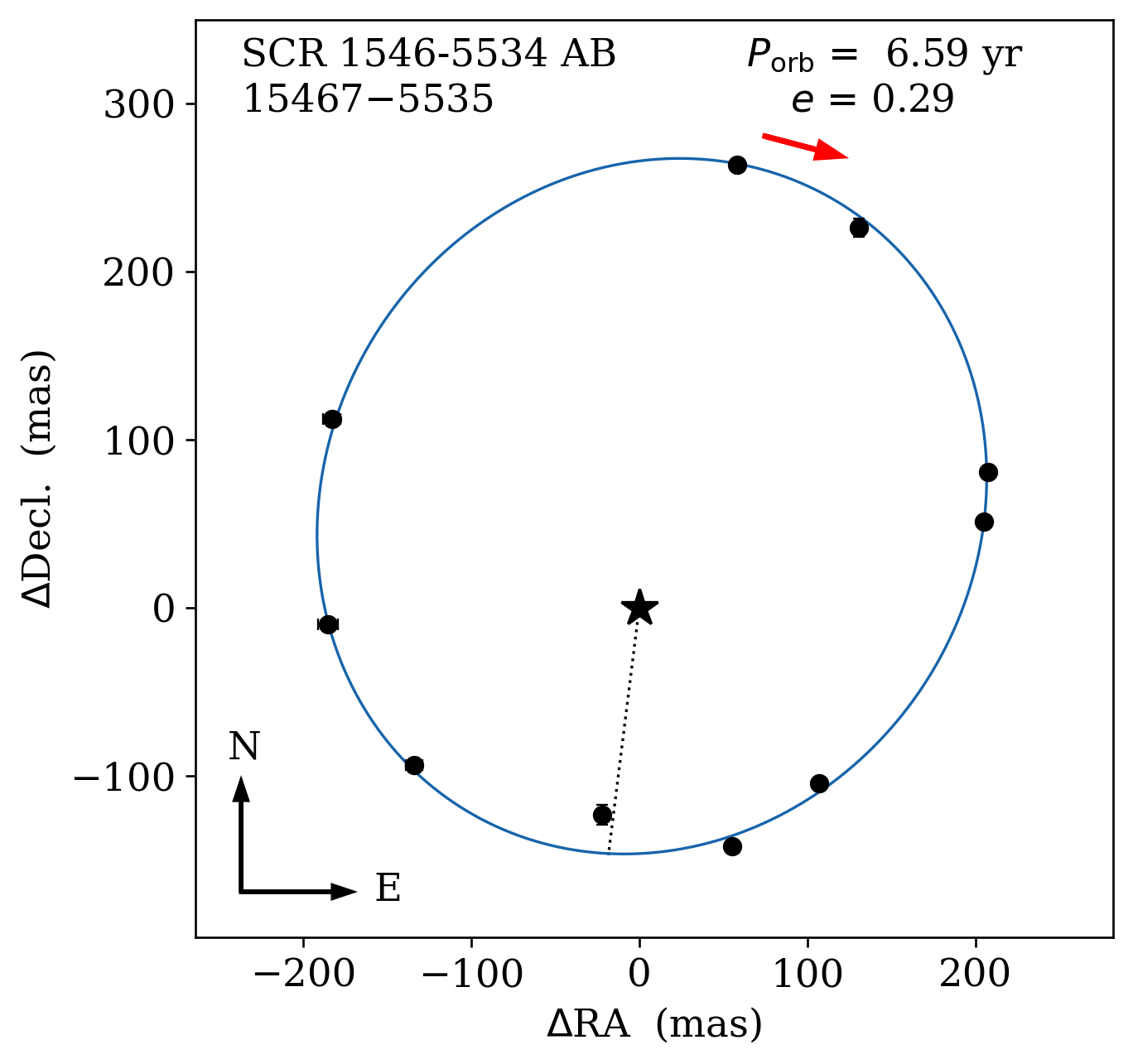}
\includegraphics[scale=0.4]{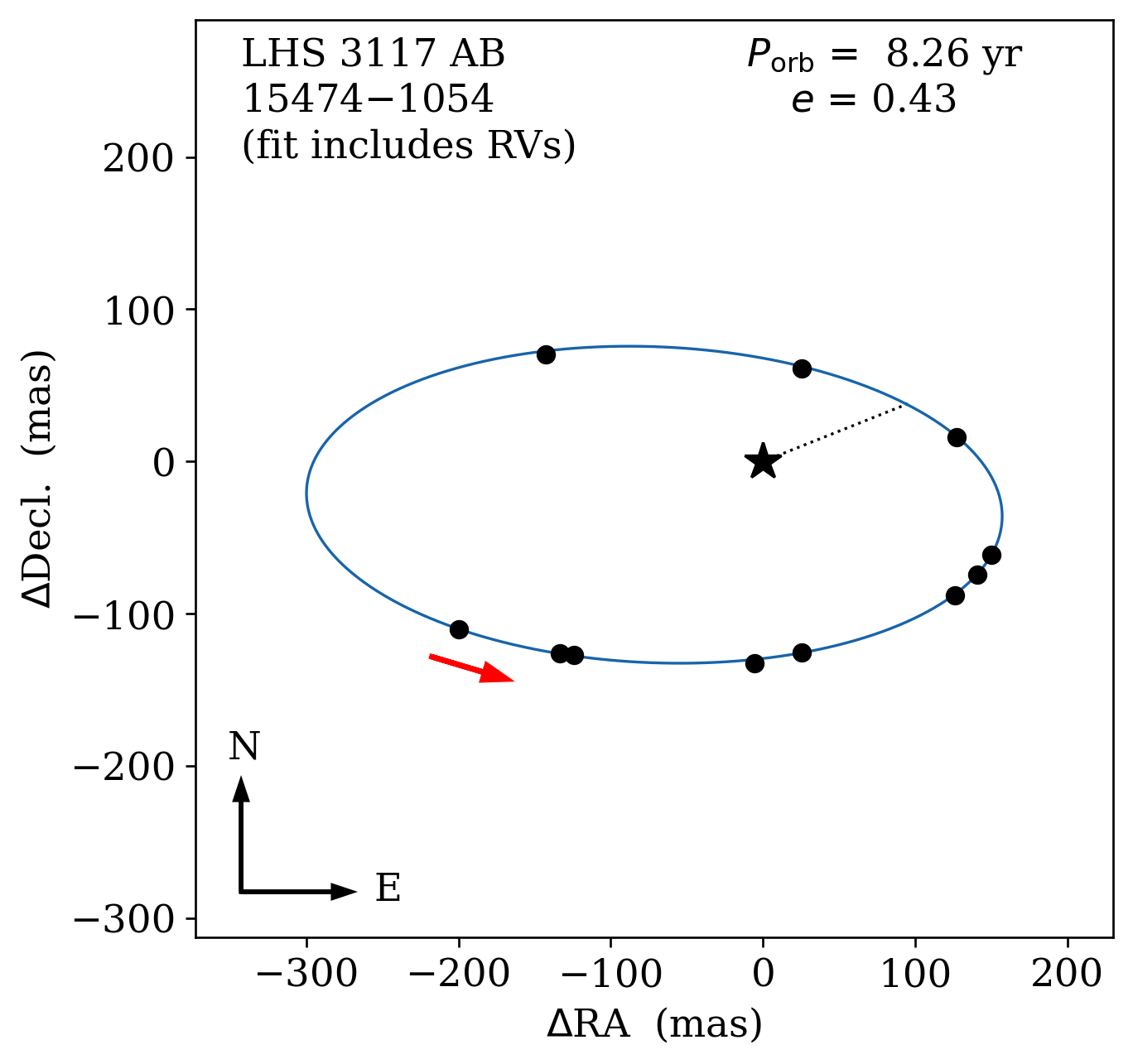}
\includegraphics[scale=0.4]{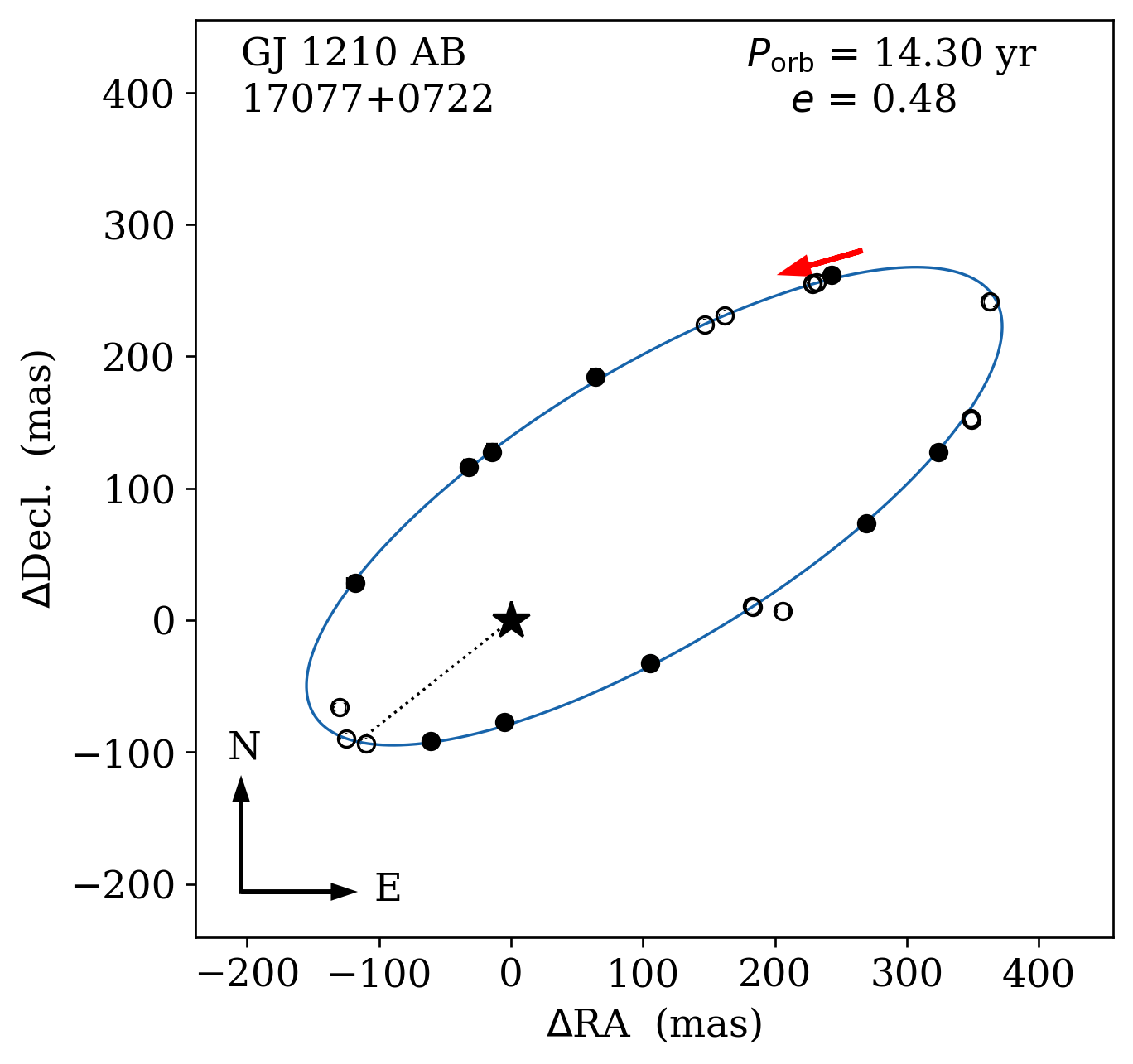}
\caption{Orbits of M dwarf systems determined from the data presented in Table~\ref{tab:results}. One system shown here (LHS~3117~AB) also incorporated the RV data in Table~\ref{tab:RVdata}; its corresponding RV fit is shown in Figure~\ref{fig:RVorbits}. For additional details, see the caption to Figure~\ref{fig:orbits}.
\label{fig:orbits3}}
\end{figure}

\begin{figure} \centering
\includegraphics[scale=0.4]{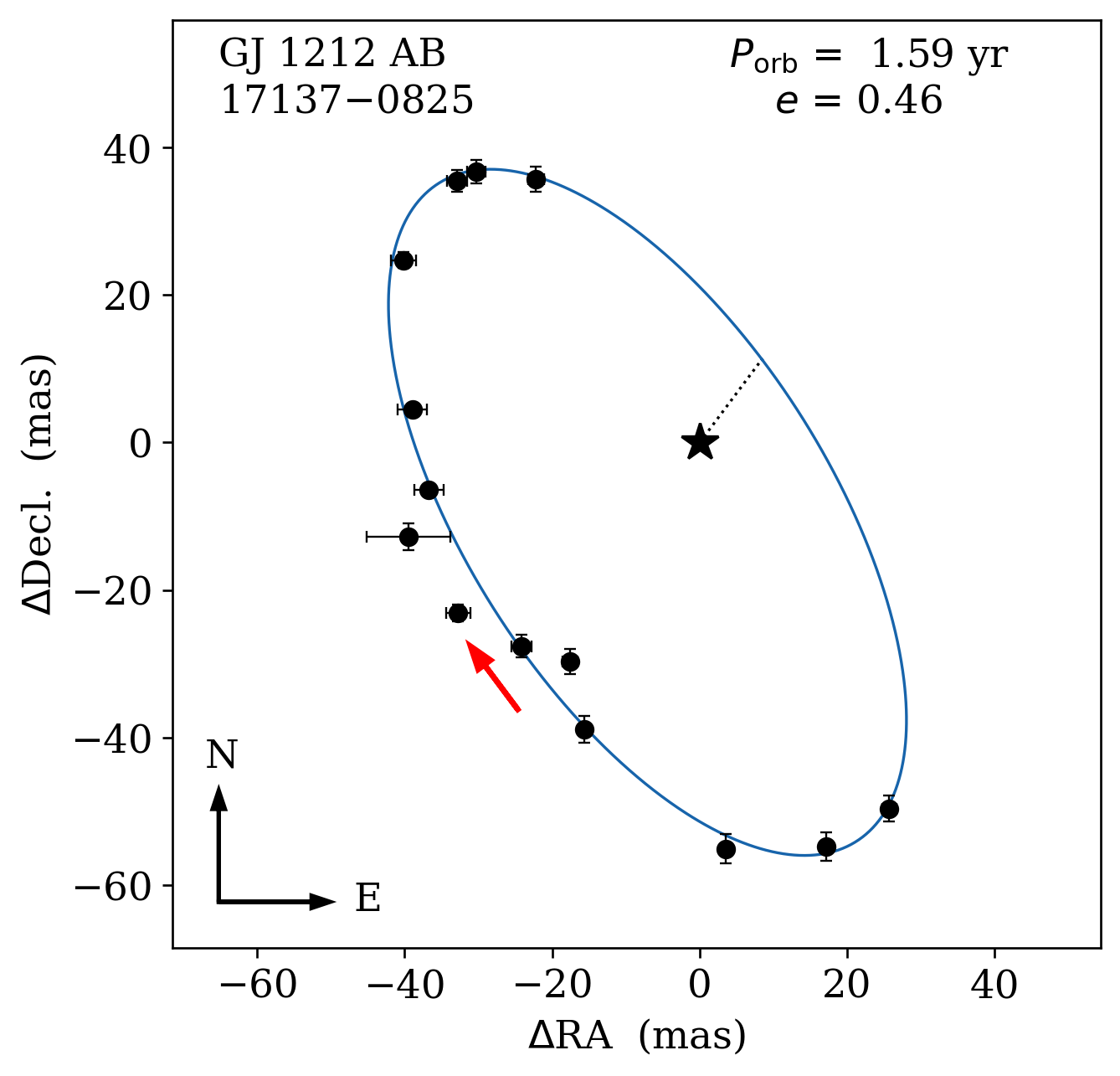}
\includegraphics[scale=0.4]{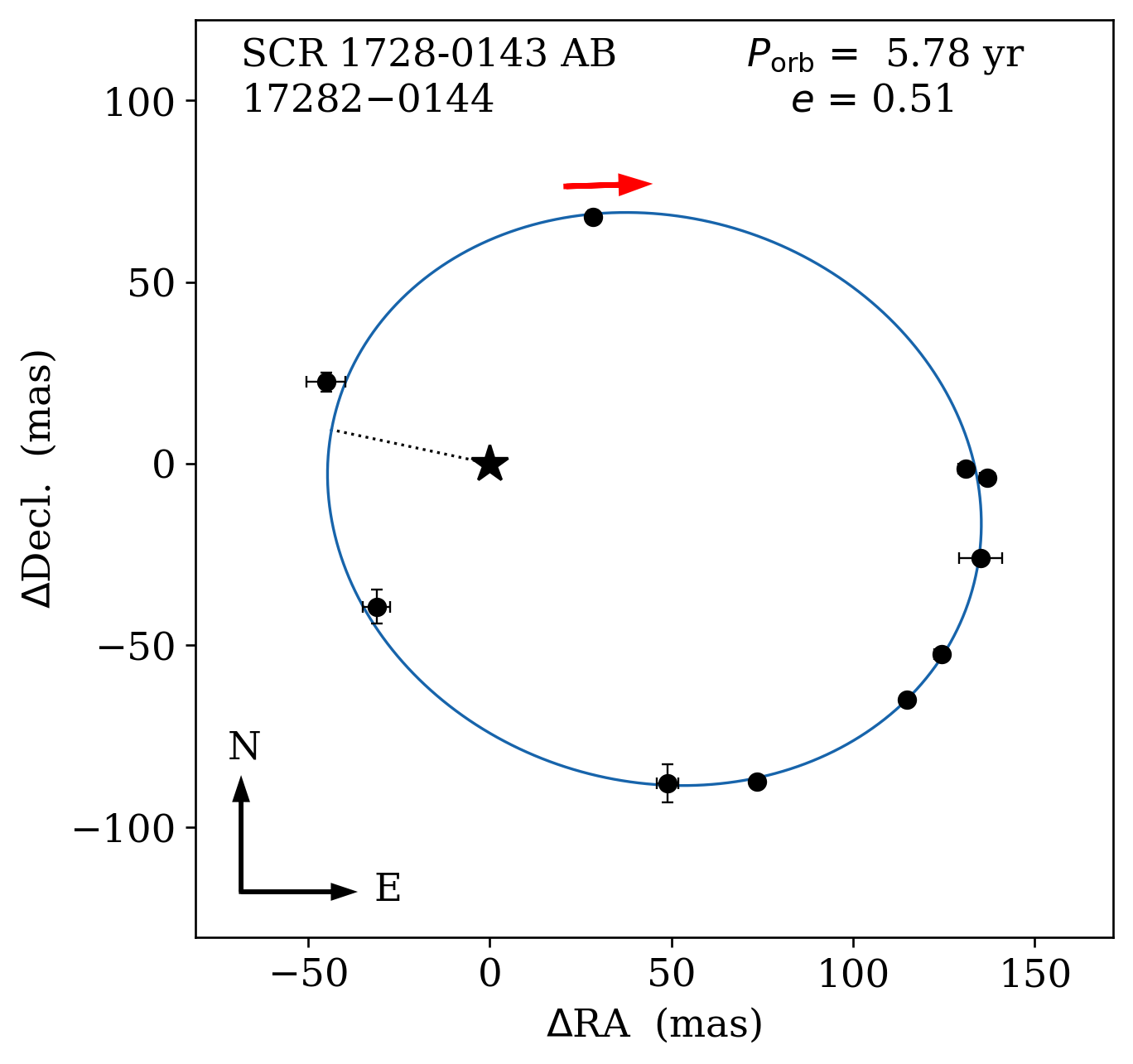}
\includegraphics[scale=0.4]{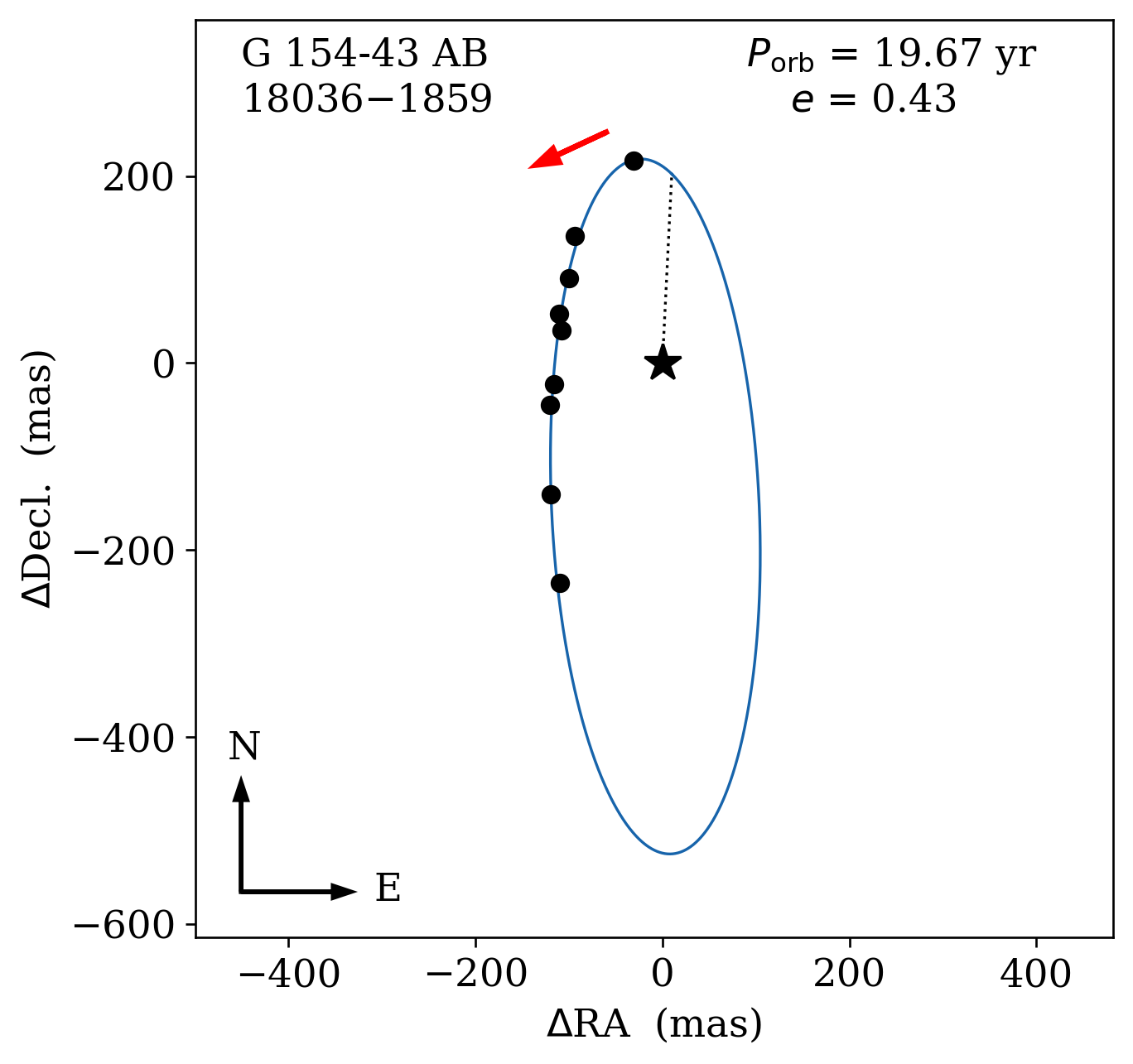}
\includegraphics[scale=0.4]{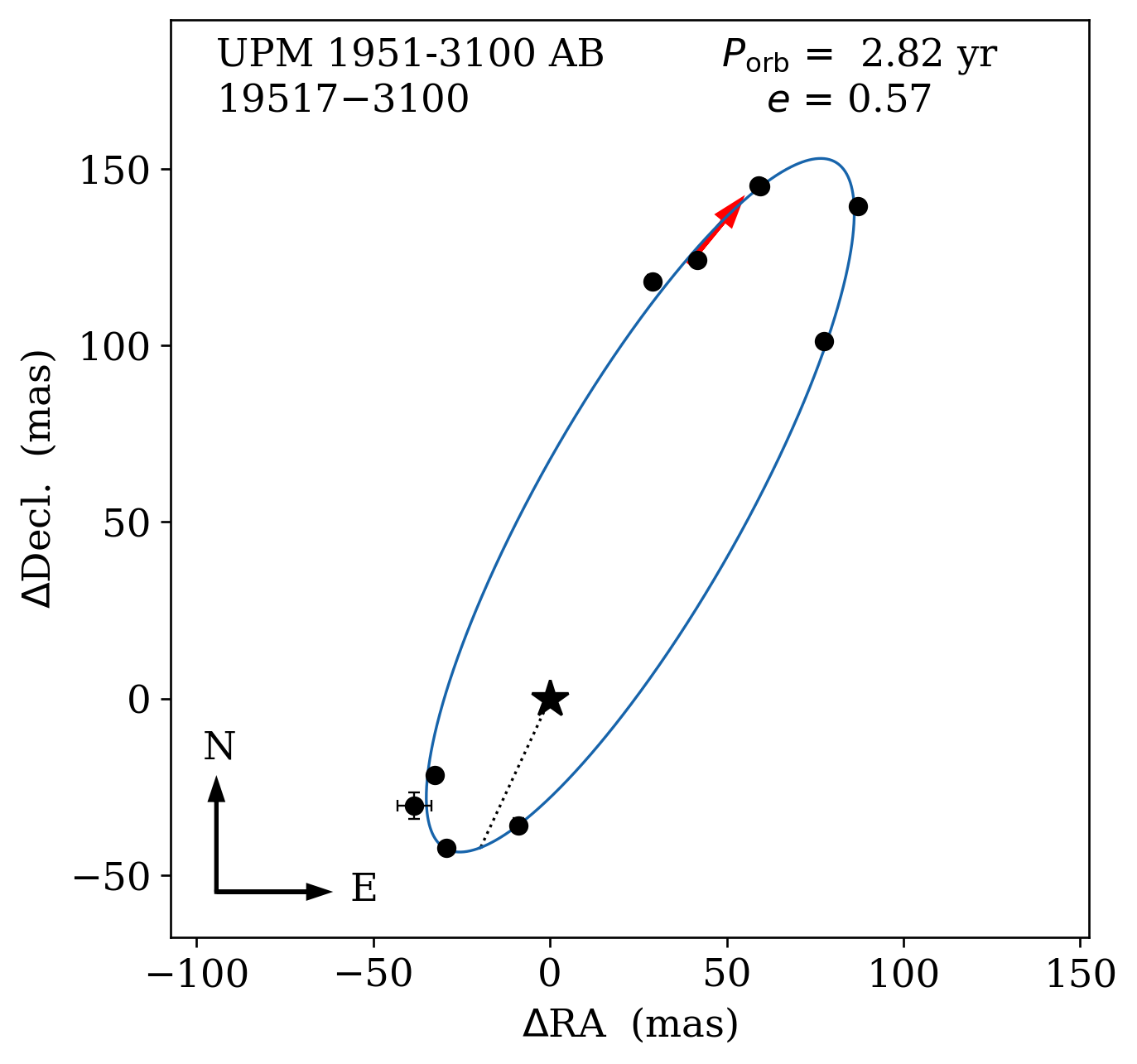}
\includegraphics[scale=0.4]{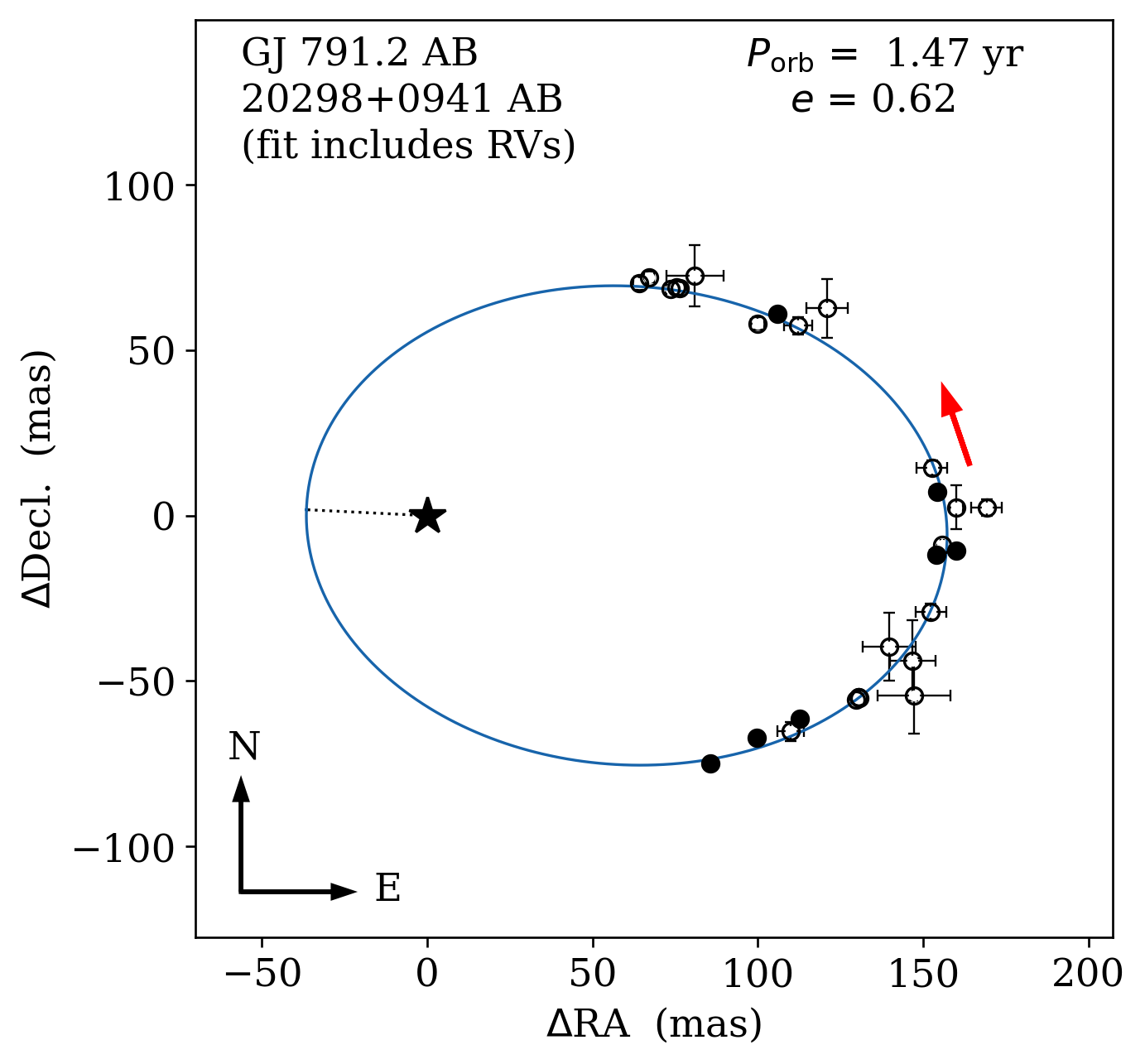}
\includegraphics[scale=0.4]{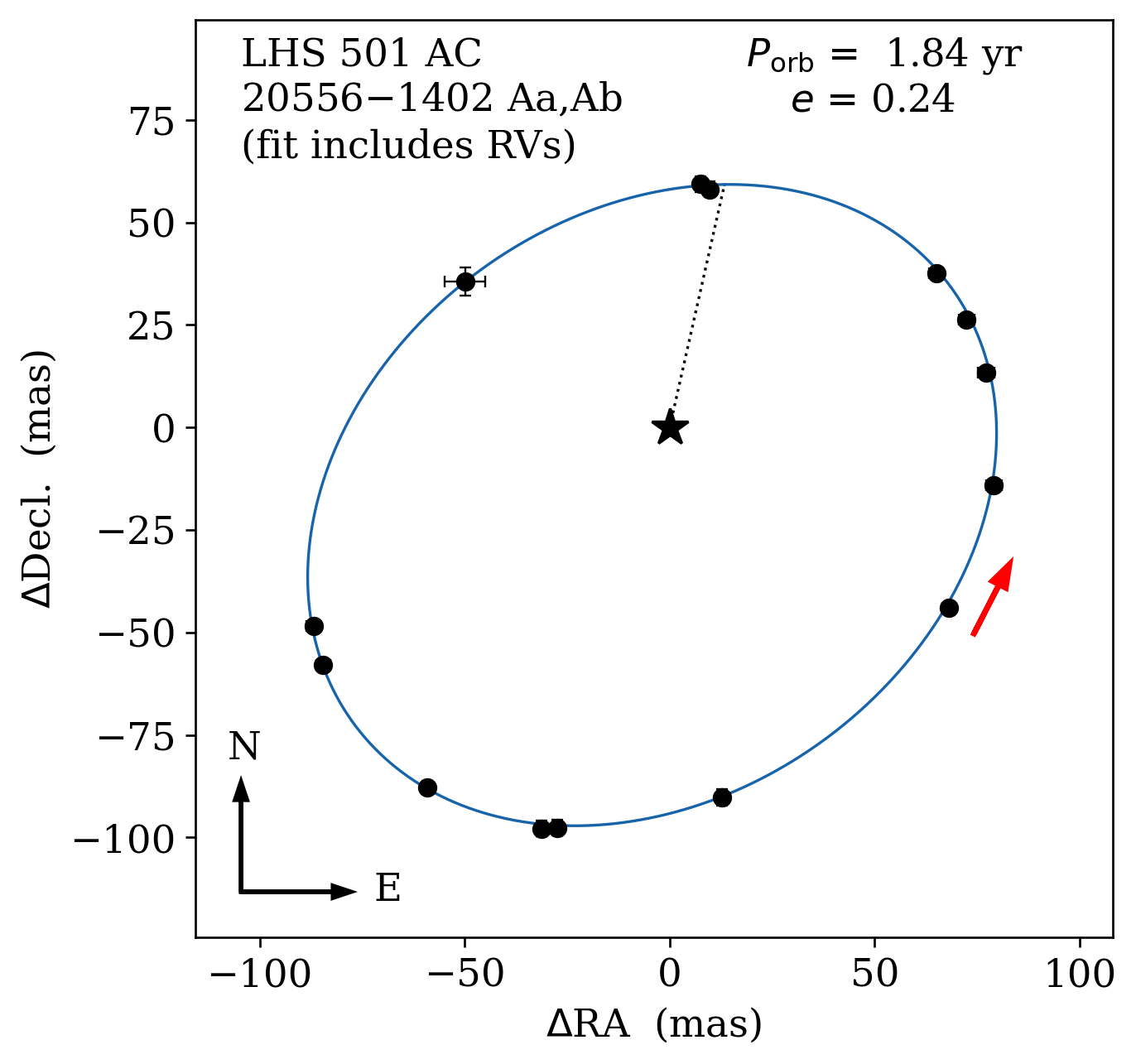}
\includegraphics[scale=0.4]{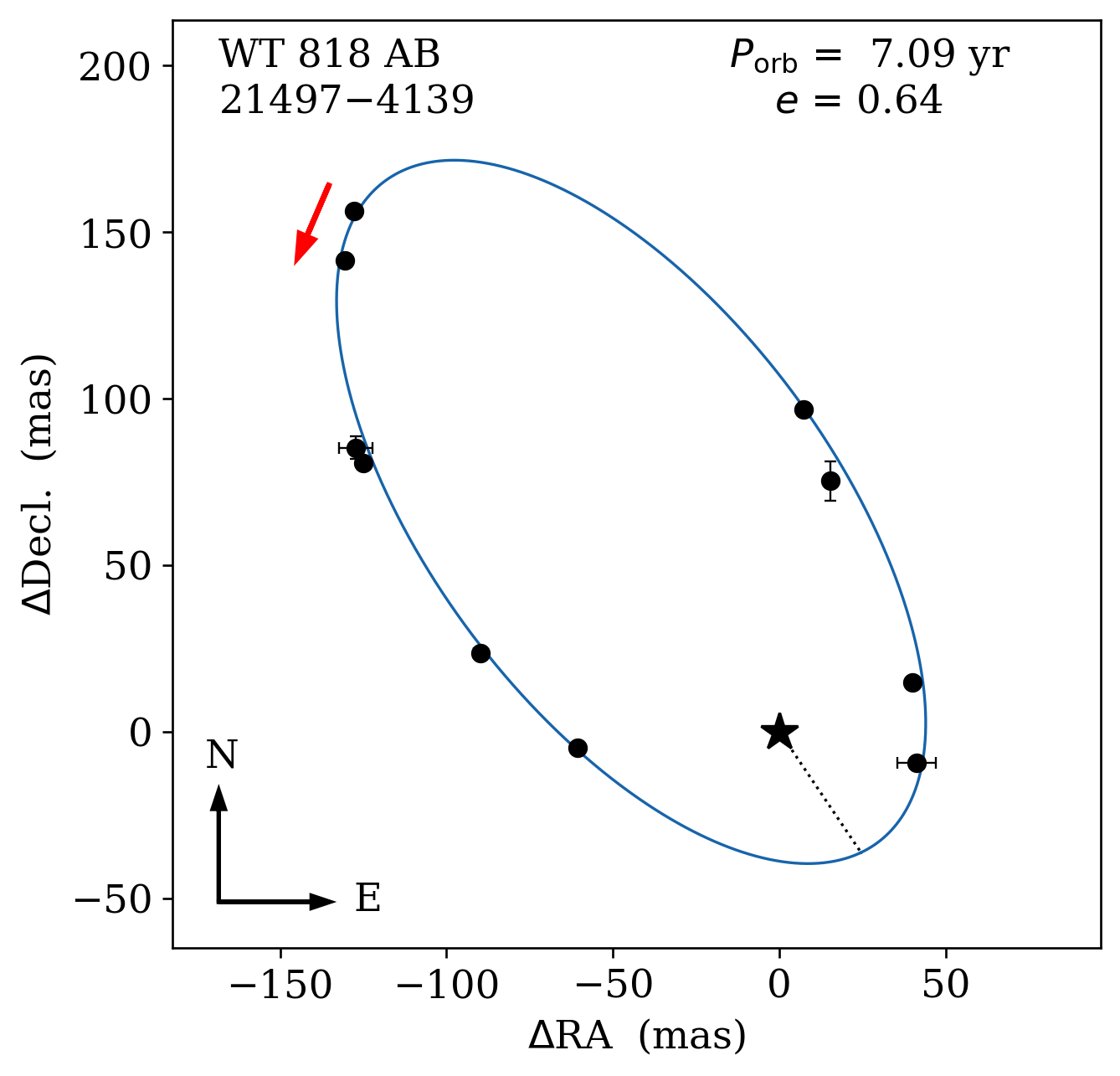}
\includegraphics[scale=0.4]{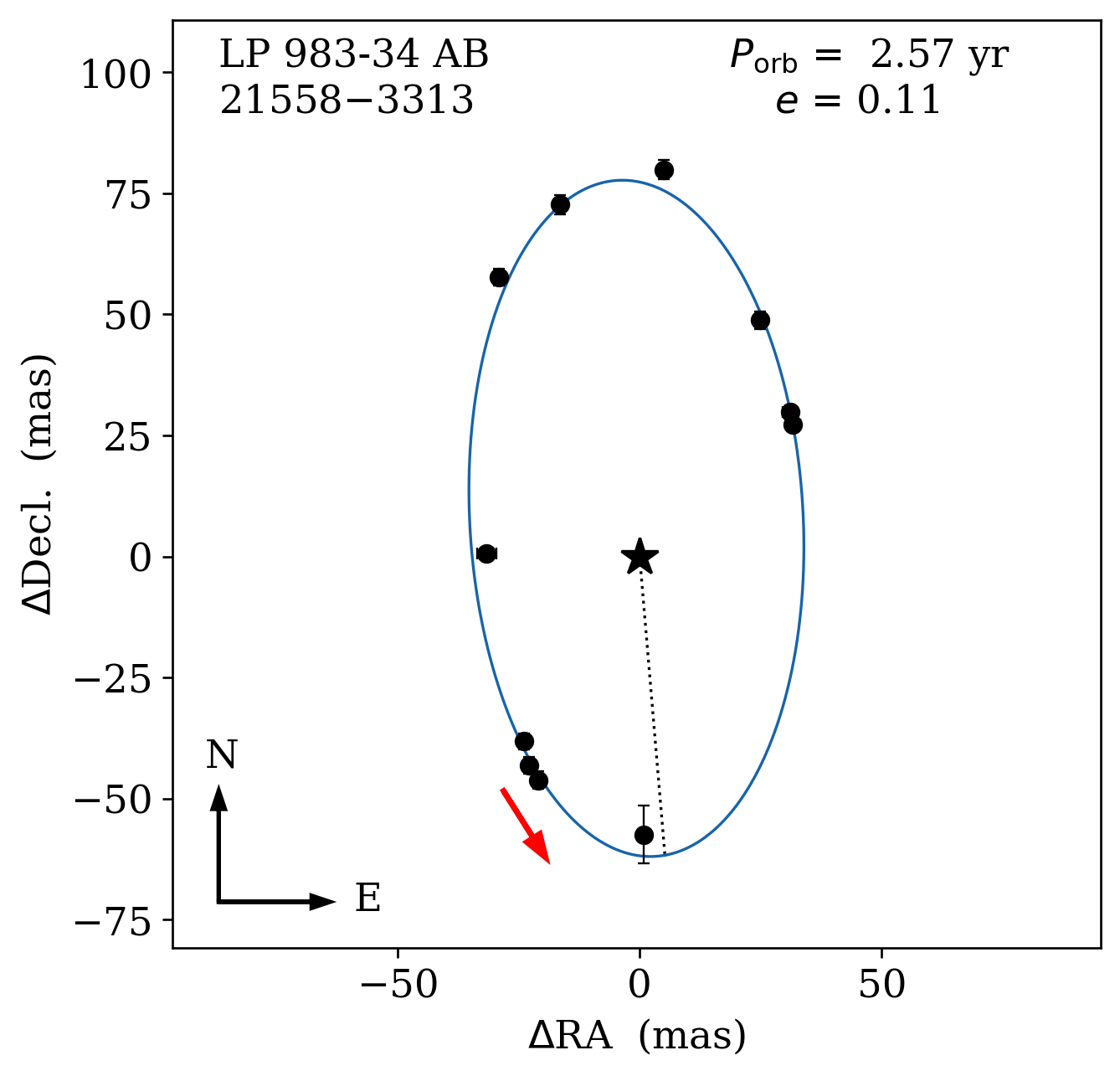}
\includegraphics[scale=0.4]{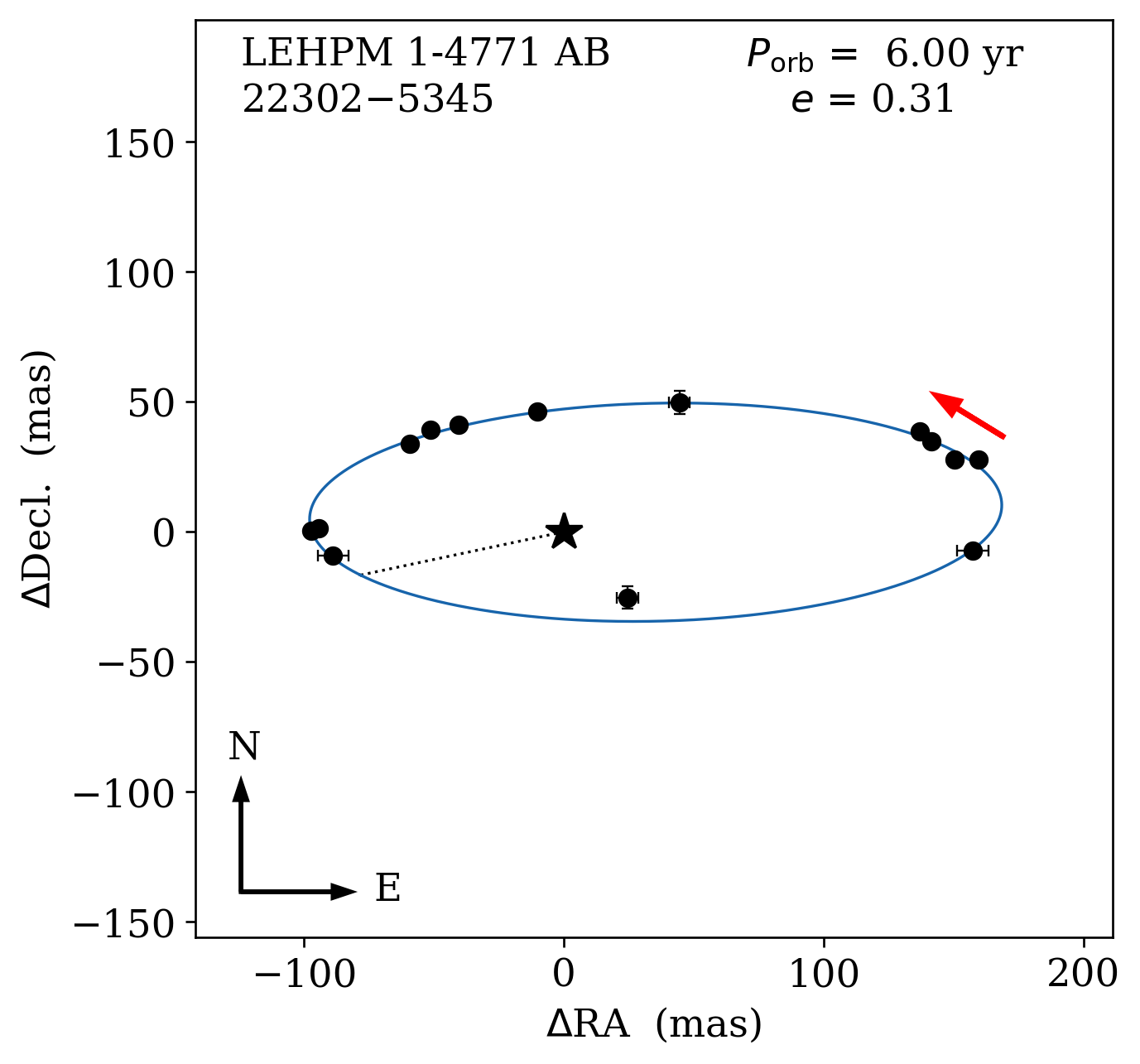}
\includegraphics[scale=0.4]{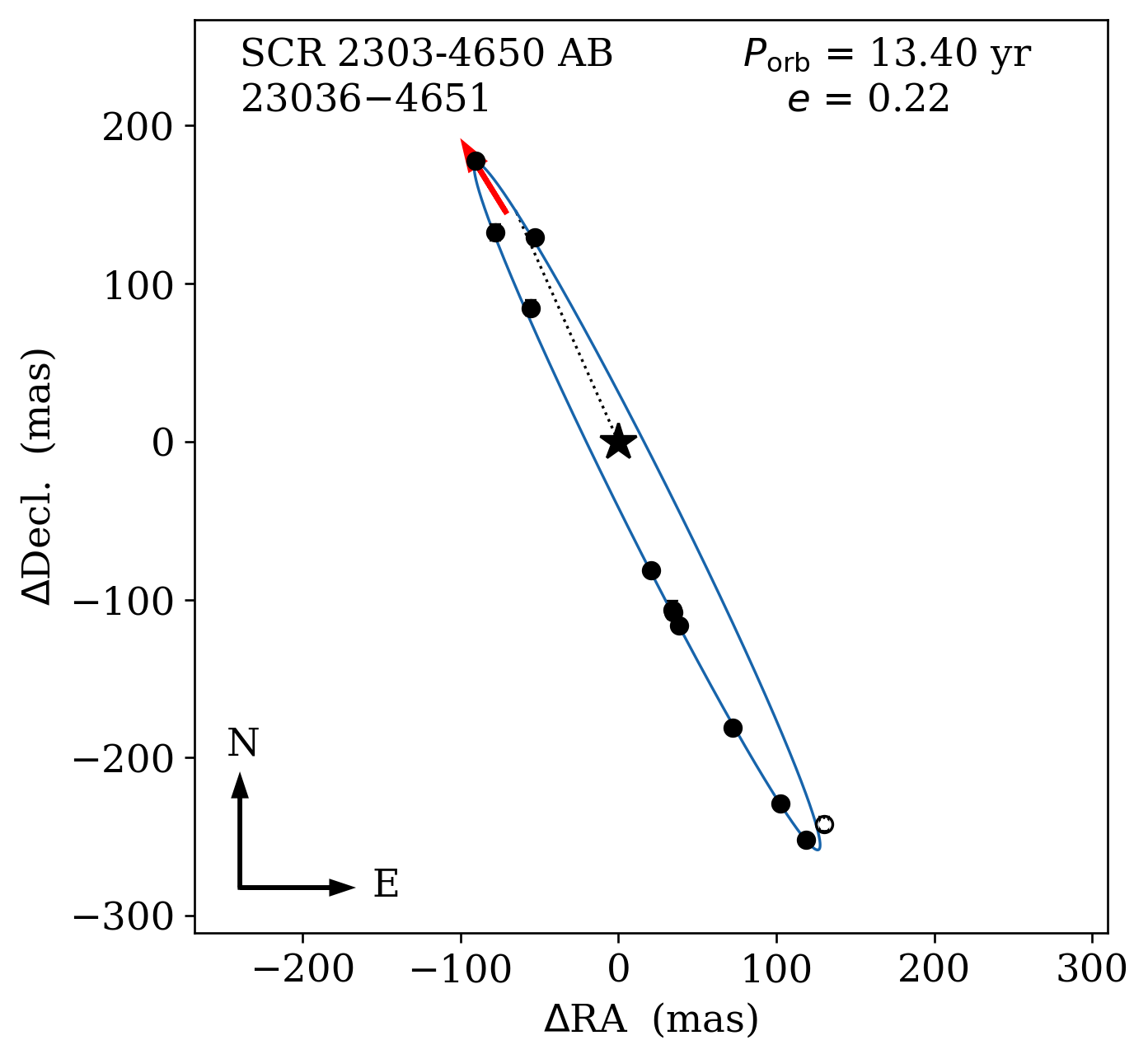}
\includegraphics[scale=0.4]{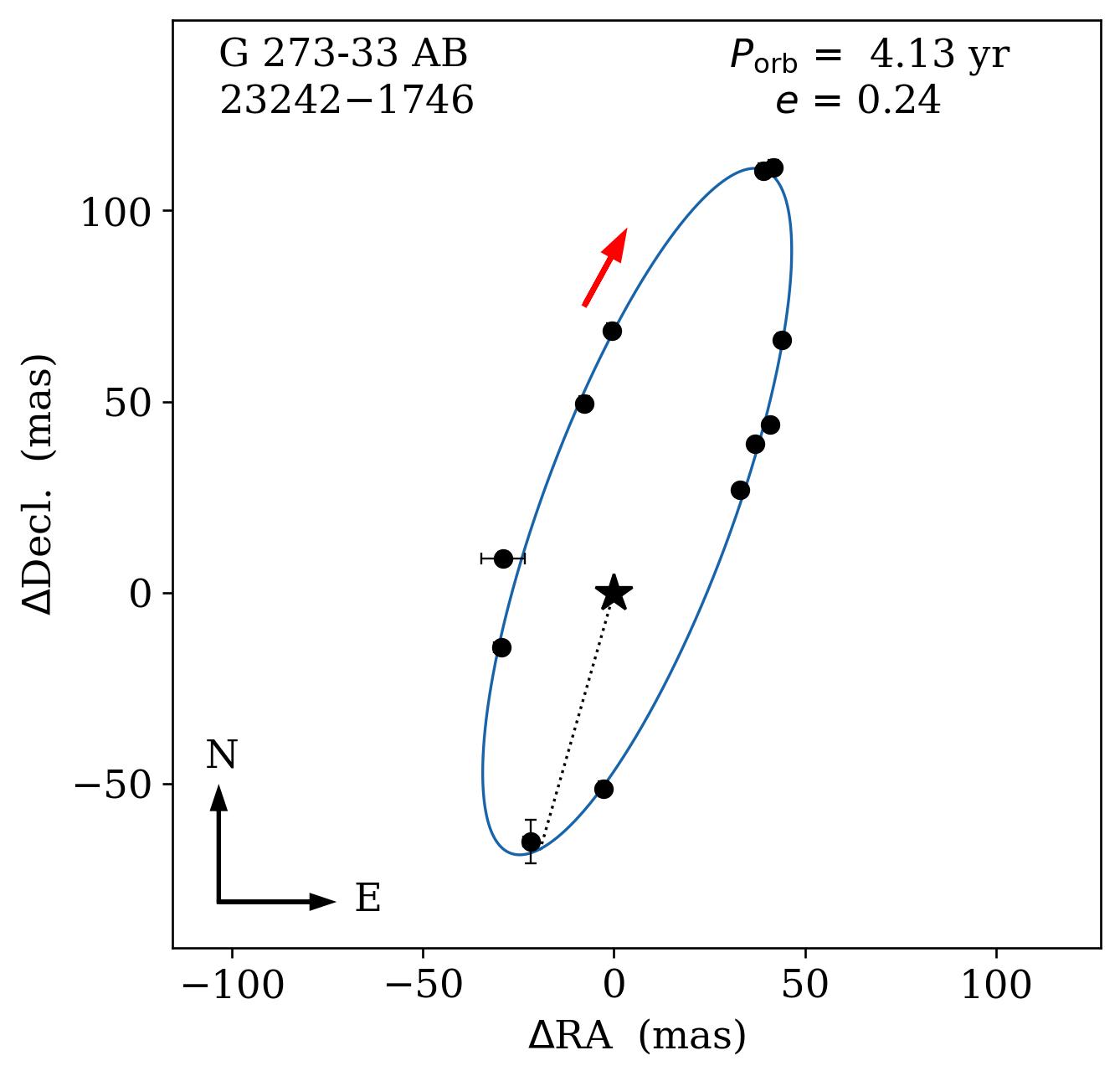}
\includegraphics[scale=0.4]{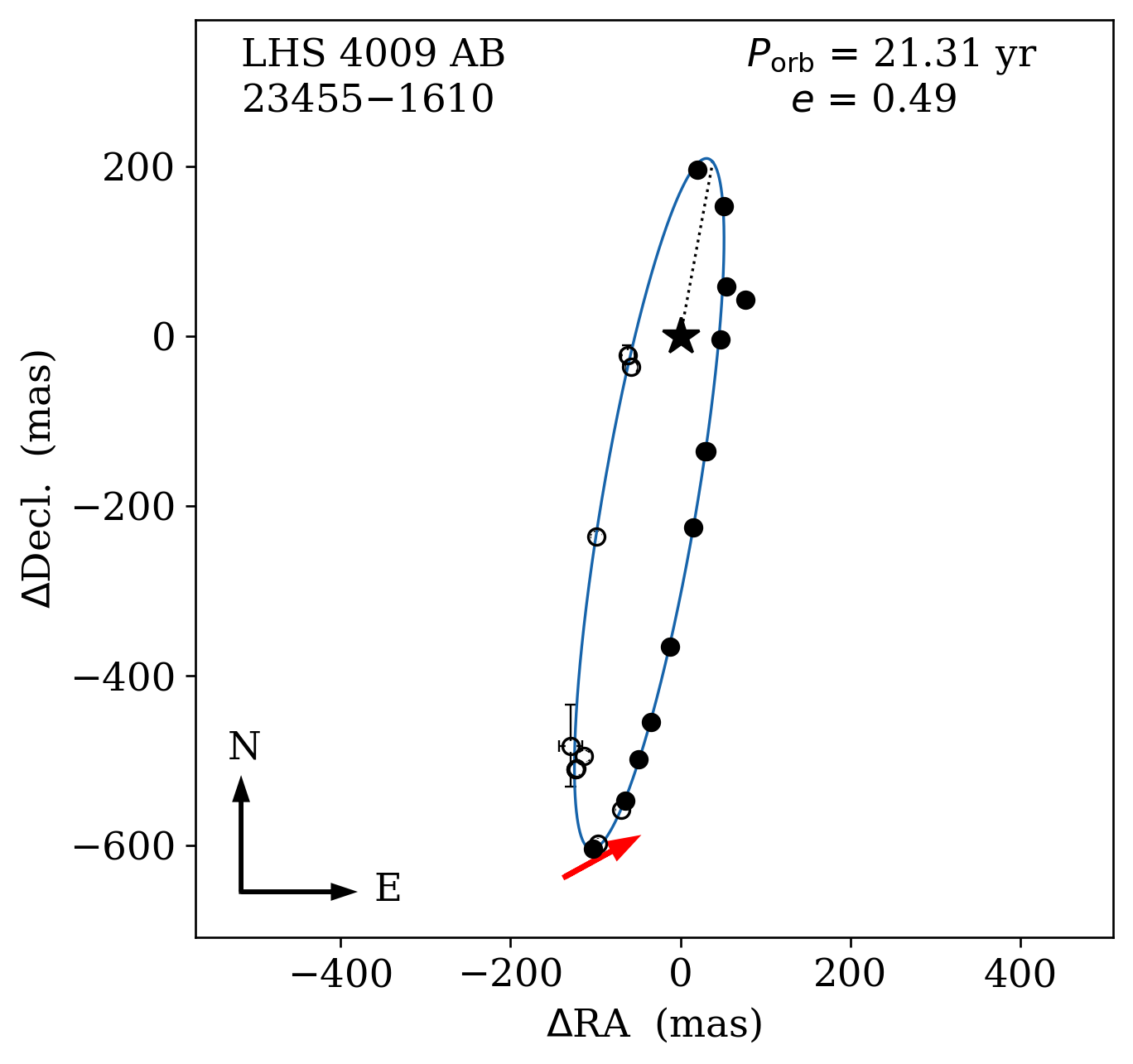}
\caption{
Orbits of M dwarf systems determined from the imaging data in Table~\ref{tab:results}. Two systems shown here (GJ~791.2~AB and LHS~501~AC) also incorporated the RV data in Table~\ref{tab:RVdata}; their corresponding RV fits are shown in Figure~\ref{fig:RVorbits}. 
For additional details, see the caption to Figure~\ref{fig:orbits}.
\label{fig:orbits4}}
\end{figure}

\begin{figure} \centering
\includegraphics[scale=0.5]{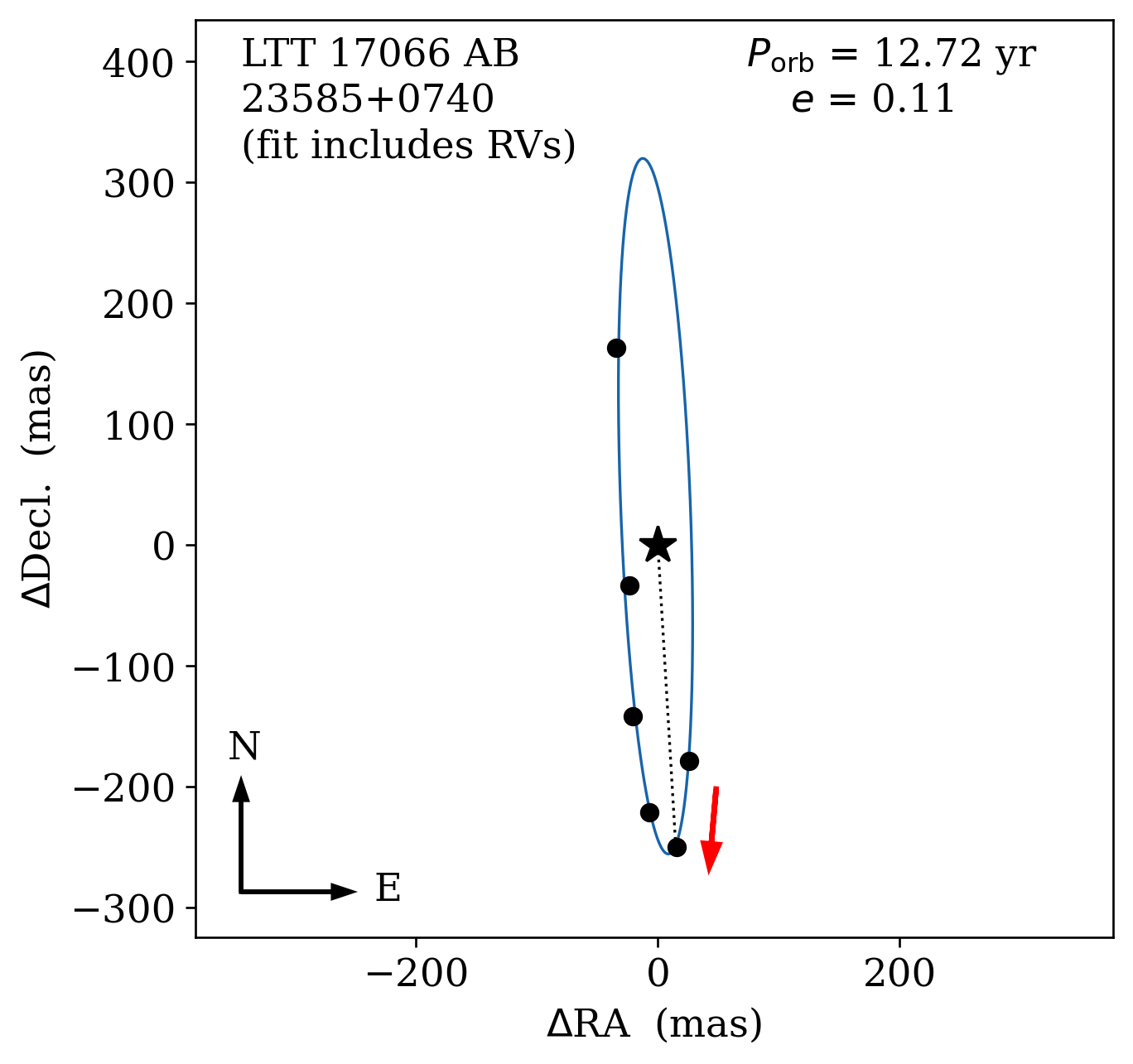}
\includegraphics[scale=0.5]{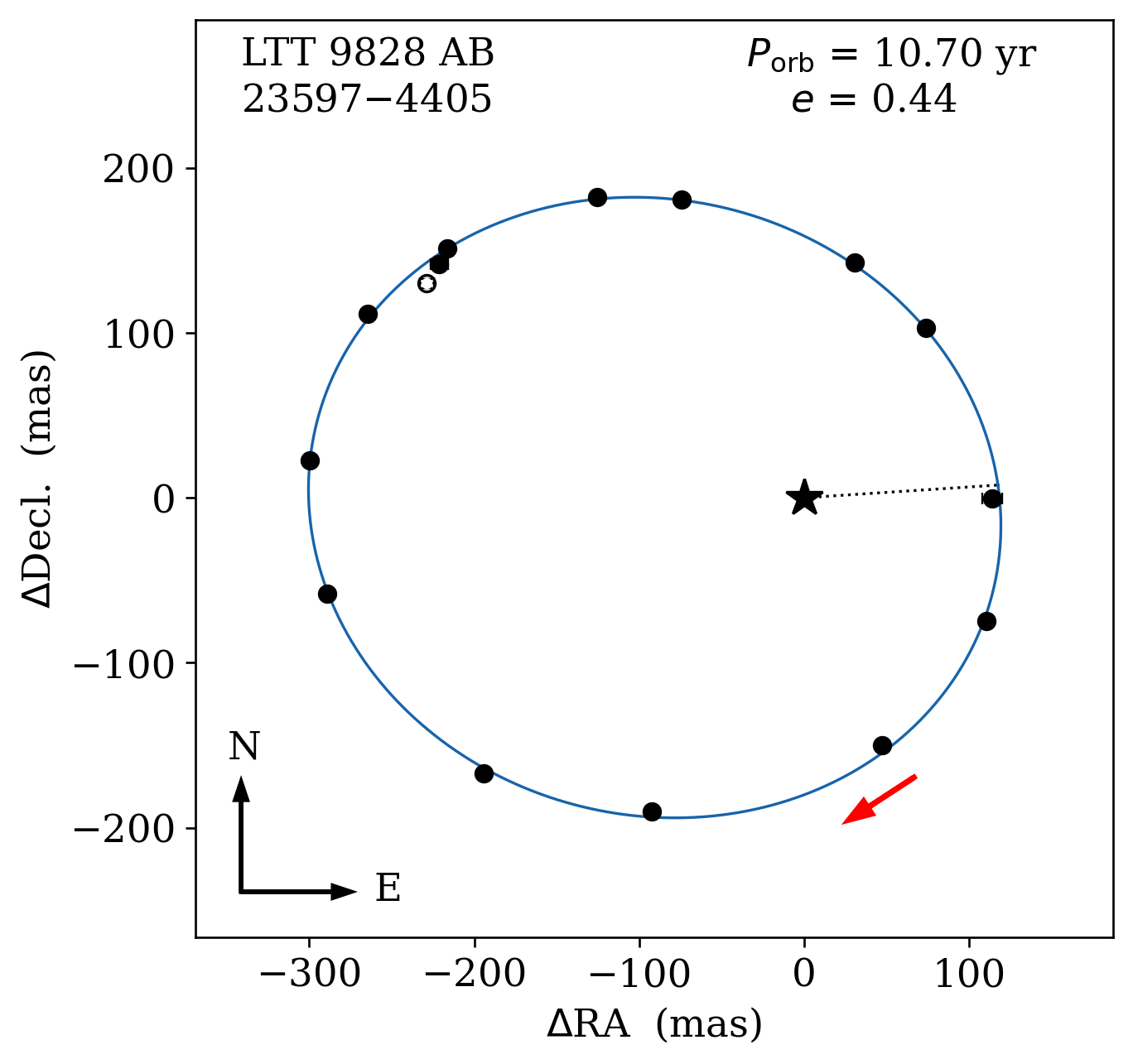}
\caption{
Orbits of M dwarf systems determined from the imaging data in Table~\ref{tab:results}. One system shown here (LTT~17066~AB) also incorporated the RV data in Table~\ref{tab:RVdata}; its corresponding RV fit is shown in Figure~\ref{fig:RVorbits}. 
For additional details, see the caption to Figure~\ref{fig:orbits}.
\label{fig:orbits5}}
\end{figure}

\begin{figure} \centering
\includegraphics[scale=0.5]{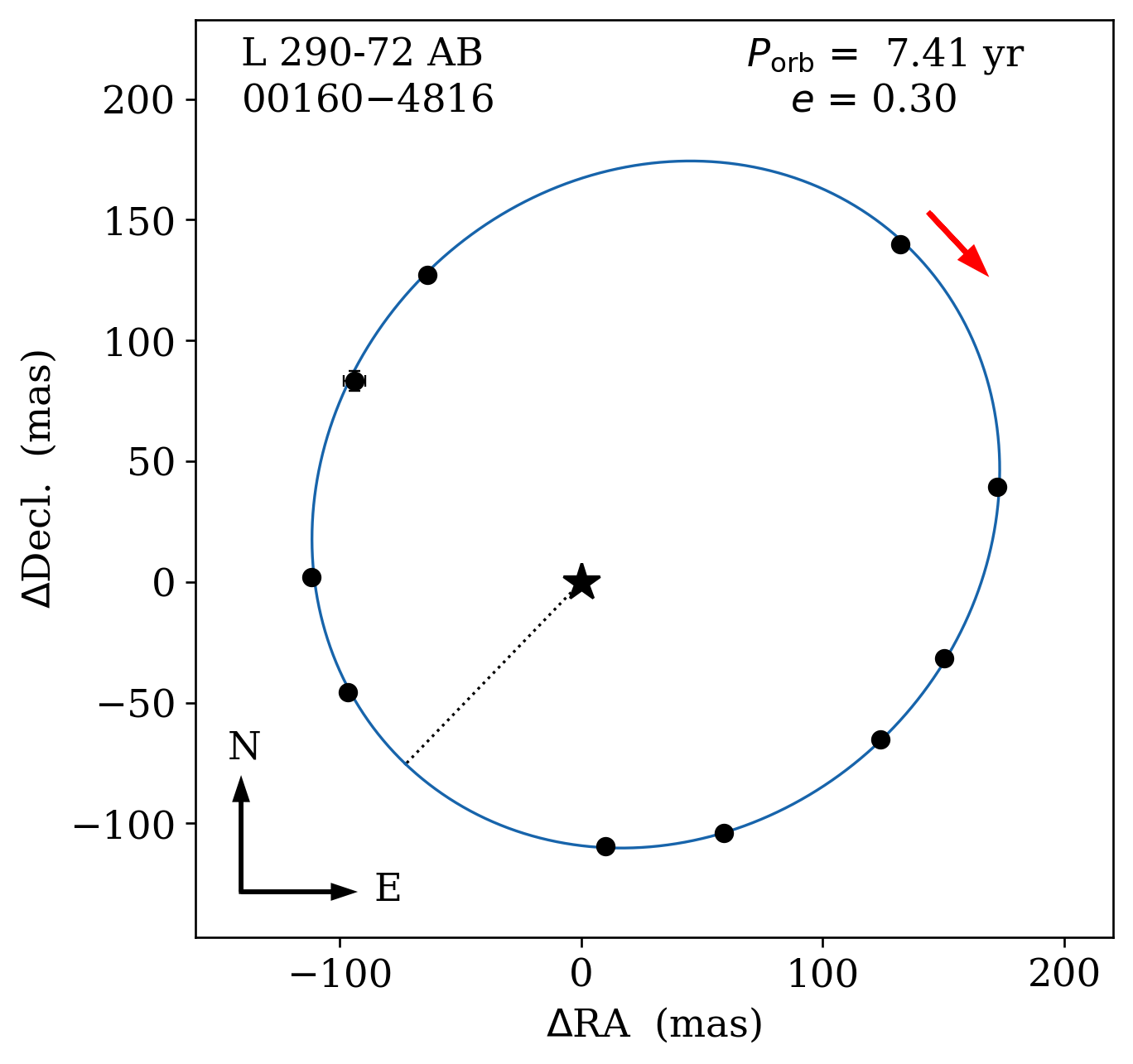}
\includegraphics[scale=0.5]{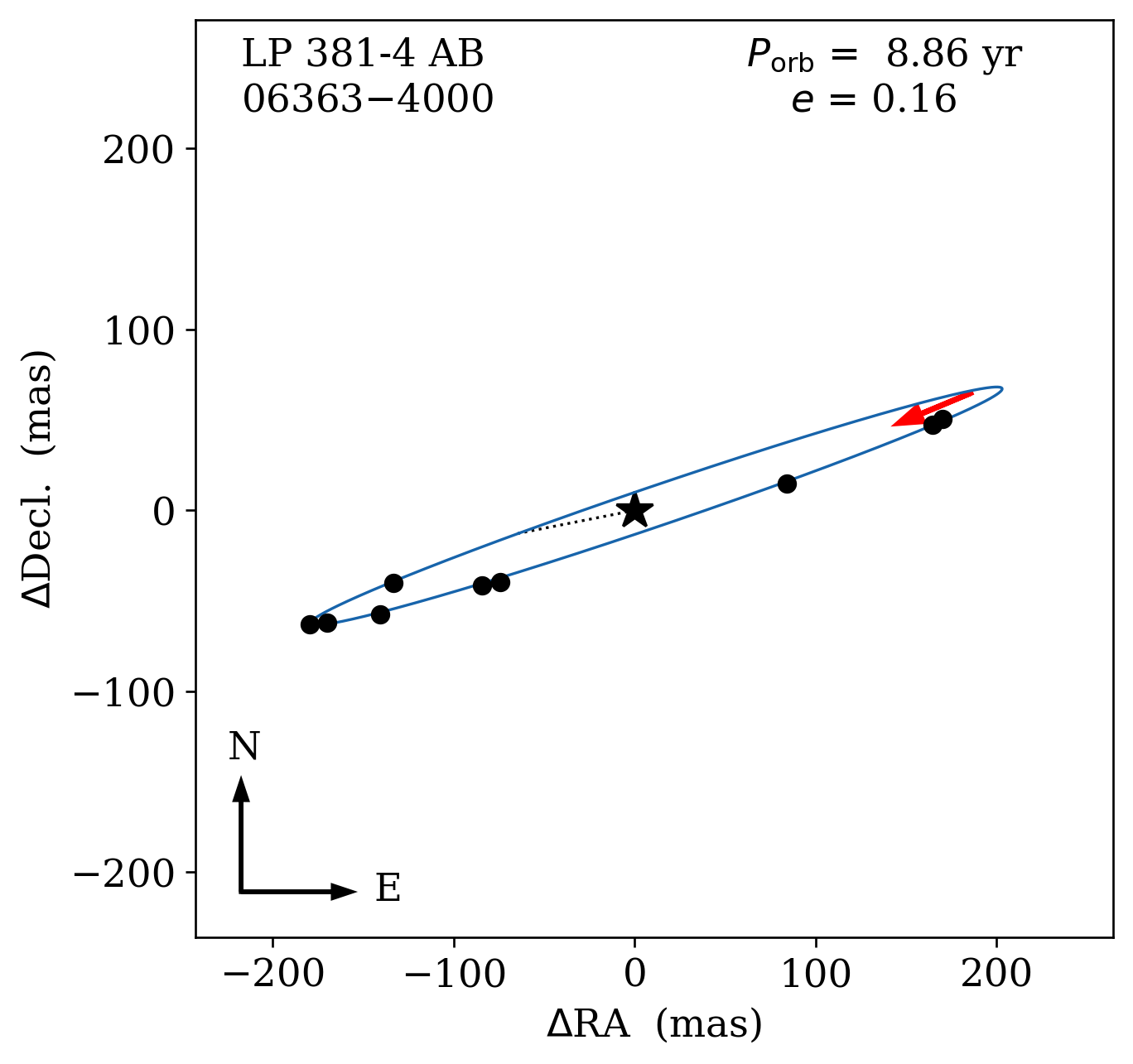}
\includegraphics[scale=0.5]{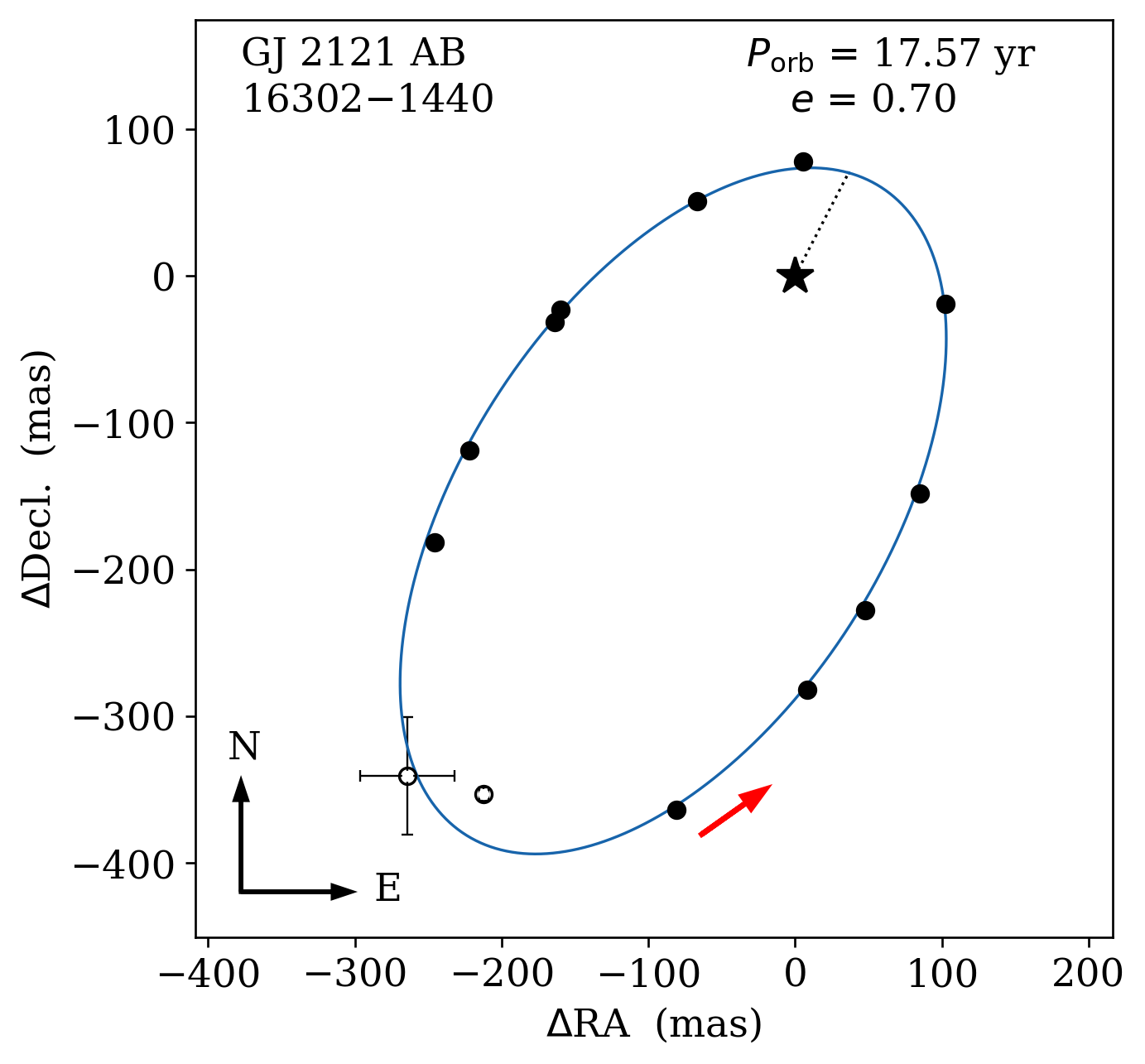}
\includegraphics[scale=0.5]{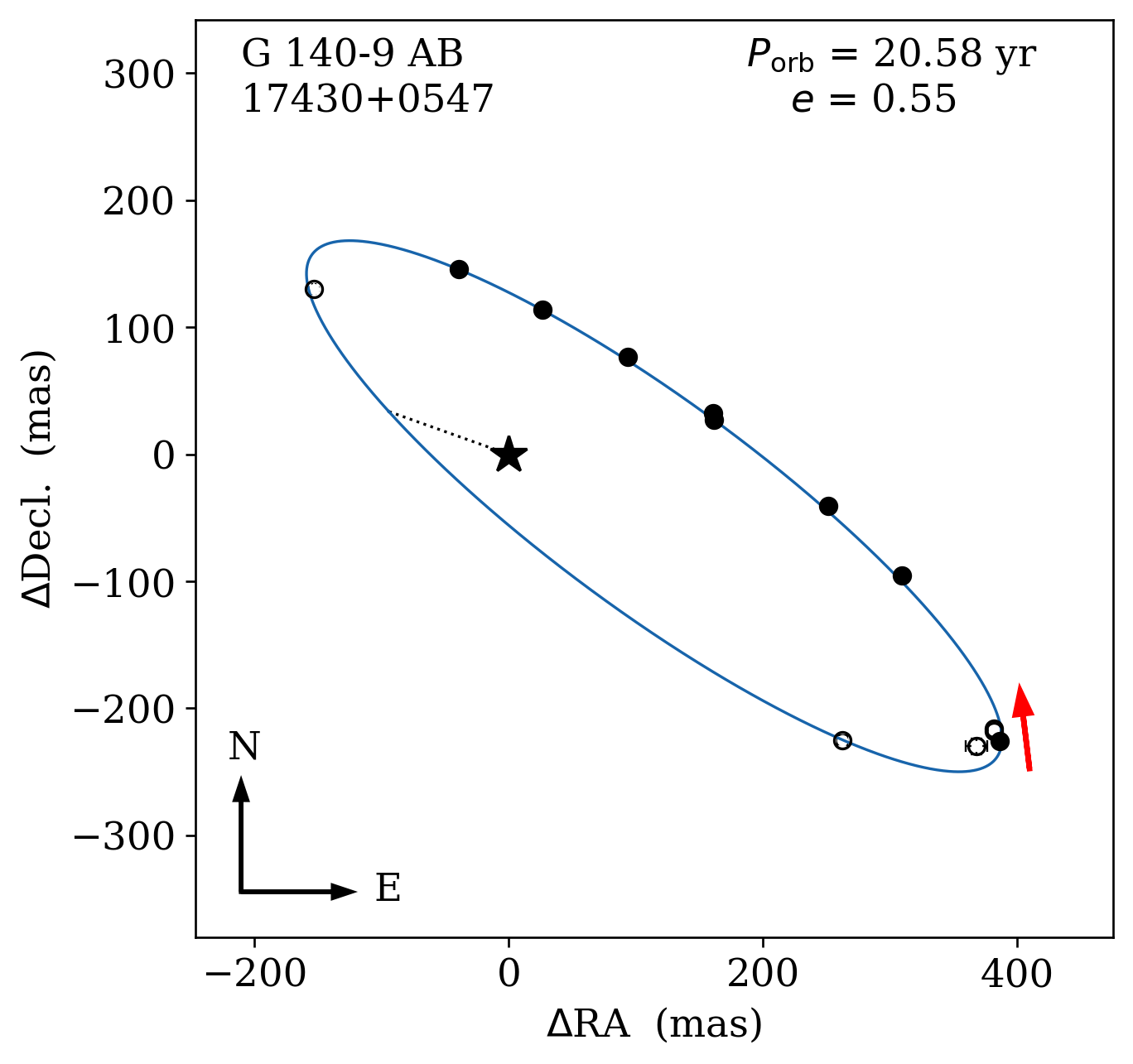}
\caption{
Orbits of M dwarf systems determined from the imaging data presented in Table~\ref{tab:results}. Each orbit's parameters are given in Table~\ref{tab:orbits}, as fit by the code \texttt{orvara} ($\S$\ref{sec:orbits}), to take advantage of \textit{Hipparcos} data available for these cases. Point types and other plot details are the same as in Figures~\ref{fig:orbits}--\ref{fig:orbits5}.
\label{fig:orbits6}}
\end{figure}

\begin{figure}[h!] \centering
\includegraphics[width=0.32\textwidth]{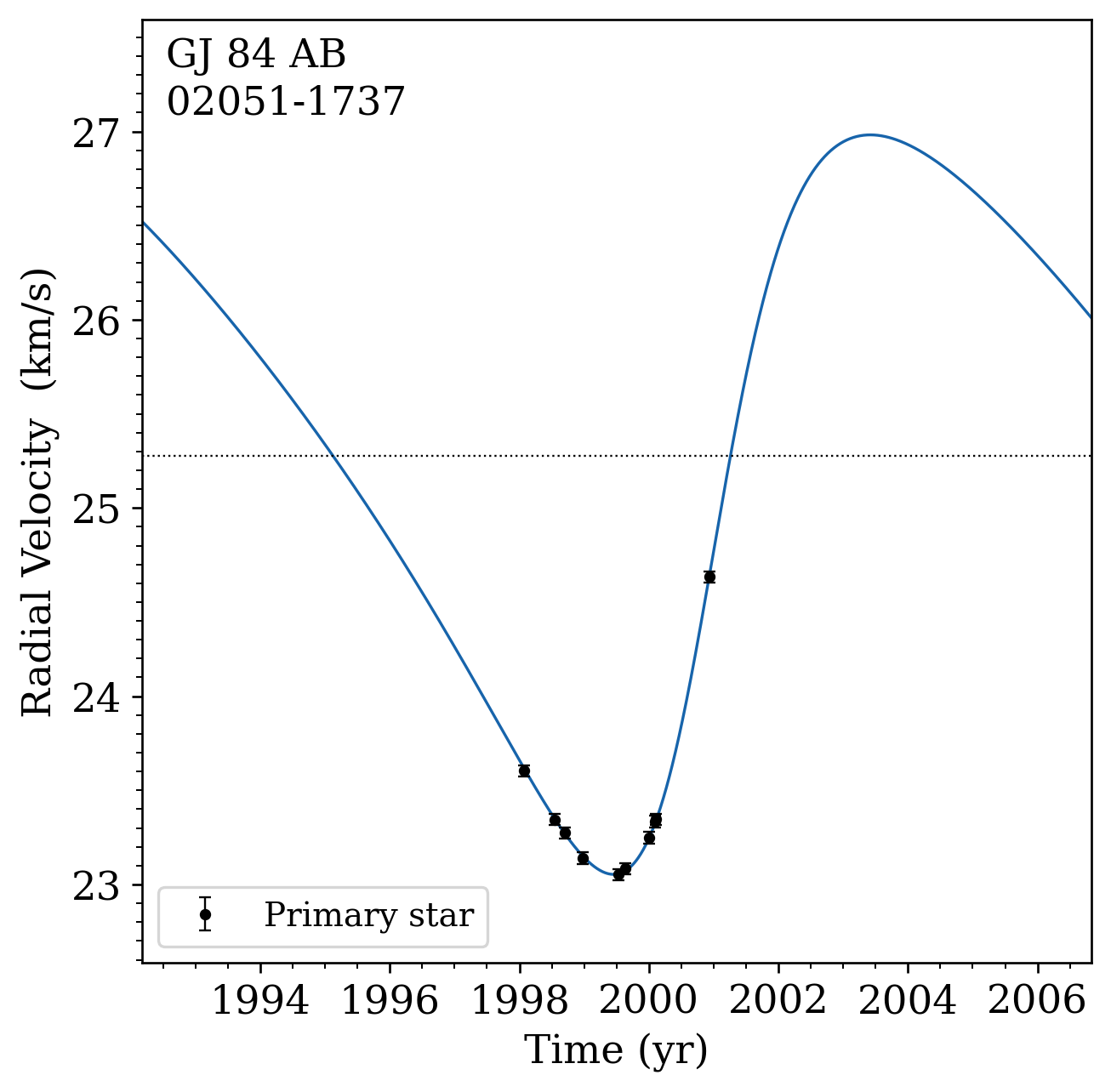}
\includegraphics[width=0.32\textwidth]{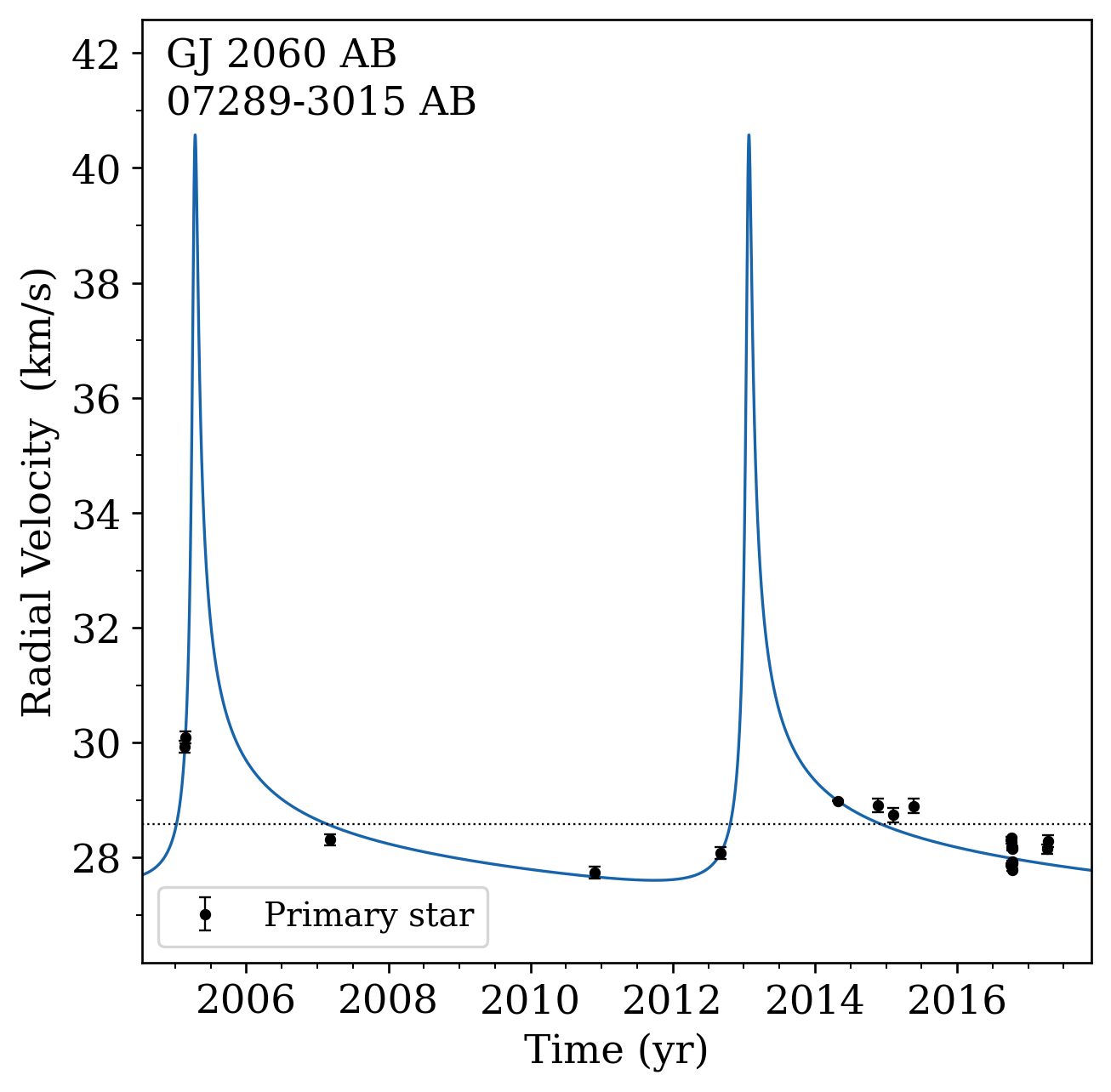}
\includegraphics[width=0.33\textwidth]{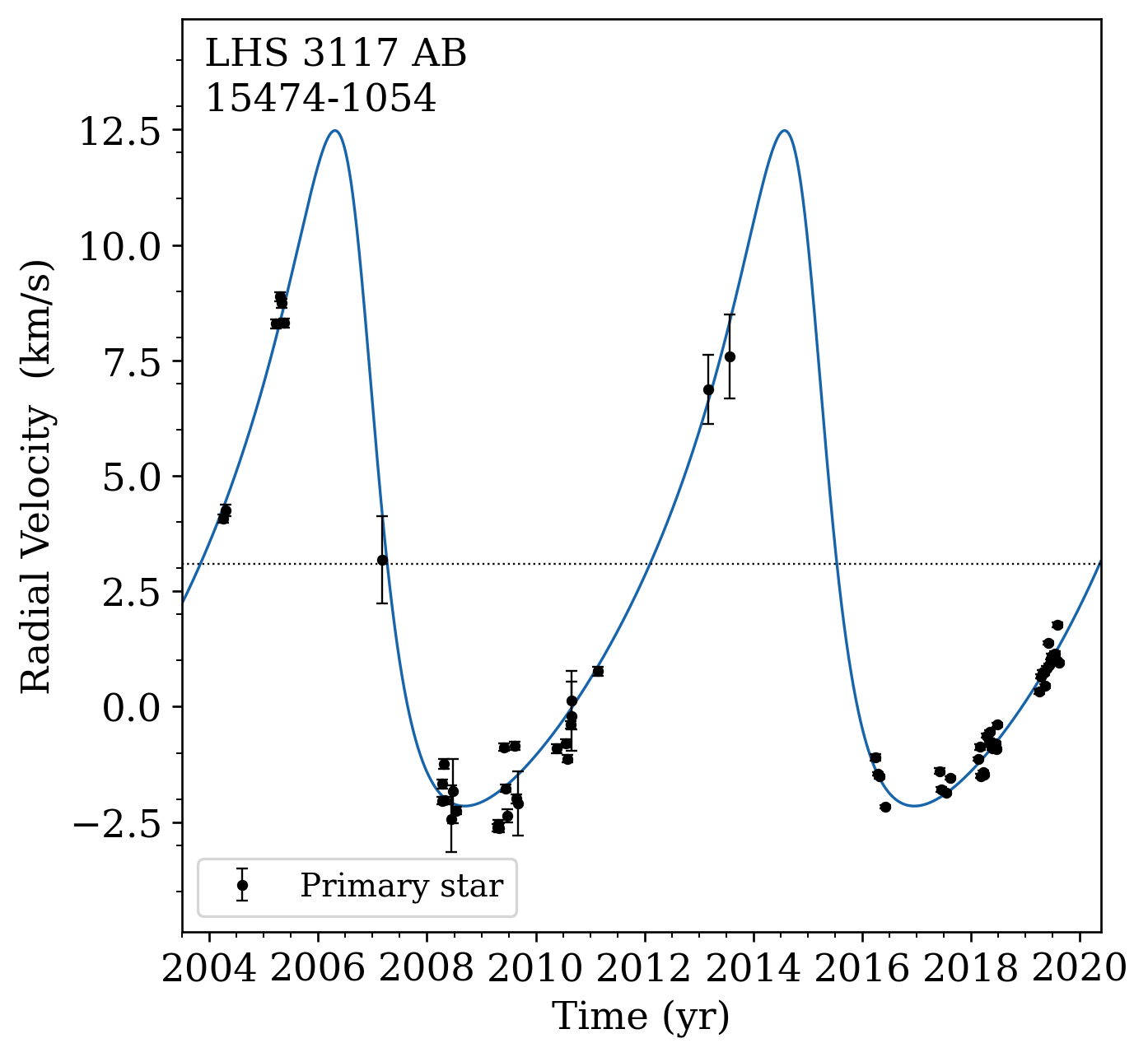}
\includegraphics[width=0.33\textwidth]{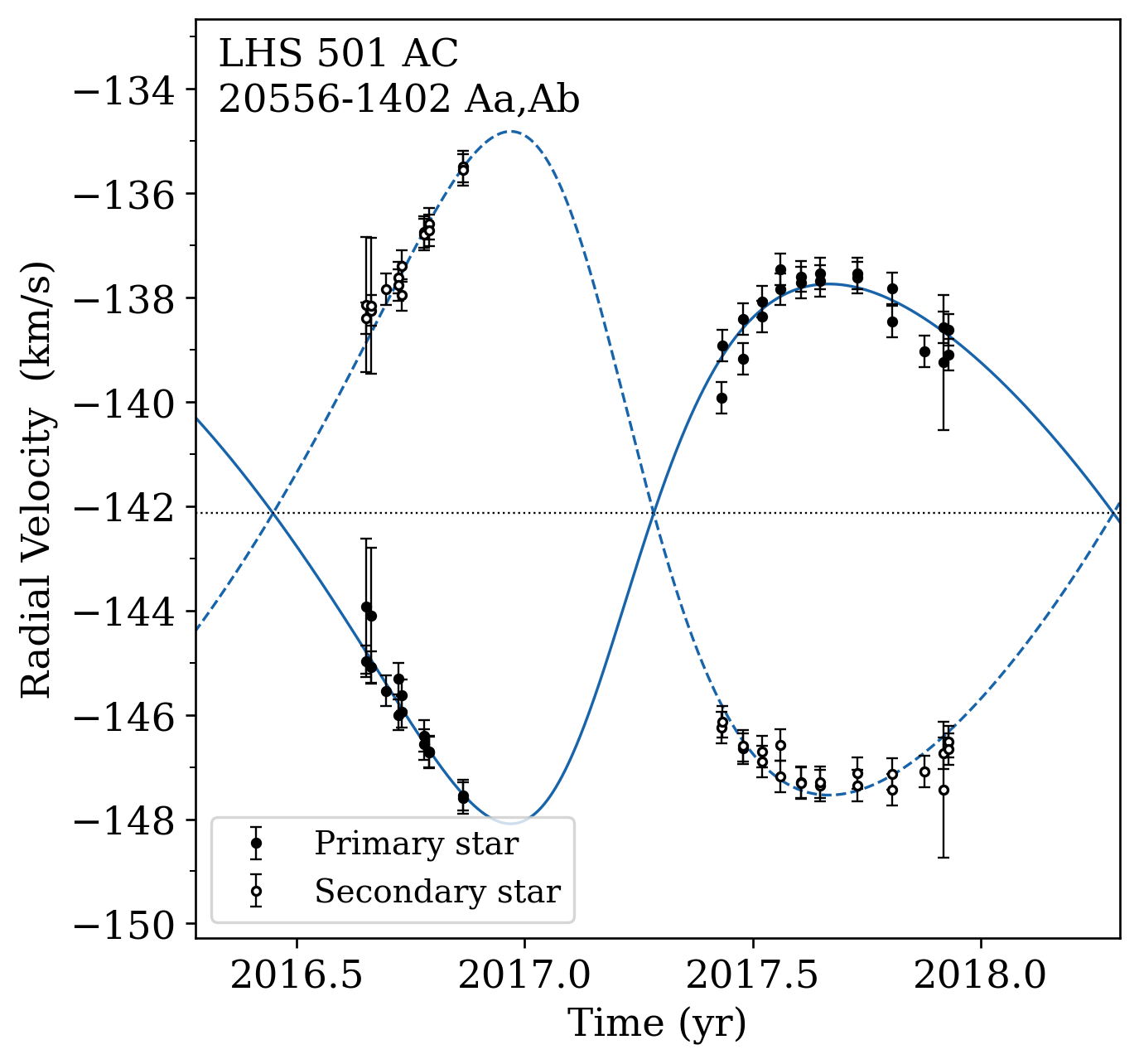}
\includegraphics[width=0.33\textwidth]{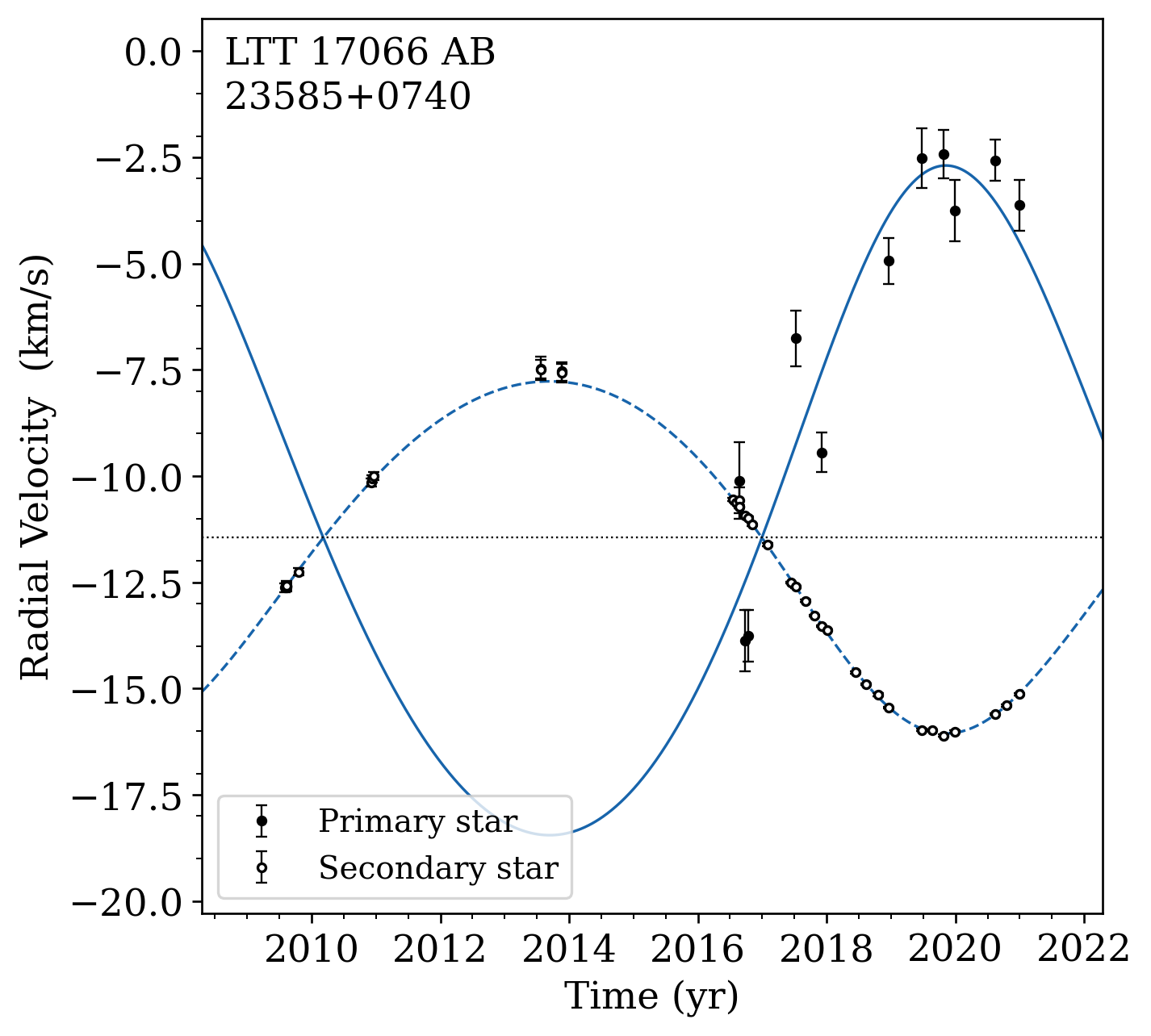}
\includegraphics[width=\textwidth]{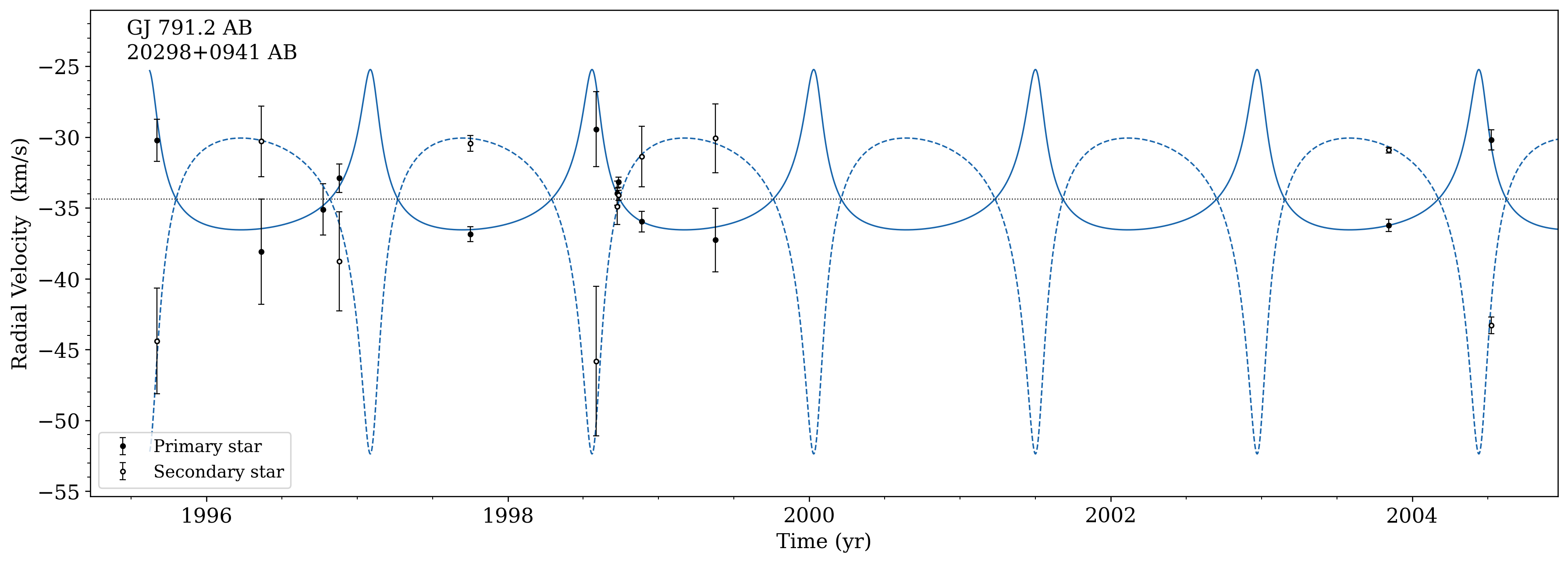}

\caption{Spectroscopic orbits for six systems with radial velocity data available in the literature; the parameters describing these orbits are given in Table~\ref{tab:RVorbits}. In all cases, these RV data were fit simultaneously with the systems' imaging data in Table~\ref{tab:results} and the corresponding visual orbit shown in Figures~\ref{fig:orbits}--\ref{fig:orbits5} represents that combined result. Filled points and solid lines mark RV data and model, respectively, of the primary star, and open points and dashed lines are RV data and model of the secondary (if available). The light dotted line indicates the systemic velocity.
\label{fig:RVorbits}}
\end{figure}

We have derived reliable orbits for \Norbits{} systems using these SOAR data in combination with high-resolution imaging published in the literature. In {\bf six} cases, we have also incorporated radial velocity data, discussed separately in $\S$\ref{sec:RVs} and $\S$\ref{sec:RVorbits}. The parameters characterizing the orbits are given in Table~\ref{tab:orbits}, and the orbits are shown visually with the data in Figures~\ref{fig:orbits}--\ref{fig:orbits6}. In addition to system identifiers (columns 1--2), orbital elements (columns 3--9), and total masses (assuming Gaia parallaxes; column~10), column~11 lists the type(s) of data used (i = imaging and HIP = \textit{Hipparcos}) and which code was used for the orbital fit presented (discussed below), and column 12 notes the reference for each system's most recently previous orbit, if any. Of the \Norbits{} presented here, \NorbitsNEW{} are new orbits, and \NorbitsNEWsoaronly{} of those new orbits are not covered by any data in the literature --- they include only SOAR data.

For a few systems shown in Figures~\ref{fig:orbits}--\ref{fig:orbits6}, the visual orbit fits match the data well for all but one data point. For example, this is the case for LP~349-25~AB (WDS~00279+2220), UPM~0838--2843~AB (WDS~08386--2843), and LTT~12366~AC (WDS~09012+0157 Aa,Ab). In those cases, the outliers represent more uncertain data from SOAR, often due to observing conditions (i.e., points flagged as ``:'' in Table~\ref{tab:results}; see $\S$\ref{sec:observations}). In general, the majority of the SOAR data presented here are well traced by their orbit fits; this ``beads on a string'' effect highlights the precision imparted by HRCam and the speckle data reduction process at SOAR.

\startlongtable
\begin{deluxetable}{llllllllllll}
\tabletypesize{\scriptsize}
\tablecaption{
Parameters of the orbits fit to the imaging data.
\label{tab:orbits}}

\tablehead{
\colhead{Name   } & \colhead{WDS} &  \colhead{\Porb} & \colhead{$a$        } & \colhead{$e$} & \colhead{$i$  } & \colhead{$\Omega$} & \colhead{$\omega$} & \colhead{$T_0$} & \colhead{$M_\mathrm{tot}$} & \colhead{Data used} & \colhead{Prev.} \\[-1em] 
\colhead{       } & \colhead{   } &  \colhead{(yr) } & \colhead{($\arcsec$)} & \colhead{   } & \colhead{(deg)} & \colhead{(deg)   } & \colhead{(deg)   } & \colhead{(yr) } & \colhead{($M_\odot$)     } & \colhead{Code used} & \colhead{orbit} \\[-1em] 
\colhead{(1)    } & \colhead{(2)} &  \colhead{(3)  } & \colhead{(4)        } & \colhead{(5)} & \colhead{(6)  } & \colhead{(7)     } & \colhead{(8)     } & \colhead{(9) }  & \colhead{(10)            } & \colhead{(11)     } & \colhead{(12) } \\[-2em] 
}
\startdata
G 131-26 AB        & 00089+2050      &      5.90571  &      0.1475   &      0.0762   &      142.27   &      82.54    &      247.20   &      2019.0111 & 0.545 & i              & Vri22  \\ 
                   &                 & $\pm$ 0.00914 & $\pm$ 0.0010  & $\pm$ 0.0045  & $\pm$ 0.95    & $\pm$ 1.32    & $\pm$ 3.51    & $\pm$ 0.0501  & $\pm$ 0.025 & {\tt ORBIT  } &  ... \\[1em] 
LEHPM 1-255 AB     & 00098-4202      &      3.33848  &      0.0952   &      0.6181   &      69.73    &      133.86   &      167.17   &      2019.9821 & 0.343 & i              & ...    \\ 
                   &                 & $\pm$ 0.01276 & $\pm$ 0.0016  & $\pm$ 0.0115  & $\pm$ 0.91    & $\pm$ 1.33    & $\pm$ 4.15    & $\pm$ 0.0303  & $\pm$ 0.018 & {\tt ORBIT  } &  ... \\[1em] 
2MA 0015-1636 AB   & 00160-1637      &      4.19102  &      0.1093   &      0.0153   &      67.64    &      91.78    &      -66.30   &      2019.3095 & 0.421 & i              & Vri22  \\ 
                   &                 & $\pm$ 0.00921 & $\pm$ 0.0012  & $\pm$ 0.0077  & $\pm$ 1.02    & $\pm$ 0.51    & $\pm$ 12.58   & $\pm$ 0.1459  & $\pm$ 0.014 & {\tt ORBIT  } &  ... \\[1em] 
L 290-72 AB        & 00160-4816      &      7.413    &      0.14933  &      0.2981   &      18.76    &      46.4     &      177.4    &      2015.93  & 0.901 & i, HIP         & Mas23  \\ 
                   &                 & $\pm$ 0.016   & $\pm$ 0.00032 & $\pm$ 0.0012  & $\pm$ 0.31    & $^{+1.8    }_{-2.0    }$ & $^{+1.7    }_{-1.9    }$ & $\pm$ 0.018 & $\pm$ 0.036 & {\tt orvara } &  ... \\[1em]
LP 349-25 AB       & 00279+2220      &      7.88077  &      0.1472   &      0.0627   &      117.38   &      35.82    &      247.60   &      2009.0417 & 0.145 & i              & Dup10b \\ 
                   &                 & $\pm$ 0.01450 & $\pm$ 0.0001  & $\pm$ 0.0018  & $\pm$ 0.06    & $\pm$ 0.07    & $\pm$ 0.85    & $\pm$ 0.0172  & $\pm$ 0.003 & {\tt ORBIT  } &  ... \\[1em] 
G 34-23 AB         & 01222+2209      &      5.76562  &      0.1968   &      0.5680   &      105.81   &      170.82   &      349.60   &      2024.7421 & 0.365 & i              & ...    \\ 
                   &                 & $\pm$ 0.02735 & $\pm$ 0.0011  & $\pm$ 0.0051  & $\pm$ 0.46    & $\pm$ 0.32    & $\pm$ 1.38    & $\pm$ 0.0165  & $\pm$ 0.023 & {\tt ORBIT  } &  ... \\[1em] 
L 88-43 AB         & 01536-6654      &      6.82153  &      0.1812   &      0.4715   &      57.72    &      113.13   &      373.87   &      2013.1641 & 0.233 & i              & ...    \\ 
                   &                 & $\pm$ 0.61272 & $\pm$ 0.0136  & $\pm$ 0.0888  & $\pm$ 4.40    & $\pm$ 1.61    & $\pm$ 7.10    & $\pm$ 0.7507  & $\pm$ 0.073 & {\tt ORBIT  } &  ... \\[1em] 
LP 770-20 AB       & 02275-1908      &      12.14530 &      0.2076   &      0.6456   &      35.37    &      204.65   &      125.66   &      2021.6120 & 0.435 & i              & ...    \\ 
                   &                 & $\pm$ 0.25882 & $\pm$ 0.0024  & $\pm$ 0.0063  & $\pm$ 0.67    & $\pm$ 1.75    & $\pm$ 1.78    & $\pm$ 0.0061  & $\pm$ 0.025 & {\tt ORBIT  } &  ... \\[1em] 
L 225-57 AB        & 02344-5306      &      6.40055  &      0.1591   &      0.2223   &      28.82    &      120.72   &      283.70   &      2020.7041 & 0.782 & i              & ...    \\ 
                   &                 & $\pm$ 0.03735 & $\pm$ 0.0010  & $\pm$ 0.0033  & $\pm$ 1.05    & $\pm$ 1.79    & $\pm$ 2.12    & $\pm$ 0.0158  & $\pm$ 0.040 & {\tt ORBIT  } &  ... \\[1em] 
LP 993-115 BC      & 02452-4344 Aa,Ab &      28.07851 &      0.6294   &      0.2394   &      116.99   &      158.87   &      302.86   &      2009.5946 & 0.431 & i              & Vri22  \\ 
                   &                 & $\pm$ 0.33121 & $\pm$ 0.0045  & $\pm$ 0.0021  & $\pm$ 0.20    & $\pm$ 0.39    & $\pm$ 1.72    & $\pm$ 0.0916  & $\pm$ 0.014 & {\tt ORBIT  } &  ... \\[1em] 
LHS 1561 AB        & 03347-0451      &      3.89669  &      0.0923   &      0.5255   &      144.21   &      173.17   &      173.55   &      2021.8910 & 0.957 & i              & ...    \\ 
                   &                 & $\pm$ 0.01401 & $\pm$ 0.0009  & $\pm$ 0.0088  & $\pm$ 1.82    & $\pm$ 2.99    & $\pm$ 3.89    & $\pm$ 0.0139  & $\pm$ 0.055 & {\tt ORBIT  } &  ... \\[1em] 
LEP 0448+1003 AB   & 04488+1003      &      2.69193  &      0.1021   &      0.2003   &      53.52    &      283.20   &      172.33   &      2018.2024 & 0.757 & i              & ...    \\ 
                   &                 & $\pm$ 0.00716 & $\pm$ 0.0008  & $\pm$ 0.0048  & $\pm$ 0.75    & $\pm$ 0.79    & $\pm$ 2.01    & $\pm$ 0.0144  & $\pm$ 0.024 & {\tt ORBIT  } &  ... \\[1em] 
SCR 0533-4257 AB   & 05335-4257      &      0.67001  &      0.0538   &      0.4603   &      149.44   &      91.06    &      21.64    &      2017.1509 & 0.393 & i              & Vri22  \\ 
                   &                 & $\pm$ 0.00051 & $\pm$ 0.0008  & $\pm$ 0.0110  & $\pm$ 3.01    & $\pm$ 5.11    & $\pm$ 5.95    & $\pm$ 0.0040  & $\pm$ 0.018 & {\tt ORBIT  } &  ... \\[1em] 
LP 837-19 AB       & 05450-2137      &      4.37456  &      0.1000   &      0.8474   &      150.11   &      345.77   &      129.90   &      2021.4509 & 0.639 & i              & ...    \\ 
                   &                 & $\pm$ 0.02689 & $\pm$ 0.0024  & $\pm$ 0.0067  & $\pm$ 3.71    & $\pm$ 6.16    & $\pm$ 6.53    & $\pm$ 0.0111  & $\pm$ 0.047 & {\tt ORBIT  } &  ... \\[1em] 
UPM 0624-2655 AB   & 06241-2655      &      3.53320  &      0.1162   &      0.5191   &      145.93   &      32.34    &      159.43   &      2021.0311 & 1.054 & i              & ...    \\ 
                   &                 & $\pm$ 0.01115 & $\pm$ 0.0007  & $\pm$ 0.0042  & $\pm$ 1.10    & $\pm$ 2.06    & $\pm$ 2.42    & $\pm$ 0.0044  & $\pm$ 0.059 & {\tt ORBIT  } &  ... \\[1em] 
G 106-45 AB        & 06242-0017      &      5.33855  &      0.0994   &      0.1498   &      77.24    &      138.38   &      217.89   &      2019.2812 & 0.363 & i              & ...    \\ 
                   &                 & $\pm$ 0.06825 & $\pm$ 0.0016  & $\pm$ 0.0231  & $\pm$ 0.79    & $\pm$ 0.61    & $\pm$ 6.06    & $\pm$ 0.0974  & $\pm$ 0.020 & {\tt ORBIT  } &  ... \\[1em] 
LP 381-4 AB        & 06363-4000      &      8.863    &      0.2040   &      0.156    &      86.890   &      71.194   &      247.9    &      2016.42  & 0.769 & i, HIP         & ...    \\ 
                   &                 & $\pm$ 0.020   & $^{+0.0017 }_{-0.0018 }$ & $^{+0.0012 }_{-0.0013 }$ & $\pm$ 0.070 & $^{+0.048  }_{-0.049  }$ & $\pm$ 1.0 & $\pm$ 0.036 & $\pm$ 0.021 & {\tt orvara } &  ... \\[1em]
SCR 0702-6102 AB   & 07028-6103      &      2.48800  &      0.0666   &      0.3467   &      163.83   &      -11.18   &      106.94   &      2020.6604 & 0.257 & i              & ...    \\ 
                   &                 & $\pm$ 0.01749 & $\pm$ 0.0013  & $\pm$ 0.0146  & $\pm$ 6.59    & $\pm$ 28.92   & $\pm$ 27.30   & $\pm$ 0.0213  & $\pm$ 0.016 & {\tt ORBIT  } &  ... \\[1em] 
APM 89 AB          & 07096-5704      &      0.69778  &      0.0342   &      0.1373   &      36.72    &      310.25   &      125.72   &      2021.3645 & 0.409 & i              & ...    \\ 
                   &                 & $\pm$ 0.00274 & $\pm$ 0.0011  & $\pm$ 0.0275  & $\pm$ 5.16    & $\pm$ 8.43    & $\pm$ 8.06    & $\pm$ 0.0196  & $\pm$ 0.040 & {\tt ORBIT  } &  ... \\[1em] 
G 89-32 AB         & 07364+0705      &      23.96533 &      0.6372   &      0.5874   &      13.43    &      75.30    &      68.67    &      2016.2144 & 0.277 & i              & Tok20b \\ 
                   &                 & $\pm$ 0.01729 & $\pm$ 0.0004  & $\pm$ 0.0003  & $\pm$ 0.28    & $\pm$ 1.01    & $\pm$ 1.06    & $\pm$ 0.0011  & $\pm$ 0.006 & {\tt ORBIT  } &  ... \\[1em] 
UPM 0838-2843 AB   & 08386-2843      &      0.70221  &      0.0448   &      0.2145   &      145.02   &      110.12   &      317.58   &      2020.8393 & 0.463 & i              & ...    \\ 
                   &                 & $\pm$ 0.00093 & $\pm$ 0.0008  & $\pm$ 0.0095  & $\pm$ 2.52    & $\pm$ 4.29    & $\pm$ 4.97    & $\pm$ 0.0047  & $\pm$ 0.025 & {\tt ORBIT  } &  ... \\[1em] 
LTT 12366 AC       & 09012+0157 Aa,Ab &      6.13568  &      0.1417   &      0.4757   &      84.63    &      145.41   &      211.31   &      2023.2128 & 0.975 & i              & ...    \\ 
                   &                 & $\pm$ 0.18675 & $\pm$ 0.0015  & $\pm$ 0.0117  & $\pm$ 0.56    & $\pm$ 0.35    & $\pm$ 3.01    & $\pm$ 0.0435  & $\pm$ 0.071 & {\tt ORBIT  } &  ... \\[1em] 
LHS 6167 AB        & 09156-1036      &      5.03365  &      0.1941   &      0.4621   &      116.95   &      112.50   &      92.09    &      2019.2052 & 0.260 & i              & Mas18  \\ 
                   &                 & $\pm$ 0.00068 & $\pm$ 0.0004  & $\pm$ 0.0024  & $\pm$ 0.10    & $\pm$ 0.17    & $\pm$ 0.10    & $\pm$ 0.0024  & $\pm$ 0.006 & {\tt ORBIT  } &  ... \\[1em] 
GJ 381 AB          & 10121-0241      &      7.57783  &      0.2898   &      0.7443   &      93.12    &      68.44    &      274.12   &      2016.8705 & 0.651 & i              & Man19  \\ 
                   &                 & $\pm$ 0.00331 & $\pm$ 0.0017  & $\pm$ 0.0033  & $\pm$ 0.04    & $\pm$ 0.03    & $\pm$ 0.05    & $\pm$ 0.0047  & $\pm$ 0.020 & {\tt ORBIT  } &  ... \\[1em] 
TWA 22 AB          & 10174-5354 Aa,Ab &      5.34562  &      0.1002   &      0.1304   &      21.48    &      125.54   &      111.00   &      2006.0853 & 0.273 & i              & Men21  \\ 
                   &                 & $\pm$ 0.00468 & $\pm$ 0.0010  & $\pm$ 0.0047  & $\pm$ 1.96    & $\pm$ 5.48    & $\pm$ 5.59    & $\pm$ 0.0290  & $\pm$ 0.009 & {\tt ORBIT  } &  ... \\[1em] 
2MA 1036+1521 BC   & 10367+1522 BC   &      8.56325  &      0.1471   &      0.3392   &      24.84    &      89.35    &      53.07    &      2011.1127 & 0.348 & i              & Cal17  \\ 
                   &                 & $\pm$ 0.01321 & $\pm$ 0.0006  & $\pm$ 0.0026  & $\pm$ 0.73    & $\pm$ 1.95    & $\pm$ 1.79    & $\pm$ 0.0174  & $\pm$ 0.005 & {\tt ORBIT  } &  ... \\[1em] 
WT 1827 AB         & 10430-0913      &      29.96684 &      0.5604   &      0.3549   &      124.49   &      103.19   &      108.97   &      2001.4251 & 0.355 & i              & Mas23  \\ 
                   &                 & $\pm$ 0.94848 & $\pm$ 0.0055  & $\pm$ 0.0337  & $\pm$ 1.39    & $\pm$ 0.93    & $\pm$ 2.18    & $\pm$ 0.3419  & $\pm$ 0.025 & {\tt ORBIT  } &  ... \\[1em] 
GAI 1058-5525 AB   & 10581-5525      &      4.45804  &      0.1000   &      0.0333   &      63.04    &      197.75   &      111.46   &      2023.9629 & 0.393 & i              & ...    \\ 
                   &                 & $\pm$ 0.07084 & $\pm$ 0.0014  & $\pm$ 0.0183  & $\pm$ 1.73    & $\pm$ 1.30    & $\pm$ 16.10   & $\pm$ 0.2051  & $\pm$ 0.021 & {\tt ORBIT  } &  ... \\[1em] 
NLTT 30359 AB      & 12206-8226      &      2.30857  &      0.1208   &      0.8174   &      121.20   &      36.99    &      151.91   &      2021.4160 & 0.605 & i              & ...    \\ 
                   &                 & $\pm$ 0.00470 & $\pm$ 0.0016  & $\pm$ 0.0054  & $\pm$ 1.00    & $\pm$ 1.19    & $\pm$ 2.16    & $\pm$ 0.0059  & $\pm$ 0.028 & {\tt ORBIT  } &  ... \\[1em] 
SCR 1230-3411 AB   & 12300-3411      &      4.50617  &      0.1038   &      0.2148   &      83.17    &      58.27    &      167.46   &      2021.5914 & 0.379 & i              & Win17  \\ 
                   &                 & $\pm$ 0.09316 & $\pm$ 0.0019  & $\pm$ 0.0229  & $\pm$ 0.86    & $\pm$ 0.58    & $\pm$ 7.38    & $\pm$ 0.0943  & $\pm$ 0.027 & {\tt ORBIT  } &  ... \\[1em] 
L 327-121 AB       & 12336-4826      &      8.72337  &      0.1858   &      0.0692   &      63.66    &      34.08    &      221.46   &      2017.8213 & 0.758 & i              & ...    \\ 
                   &                 & $\pm$ 0.51326 & $\pm$ 0.0073  & $\pm$ 0.0497  & $\pm$ 1.14    & $\pm$ 1.39    & $\pm$ 8.59    & $\pm$ 0.1702  & $\pm$ 0.132 & {\tt ORBIT  } &  ... \\[1em] 
NLTT 33370 AB      & 13143+1320      &      9.49677  &      0.1458   &      0.6023   &      48.81    &      60.52    &      205.24   &      2013.5644 & 0.151 & i              & Dup16  \\ 
                   &                 & $\pm$ 0.02213 & $\pm$ 0.0004  & $\pm$ 0.0036  & $\pm$ 0.34    & $\pm$ 0.62    & $\pm$ 1.03    & $\pm$ 0.0094  & $\pm$ 0.021 & {\tt ORBIT  } &  ... \\[1em] 
LHS 3056 AB        & 15192-1245      &      13.77409 &      0.2343   &      0.4553   &      52.78    &      73.42    &      27.74    &      2022.0006 & 0.677 & i              & ...    \\ 
                   &                 & $\pm$ 0.11057 & $\pm$ 0.0018  & $\pm$ 0.0035  & $\pm$ 0.42    & $\pm$ 0.63    & $\pm$ 0.75    & $\pm$ 0.0143  & $\pm$ 0.024 & {\tt ORBIT  } &  ... \\[1em] 
SCR 1546-5534 AB   & 15467-5535      &      6.58595  &      0.2152   &      0.2920   &      21.99    &      227.12   &      317.84   &      2022.6505 & 0.136 & i              & ...    \\ 
                   &                 & $\pm$ 0.07030 & $\pm$ 0.0014  & $\pm$ 0.0060  & $\pm$ 1.86    & $\pm$ 4.33    & $\pm$ 5.29    & $\pm$ 0.0180  & $\pm$ 0.005 & {\tt ORBIT  } &  ... \\[1em] 
GJ 2121 AB         & 16302-1440      &      17.565   &      0.26741  &      0.69508  &      136.97   &      42.88    &      20.72    &      2020.98  & 0.720 & i, HIP         & Tok21  \\ 
                   &                 & $^{+0.074  }_{-0.075  }$ & $^{+0.00073}_{-0.00074}$ & $^{+0.00093}_{-0.00094}$ & $\pm$ 0.12   & $\pm$ 0.18 & $\pm$ 0.20 & $\pm$ 0.00090 & $\pm$ 0.011 & {\tt orvara } &  ... \\[1em]
GJ 1210 AB         & 17077+0722      &      14.30005 &      0.3079   &      0.4838   &      112.33   &      60.69    &      203.07   &      2006.5559 & 0.259 & i              & Mas18  \\ 
                   &                 & $\pm$ 0.00609 & $\pm$ 0.0002  & $\pm$ 0.0003  & $\pm$ 0.06    & $\pm$ 0.06    & $\pm$ 0.11    & $\pm$ 0.0013  & $\pm$ 0.022 & {\tt ORBIT  } &  ... \\[1em] 
GJ 1212 AB         & 17137-0825      &      1.59157  &      0.0592   &      0.4562   &      65.37    &      146.42   &      261.55   &      2020.3375 & 0.719 & i              & ...    \\ 
                   &                 & $\pm$ 0.00541 & $\pm$ 0.0012  & $\pm$ 0.0197  & $\pm$ 1.07    & $\pm$ 1.49    & $\pm$ 0.96    & $\pm$ 0.0117  & $\pm$ 0.046 & {\tt ORBIT  } &  ... \\[1em] 
SCR 1728-0143 AB   & 17282-0144      &      5.78256  &      0.0920   &      0.5097   &      13.81    &      141.78   &      139.51   &      2024.6362 & 0.446 & i              & ...    \\ 
                   &                 & $\pm$ 0.07130 & $\pm$ 0.0016  & $\pm$ 0.0209  & $\pm$ 8.61    & $\pm$ 36.78   & $\pm$ 35.89   & $\pm$ 0.0278  & $\pm$ 0.029 & {\tt ORBIT  } &  ... \\[1em] 
G 140-9 AB         & 17430+0547      &      20.576   &      0.3750   &      0.5479   &      103.58   &      308.86   &      56.04    &      2028.02  & 1.138 & i, HIP         & Tok19a \\ 
                   &                 & $\pm$ 0.032   & $^{+0.0013 }_{-0.0014 }$ & $\pm$ 0.0035 & $\pm$ 0.12 & $^{+0.16   }_{-0.12   }$ & $^{+0.34   }_{-0.45   }$ & $\pm$ 0.047 & $\pm$ 0.049 & {\tt orvara } &  ... \\[1em]
G 154-43 AB        & 18036-1859      &      19.66859 &      0.3761   &      0.4316   &      108.89   &      356.15   &      340.12   &      2018.8688 & 0.318 & i              & ...    \\ 
                   &                 & $\pm$ 2.85403 & $\pm$ 0.0461  & $\pm$ 0.0737  & $\pm$ 2.15    & $\pm$ 1.14    & $\pm$ 8.54    & $\pm$ 0.1586  & $\pm$ 0.149 & {\tt ORBIT  } &  ... \\[1em] 
UPM 1951-3100 AB   & 19517-3100      &      2.81655  &      0.1141   &      0.5651   &      72.70    &      31.69    &      158.48   &      2021.4450 & 0.419 & i              & ...    \\ 
                   &                 & $\pm$ 0.01077 & $\pm$ 0.0009  & $\pm$ 0.0070  & $\pm$ 0.73    & $\pm$ 0.57    & $\pm$ 1.37    & $\pm$ 0.0068  & $\pm$ 0.018 & {\tt ORBIT  } &  ... \\[1em] 
WT 818 AB          & 21497-4139      &      7.08866  &      0.1244   &      0.6449   &      129.20   &      140.07   &      350.71   &      2022.4625 & 0.665 & i              & ...    \\ 
                   &                 & $\pm$ 0.18969 & $\pm$ 0.0011  & $\pm$ 0.0103  & $\pm$ 1.09    & $\pm$ 1.77    & $\pm$ 2.59    & $\pm$ 0.0191  & $\pm$ 0.050 & {\tt ORBIT  } &  ... \\[1em] 
LP 983-34 AB       & 21558-3313      &      2.56572  &      0.0699   &      0.1134   &      119.72   &      176.89   &      3.49     &      2022.7285 & 0.653 & i              & ...    \\ 
                   &                 & $\pm$ 0.00874 & $\pm$ 0.0012  & $\pm$ 0.0066  & $\pm$ 1.53    & $\pm$ 1.34    & $\pm$ 8.00    & $\pm$ 0.0547  & $\pm$ 0.040 & {\tt ORBIT  } &  ... \\[1em] 
LEHPM 1-4771 AB    & 22302-5345      &      5.99805  &      0.1351   &      0.3118   &      108.78   &      89.64    &      212.67   &      2022.7173 & 0.257 & i              & Vri20  \\ 
                   &                 & $\pm$ 0.16246 & $\pm$ 0.0018  & $\pm$ 0.0120  & $\pm$ 0.83    & $\pm$ 0.70    & $\pm$ 3.38    & $\pm$ 0.0412  & $\pm$ 0.018 & {\tt ORBIT  } &  ... \\[1em] 
SCR 2303-4650 AB   & 23036-4651      &      13.40008 &      0.2448   &      0.2197   &      93.97    &      153.39   &      146.39   &      2014.9871 & 0.267 & i              & Tok20b  \\ 
                   &                 & $\pm$ 0.25457 & $\pm$ 0.0018  & $\pm$ 0.0118  & $\pm$ 0.28    & $\pm$ 0.29    & $\pm$ 4.59    & $\pm$ 0.1287  & $\pm$ 0.016 & {\tt ORBIT  } &  ... \\[1em] 
G 273-33 AB        & 23242-1746      &      4.12703  &      0.0956   &      0.2402   &      74.57    &      20.59    &      163.26   &      2022.6730 & 0.607 & i              & ...    \\ 
                   &                 & $\pm$ 0.02601 & $\pm$ 0.0011  & $\pm$ 0.0081  & $\pm$ 0.64    & $\pm$ 0.69    & $\pm$ 4.29    & $\pm$ 0.0463  & $\pm$ 0.027 & {\tt ORBIT  } &  ... \\[1em] 
LHS 4009 AB        & 23455-1610      &      21.31343 &      0.4133   &      0.4912   &      98.91    &      9.29     &      352.09   &      2024.2706 & 0.304 & i              & Mas23  \\ 
                   &                 & $\pm$ 0.09191 & $\pm$ 0.0014  & $\pm$ 0.0045  & $\pm$ 0.12    & $\pm$ 0.05    & $\pm$ 0.52    & $\pm$ 0.0224  & $\pm$ 0.016 & {\tt ORBIT  } &  ... \\[1em] 
LTT 9828 AB        & 23597-4405      &      10.70471 &      0.2178   &      0.4355   &      22.66    &      132.64   &      311.32   &      2003.0292 & 0.456 & i              & Tok22  \\ 
                   &                 & $\pm$ 0.01773 & $\pm$ 0.0009  & $\pm$ 0.0022  & $\pm$ 0.91    & $\pm$ 2.14    & $\pm$ 2.14    & $\pm$ 0.0351  & $\pm$ 0.014 & {\tt ORBIT  } &  ... \\[1em] 
\enddata
 
\tablecomments{
The astrometry used in each of these fits is given in Table~\ref{tab:results}, and illustrations of those data and orbits are shown in Figures~\ref{fig:orbits}--\ref{fig:orbits5}.
The six orbits fit using both imaging and RV data are listed separately in Table~\ref{tab:RVorbits}.
For ``data used'' (column~11), ``i'' indicates imaging data (from this work at SOAR as well as others published in the literature) and ``HIP'' indicates data from \textit{Hipparcos}.}
\end{deluxetable}

\begin{longrotatetable}
\startlongtable
\begin{deluxetable}{lllllllllllllll}
\tabletypesize{\scriptsize}
\tablecaption{
Parameters of the orbits fit to the RV and imaging data.
\label{tab:RVorbits}}

\tablehead{
\colhead{Name   } & \colhead{WDS} &  \colhead{\Porb} & \colhead{$a$        } & \colhead{$e$} & \colhead{$i$  } & \colhead{$\Omega$} & \colhead{$\omega$} & \colhead{$T_0$} & \colhead{$K_1$        } & \colhead{$K_2$        } & \colhead{$\gamma$     } & \colhead{$M_\mathrm{tot}$} & \colhead{Data used} & \colhead{Prev.} \\[-1em] 
\colhead{       } & \colhead{   } &  \colhead{(yr) } & \colhead{($\arcsec$)} & \colhead{   } & \colhead{(deg)} & \colhead{(deg)   } & \colhead{(deg)   } & \colhead{(yr) } & \colhead{(km s$^{-1}$)} & \colhead{(km s$^{-1}$)} & \colhead{(km s$^{-1}$)} & \colhead{($M_\odot$)     } & \colhead{Code used} & \colhead{orbit} \\[-1em] 
\colhead{(1)    } & \colhead{(2)} &  \colhead{(3)  } & \colhead{(4)        } & \colhead{(5)} & \colhead{(6)  } & \colhead{(7)     } & \colhead{(8)     } & \colhead{(9) }  & \colhead{(10)         } & \colhead{(11)         } & \colhead{(12)         } & \colhead{(13)            } & \colhead{(14)     } & \colhead{(15) } \\[-2em] 
}
\startdata
GJ 84 AB           & 02051-1737             &      13.32550 &      0.5008   &      0.3470   &      91.26    &      102.86   &      247.72   &      2000.7449 &      1.964    &      ...      &      25.276   & 0.573 & i, RV          & Man19  \\ 
                   &                 & $\pm$ 0.01039 & $\pm$ 0.0006  & $\pm$ 0.0017  & $\pm$ 0.04    & $\pm$ 0.02    & $\pm$ 0.21    & $\pm$ 0.0102  & $\pm$ 0.048   & ...     & $\pm$ 0.049   & 0.004 & {\tt ORBIT  } &  ... \\[1em] 
GJ 2060 AB         & 07289-3015 AB          &      7.78926  &      0.2552   &      0.8972   &      31.96    &      368.17   &      341.31   &      2013.0616 &      6.485    &      ...      &      28.579   & 1.038 & i, RV          & Cal22  \\ 
                   &                 & $\pm$ 0.00336 & $\pm$ 0.0011  & $\pm$ 0.0021  & $\pm$ 1.55    & $\pm$ 0.98    & $\pm$ 1.28    & $\pm$ 0.0094  & $\pm$ 0.286   & ...     & $\pm$ 0.027   & 0.027 & {\tt ORBIT  } &  ... \\[1em] 
LHS 3117 AB        & 15474-1054             &      8.26378  &      0.2414   &      0.4307   &      116.81   &      95.72    &      49.02    &      2015.0081 &      7.315    &      ...      &      3.096    & 0.843 & i, RV          & ...    \\ 
                   &                 & $\pm$ 0.00610 & $\pm$ 0.0013  & $\pm$ 0.0020  & $\pm$ 0.28    & $\pm$ 0.38    & $\pm$ 0.23    & $\pm$ 0.0090  & $\pm$ 0.051   & ...     & $\pm$ 0.026   & 0.018 & {\tt ORBIT  } &  ... \\[1em] 
GJ 791.2 AB        & 20298+0941 AB          &      1.47080  &      0.0971   &      0.6232   &      162.09   &      281.96   &      9.60     &      2019.1567 &      5.658    &      11.147   &      $-$34.363 & 0.177 & i, RV          & Man19  \\ 
                   &                 & $\pm$ 0.00031 & $\pm$ 0.0006  & $\pm$ 0.0059  & $\pm$ 3.12    & $\pm$ 1.96    & $\pm$ 2.31    & $\pm$ 0.0033  & $\pm$ 0.606   & $\pm$ 0.634   & $\pm$ 0.144   & 0.006 & {\tt ORBIT  } &  ... \\[1em] 
LHS 501 AC         & 20556-1402 Aa,Ab       &      1.84422  &      0.0918   &      0.2424   &      141.54   &      237.35   &      231.58   &      2017.1339 &      5.172    &      6.359    &      $-$142.138 & 0.372 & i, RV          & Vri22  \\ 
                   &                 & $\pm$ 0.00201 & $\pm$ 0.0007  & $\pm$ 0.0046  & $\pm$ 1.14    & $\pm$ 1.03    & $\pm$ 1.25    & $\pm$ 0.0058  & $\pm$ 0.078   & $\pm$ 0.084   & $\pm$ 0.038   & 0.011 & {\tt ORBIT  } &  ... \\[1em] 
LTT 17066 AB       & 23585+0740             &      12.71810 &      0.2880   &      0.1145   &      84.22    &      177.90   &      347.28   &      2019.4954 &      7.880    &      4.144    &      $-$11.454 & 0.716 & i, RV          & ...    \\ 
                   &                 & $\pm$ 0.01284 & $\pm$ 0.0016  & $\pm$ 0.0027  & $\pm$ 0.24    & $\pm$ 0.26    & $\pm$ 1.26    & $\pm$ 0.0390  & $\pm$ 0.226   & $\pm$ 0.028   & $\pm$ 0.021   & 0.016 & {\tt ORBIT  } &  ... \\[1em] 
\enddata
 
\tablecomments{
The RV data used in each of these fits is given in Table~\ref{tab:RVdata}, and illustrations of those data and orbits are shown in Figures~\ref{fig:orbits}--\ref{fig:orbits5} for the imaging components, and Figure~\ref{fig:RVorbits} for their RV components.
For ``data used'' (column~14), ``i'' indicates imaging data (from this work at SOAR as well as others published in the literature) and ``RV'' indicates radial velocity data (published in the literature).}
\end{deluxetable}
\end{longrotatetable}

In this work, an orbit fit is deemed reliable enough for publication when its orbital period is determined to better than 20\% and its eccentricity to better than 8\%. In several cases, the imaging data alone initially could not constrain the orbit well, but the RECONS long-term astrometry of those systems \citep[e.g.,][]{Hen18,Vri20} ruled out a range of potential orbital periods and allowed a convincing orbit fit under those constraints. A comprehensive assessment of the RECONS astrometry and how it can be used with resolved data such as that from SOAR will be presented in a forthcoming publication.

Each system's data were fit initially with the {\tt ORBIT} code \citep{Tok-orbit}. If {\it Hipparcos} data were also available, then another fit was attempted using {\tt orvara} \citep{orvara}, which incorporates {\it Hipparcos} and Gaia proper motions as additional constraints to the data. When using {\tt ORBIT} or {\tt orvara}, the uncertainties on each data point were set to the typical value of 2~mas for HRCam observations (or 6~mas for data flagged as noisy as described in $\S$\ref{sec:observations}), or set to the published errors for literature data. Both codes convert the data's position angles to equinox J2000 when commencing their fits.

In addition to using different combinations of data, the two orbit-fitting codes use different methodologies for selecting their orbits. 
Briefly, {\tt ORBIT} weights the data inversely to measurement uncertainties and uses a Levenberg-Marquardt least-squares algorithm to determine the best-fitting model. {\tt orvara} takes the same input as {\tt ORBIT} but also incorporates any change in the system's proper motion measured between the {\it Hipparcos} and {\it Gaia} EDR3 catalogs, then reports the distribution of models it generated using a Markov chain Monte Carlo (MCMC) algorithm. We take the medians of those parameter posteriors as the best fit. An advantage of the simpler {\tt ORBIT} algorithm is that it only takes seconds to converge, allowing for repeated runs with various starting values to inform the quality of the fit. In contrast, {\tt orvara} takes 15--30 minutes per target, depending on the size of the dataset and requested Markov chains, but each run's MCMC visualizations provide a more comprehensive view of the models' suitability. When both {\tt ORBIT} and {\tt orvara} agree on a fit, we report the {\tt orvara} result here because it includes the additional data from {\it Hipparcos} and {\it Gaia}. If the two codes did not arrive at the same result, we conclude that the system's orbit is too poorly constrained to publish at this time. 

Only four {\tt orvara} orbits survived the above process and are presented here. This is primarily because only a small fraction of our targets were bright enough to be included in {\it Hipparcos}, and thus they cannot have {\it Hipparcos}-Gaia accelerations. Of those that did have the data to attempt an {\tt orvara} fit, most had orbital periods greater than 30~years, which is too long to be considered for our project.

\subsection{Orbits with High-Resolution Imaging and Radial Velocity Data}
\label{sec:RVorbits}

For the six systems using combined high-resolution and RV data, we completed fits using the IDL code \texttt{ORBIT} \citep{Tok-orbit}. As outlined in $\S$\ref{sec:RVs}, the systems are SB1s and SB2s (and SB3s that we fit as SB1s and SB2s), and the orbital coverage varies from a small portion of one orbit to six orbital cycles. Thus, the available RV coverage permits different levels of improvement. One measure that characterizes how much the RV data improves orbits is the relative drop in uncertainties of the orbital parameters when using only the imaging data vs.\ incorporating the RV data. For the six cases here, we ran \texttt{ORBIT} with and without the RV data, and found that including that supplemental RV data had effects ranging from negligible to significant. For example, GJ~84~AB and LHS~3117~AB had $e$ errors $\ll1$\% regardless of RV data, while the $e$ error for GJ~791.2~AB dropped from 11\% (without RVs) to 0.9\% (with RVs), and for LTT~17066~AB it dropped from 27\% to 2.4\%.

\section{Discussion} 
\label{sec:discussion}

With \Nobservations{} observations of \Ntargets{} M dwarf pairs in this program \citep[combining this paper and][]{Vri22}, plus orbits derived for \Norbits{} of these systems, this work constitutes a significant contribution to the growing field of low-mass multiple star astronomy. Here we add \NorbitsNEW{} new orbits to the 257 orbits for nearby M dwarfs (within 25 pc) in the Sixth Catalog of Visual Binary Orbits \citep{ORB6}, an increase of 11\%.  We also update \NorbitsUPDATED{} orbits, providing improvements for another 8.5\% of those cataloged low-mass systems. Even for the 161 resolved systems that still do not have enough data to determine orbits, the astrometry points presented here represent critical steps toward determining their orbits in the coming years. 

These orbits, both new and improved, will contribute toward our program's ultimate goal to determine the properties of low-mass systems' dynamics as described in \cite{Vri20}. In particular, we aim to plot $P_\mathrm{orb}$ vs.\ $e$ for M dwarf multiples to draw conclusions regarding how these systems formed and dynamically evolved. This project requires the union of several observing techniques in order to cover $P_\mathrm{orb}$ from days to decades. Our SOAR speckle observing strategy is optimized for $P_\mathrm{orb} \gtrsim 1$~yr, and longer multi-year orbits will be supplied by the RECONS long-term astrometry survey \citep{Hen18,Vri20} and, ultimately, \Gaia. In contrast, days- to months-long orbits are accessible mainly to radial velocity work. Only with a more comprehensive set of short- (RV), medium- (SOAR), and long-period (RECONS/\Gaia) systems can we assess the overall orbital period distribution. Ultimately, our consideration of a volume-complete sample of low-mass binaries within 25 pc will provide new insight into the {\it statistical} nature of low-mass stellar systems, e.g., answering key questions related to how often various types of orbits occur.

Although \Norbits{} orbits is a substantial amount, we defer full discussion of their $P_\mathrm{orb}$ vs.\ $e$ distribution to a later publication, when we can consider the biases and sensitivity of the many contributing techniques. 
In the work presented here we report 40 orbits with \Porb{} $< 10$~yr, nine with \Porb{} = 10--20~yr, and five with \Porb{} = 20--30~yr, indicating that a key area of focus is to determine accurate orbits for the longest period systems. From this SOAR effort, there are only three systems with \Porb{} $<$~1~yr: SCR~0533-4257~AB (WDS~05335$-$4257), APM~89~AB (WDS~07096$-$5704), and UPM~0838-2843~AB (WDS~08386$-$2843), as they are particularly challenging due to our sparse observing cadence and the speckle data's 180$^\circ$ position angle ambiguity ($\S$\ref{sec:observations}) at small separations. Overall, we observe an extreme range in eccentricities from 0.0082 for G~34-23~AB (WDS~01222$+$2209) to 0.8972 for GJ~2060~AB (WDS~07289$-$3015 AB), with $e \leq 0.5$ for 70\% of our systems.

Finally, it is worth noting that two of the systems presented here have orbits in the \Gaia{} DR3 Non-Single Star (NSS) catalog \citep{Gaia-NSS-documentation,Gaia-NSS-processing}: APM~89~AB (WDS~07096$-$5704, Gaia~DR3~5486916932205092352, \Porb{} = 0.7~yr) with  and LHS~3056~AB (WDS~15192$-$1245, Gaia~DR3~6314466489153193344, \Porb{} = 13.8~yr). Those \Gaia{} results were derived from that mission's unprecedentedly precise space-based astrometry taken over 34 months.  Although the time-series data and residuals to those fits are not yet publicly available, the SOAR data can be used as independent checks of the \Gaia{} orbital fits. The two orbits match reasonably well for the short-period system APM~89~AB but are quite different for longer-period system LHS~3056~AB; these comparisons are discussed in detail in the Systems of Note section below ($\S$\ref{sec:systems}). Given only two overlapping systems in this paper, it is premature here to discuss at length the consistency of \Gaia{} NSS orbit results, but future work will bring the reliability of those orbits into focus.

\subsection{Systems of Note}
\label{sec:systems}

Here we describe several systems for which these SOAR results add significant new information or shed light on unusual observational histories.

\textit{2MA~0015$-$1636~AB (00160$-$1637)} --- In \citet{Vri22} we presented an orbit for this system that suggested a total mass higher than implied by the components' absolute magnitudes. Here we present an updated orbit, now covering most orbital phases, that significantly adjusts the determined eccentricity (previously 0.433, now 0.0153) and improves the orbital period by a factor of four. Despite this update, the discrepancy between total mass (from orbit) and estimated component masses (from magnitudes) remains nearly unchanged. This issue would be resolved if the parallax were smaller than reported by Gaia DR3 --- and, indeed, the Gaia reduced unit weight error of 4.1 indicates that parallax could be inaccurate. Alternatively, the components' masses may be estimated as too large because the stars may be brightened by youth, as \citet{Shk09} estimates an age of 35--300~Myr for the system.

\textit{GJ~1006~AC (00162$+$1952 Aa,Ab)} --- 
This hierarchical multiple was first revealed as triple by \citet{Bar18}, who reported that component A was an SB2 and determined the orbit with a period of $\sim$4~d and nearly-zero eccentricity. Our effort at SOAR tentatively resolved the AC pair in 2020 \citep{Vri22} at 31 mas separation, with three unsuccessful attempts in 2019 and 2020. The acceleration detected from {\it Hipparcos} and Gaia DR2/DR3 data provides a strong evidence that this system is quadruple (SB2 inner pair, its companion producing acceleration, and   the outer companion at 25\farcs2).

\textit{LP 993-115 BC (02452$-$4344 Aa,Ab)} Also known as LP~993-116~AB, this triple system has a third component (LP 993-115~A) at 44\farcs7. We assign the designations of ``A,'' ``B,'' and ``C'' here based on the absolute magnitudes of those components in Gaia's $G$ filter (as a proxy for their masses). Our orbit here updates that of our previous publication \citep{Vri22} using more than twice as much SOAR data, and though the orbital elements are consistent with that earlier fit, their precision is now significantly improved.

\textit{LHS~178~AB (03425$+$1232)} --- 
This well-known subdwarf \citep{Giz97} has been the target of several imaging searches with no companion detected \citep{Jao09,Lod09,Zha13}, and in the 2018 and 2019 observations on our program we were likewise unsuccessful \citep{Vri22}, but the high RUWE values in Gaia DR2 and DR3 motivated our continued follow-up. These revealed the companion once per year from 2020--2024 at steadily increasing separations until the last year ($0\farcs10$--$0\farcs14$ and then $0\farcs07$). This separation trend explains its lack of detection by \citet{Zha13}, who were limited to $\gtrsim 0\farcs5$, and most likely also the non-detection by \citet{Lod09}, who required $\gtrsim 0\farcs1$. Finally, the companion's $\Delta I$ of $\sim$1.7~mag also explains its lack of detection in \citet{Jao09}'s $V$-band observations.

\textit{LHS~1582~AB (03434$-$0934)} --- 
This system was well-known as binary through absolute astrometry \citep{Rie10} and its 5-yr orbit has been well-characterized \citep{Vri20}. Here we have successfully imaged the companion for the first time. Previously we presented the companion's absence in 2018 and 2020 \citep{Vri22} and suggested that its mass was likely $\lesssim$0.15~\Msun{}, which is consistent with our new data showing $\Delta I \sim 2.3$~mag.

\textit{UPM~0611-3433~AB (06112$-$0036)} --- 
This newly detected binary showed signs of non-single star astrometry in Gaia DR2 and DR3, which is consistent with the  presence of the orbital motion in our 2022--2023 data. This target was included in the binary search of \citet{Wan22}, but the companion was not detected, most likely due to its tight orbit.

\textit{SCR 0702-6102~AB (07028$-$6103)} --- This system has the largest residuals with respect to its orbit size. Although it has a higher than average amount of noisy SOAR data ($\S$\ref{sec:observations}), we also cannot rule out the possibility that the secondary is actually the photocenter of a close pair. Continued astrometric monitoring is necessary to accept or reject the presence of a third companion.

\textit{APM~89~AB (07096$-$5704)} --- 
First resolved as multiple in \citet{Vri22}, APM~89~AB is one of only two systems in our sample with a counterpart in the Gaia NSS table (the other is LHS~3056~AB a.k.a.\ 15192-1245). With \Porb $\sim 0.7$~yr, the orbit is well-mapped by both our program and (presumably) Gaia, and yet only the orbital periods and eccentricities match to within their uncertainties --- the angles of inclination ($i$), ascending node ($\Omega$), and periastron ($\omega$) each disagree by 2.5--7.4$\sigma$. Mismatches in orientation angles of Gaia NSS orbits were also noted by \citet{Mar23}; it remains to be seen whether this is a broader issue affecting all Gaia NSS results.

\textit{LEP~0808-7301~AC (08083$-$7302 Aa,Ab)} --- 
The well-known AB pair of this system, denoted in WDS with discoverer code KPP2882 and separation $\sim9\farcs9$, was revealed as triple via SOAR in \citet{Vri22} with two resolutions of a companion orbiting component A. Here we report four additional observations showing, overall, that C is advancing in position angle (343.5$^\circ$ through due N to 5.1$^\circ$) while moving first outward (279.8~mas to 313.8~mas) and then slightly inward (to 289.6~mas). If the orbit is circular, then this rate implies a period of $\gtrsim$80~years, although continued sparse monitoring over the next decade will be essential for refining this estimate.

\textit{GJ~2069~AC-E (08317$+$1924 Aa,Ab)} is a 1\arcsec{} pair in a quintuple system where all stars are M dwarfs.  Its visual primary is a double-lined spectroscopic binary with a period of 2.7715~d and an estimated separation of 2\,mas \citep{Del99}, too tight to be resolved by HRCam. At 10\farcs4, there is another pair (DEL~1 Ba,Bb) with a semi-major axis of 0\farcs76 and a period of 63 yr \citep{Tok21c}.

\textit{LTT~12366~AC (09012$+$0157 Aa,Ab)} --- This is a resolved triple system also known as Ross~625, identified by \citet{Cor17}. The outer star B is located at 3\farcs94 from the inner pair AC, for which we compute the first-time orbit with a period of 6.13~yr. The period of AB estimated from its separation is under 1~kyr.

\textit{L~392-39~BC (10199$-$4149 Ba,Bb)} --- 
The B component of this system's well-known AB pair (WDS discoverer code LDS~302) joined our program because its lack of \Gaia\ DR2 solution rendered it suspect, and in \citet{Vri22} we reported discovery of the C companion to B at 168.3--230.4~mas (in 2019.948 and 2020.996, respectively). Here we present observations of its continued motion to a separation of 324.1~mas in 2024.238. At this rate, the BC pair may have an orbital period $\sim$30~yr, which could be confirmed or refined in the coming years with continued follow-up.

\textit{2MA~1036+1521 (10367$+$1522 AB)} --- A low-mass triple system identified by \citet{Dae07}. We update here the orbit of the inner pair BC with a period of 8.56~yr, updating it from \citet{Cal17}. These authors also proposed a tentative 234-yr outer orbit of the BC pair around the primary star A at 1\arcsec separation. Wobble in the outer orbit and equal magnitudes of B and C indicate equal masses, so the accurately measured mass sum leads to the masses of 0.14~\Msun{} for B and C.

\textit{WIS~1055-7356~AB (10553$-$7356)} --- 
The faintness and close separation of this late M binary have made it challenging to resolve consistently; its companion was unseen in both 2020 \citep{Vri22} and 2023 (Table~\ref{tab:results}). Its motion to date has been roughly linear, suggesting an orbit that is more edge-on than face-on. With signs of poor astrometric fit to the single-star model in both Gaia DR2 and DR3, we are optimistic that DR4's time-series data will show orbital motion to supplement our SOAR data.

\textit{LHS~3056~AB (15192$-$1245)} --- 
We resolve two stars for this target, but compelling evidence in the literature show it is likely triple, with its third star too tightly bound for SOAR to resolve. \citet{Bon13} initially revealed LHS~3056 as AB spectroscopically, and \citet{Cor17} first resolved that pair through lucky imaging. Later, \citet{Tri20} updated its RV data and noted an additional fast variation on top of its long-term trend, hinting at a third close-in companion. Photometrically, overluminosity has been noted by \citet{Wei16} and \citet{Bar17} despite lack of evidence of youth. We also note that it is elevated above the main sequence by $\sim$1.4~mag in $M_V$ and even more in $M_K$, consistent with more than one redder companion.

Finally, although this system has a counterpart in Gaia NSS (the other such system in our sample is APM~89~AB), the orbit presented by Gaia has a much shorter period than ours: 158~d vs.\ 13~yr. This Gaia orbit is consistent with that of a third companion --- confirming what the photometry and RVs suggested above. Our orbit indicates even the outer pair is too closely separated for Gaia to resolve, thus it is not yet possible to determine the exact configuration of the three stars.

\textit{UPM~1547-2755~CD (15476$-$2754 CD)} --- 
The fourth companion in this hierarchical system was noted by our SOAR program in \citet{Vri22}, after its \Gaia\ DR2 results indicated evidence of non-single-star motion. The two resolved measurements published in \citet{Vri22} are here joined by six subsequent measurements, enough to see it move inward and outward again (105.1~mas to 64.0~mas to 112.2~mas) while moving through half a revolution of position angle (150.5$^\circ$ to 330.6$^\circ$). These data do not yet permit an orbit fit reliable enough to publish, but so far they indicate the orbit is $\sim$9~yr in period.

\textit{LSR~1809-0219~AB (18097$-$0220)} --- 
Our SOAR data reveal for the first time that this target is an unequal-mass binary. Although no previously published works noted hints of its multiplicity, the system had high errors in Gaia DR2 and no 5-parameter solution in Gaia DR3\footnote{Gaia does have an entry that we are confident is LSR~1809-0219~AB despite its lack of 5-parameter solution; its ID is Gaia DR3 4177855052651068928.}, indicating its astrometry was poorly fit by their single-star models. Our recent SOAR data have resolved it after two previous attempts yielded non-resolutions \citep[in 2019 and 2021;][and this work]{Vri22}. Based on the increasing separation since then, the pair was probably too close to resolve in those earlier epochs. If the median separation observed here were the semi-major axis, we would expect \Porb $\lesssim 6$~yr, which is consistent with Gaia's inability to get a good fit in both their 22-month (DR2) and 34-month (DR3) data sets.

\textit{L~43-72~ABC (18113$-$7859 Aa,Ab and AB)} --- A triple system where the outer 1\farcs6 pair (AC-B) was identified by Gaia (Gaia DR3 6364795328246346112 is AC, and Gaia DR3 6364795328244240256 is B), and the inner subsystem AC was resolved for the first time in these SOAR data. Orbits of the inner (1.78~yr) and outer (110~yr) subsystems are in preparation.

\textit{SCR~1826-6542~AB (18268$-$6543)} --- 
\citet{Win19} noted this target as a suspected binary that they had recently confirmed in follow-up observations, thus we consider this a known binary resolved here for the first time. We observed this system several times previously at SOAR \citep{Vri22}, and its path over our more recent resolved astrometry indicate that at those earlier epochs it was too closely separated to resolve. RECONS astrometry is consistent with linear motion and low $\Delta m$, hence SCR~1826-6542~AB was not marked in \citet{Vri20} as showing any astrometric perturbation.

\textit{L~209-71~AC (20154$-$5646 Aa,Ab)} --- 
This system's two more massive M dwarf components (AB) are separated by $\sim$8$\arcsec$, while our program resolved the less massive C component $\sim$0$\farcs8$ from A in 2019 \citep{Vri22} and again in 2022 (this paper). Motion over those years has been relatively slow --- consistent with the RECONS astrometry data from the CTIO/SMARTS 0.9m \citep{Vri20}, which over 2012--2024 has showed roughly straight-line motion punctuated by a short perturbation in mid-2014. This behavior indicates its orbit is eccentric, with periastron passage in 2014.

\textit{SCR~2025-2259~ABC (20253$-$2259 Aa,Ab and BC)} --- A triple system consisting of the inner pair BC at 0\farcs1 showing considerable motion in 2022--2025 (estimated period $\sim$15~yr) and the outer component A at 0\farcs9. The SOAR data presented here are the first to detect and resolve BC, although Gaia DR3 hinted at its potential higher-order multiplicity by omitting its parallax (Gaia DR3 6849857217985836928).

\textit{LHS~501~ABC (20556$-$1402 Aa,Ab)} --- A triple system where the orbit of the inner pair AC with a period of 1.844~yr is published in \citet{Vri22}, while the distant companion B is located at 107\arcsec.

\textit{LHS~3739~BC (21588$-$3226)} ---  
This pair is part of the well-known hierarchical triple (WDS code 21589$-$3227 for the full system) composed of the earlier M dwarf LHS~3738 as component A, separated $\sim\!112\arcsec$ from the BC pair known as LHS~3739 \citep{Jao03,GaiaEDR3-cat} and having an astrometric orbit with a period of 6.1~yr by \citet{Lur14} that corresponds to a semimajor axis of 0\farcs1. We have tentatively resolved BC in 2021.75 (the pair is too faint for HRCam). Note that the system was previously mentioned to be resolved on Gemini-N by \citet{Rie10}, but no astrometry was given there. Based on two previous unsuccessful attempts, in \citet{Vri22} we concluded the C component was less massive than 0.1~\Msun{}; the new resolved photometry are consistent with 0.1~\Msun{} for C, and concurrent astrometry from the RECONS program also indicates the pair was more closely separated (thus not resolvable) during the earlier attempts in 2019.

\textit{2MA~2307$-$0415~AC (23073$-$0416 Aa,Ab)} --- 
This inner pair of a hierarchical triple was presented in \citet{Vri22}, and motion since then has been slow but steady: it  has moved in separation from 451.4~mas to 278.3~mas over 2019.5--2024.9, and in position angle from 111.6$^\circ$ to 68.9$^\circ$ in that time. With a likely orbital period of several decades, this system is worthy of sparse but consistent follow-up.

\section{Conclusions} 
\label{sec:conclusion}

Since 1 January 2021, we have taken \Nobservations{} observations of \Ntargets{} targets at SOAR with the speckle instrument HRCam. These data previously revealed nearly 100 new binaries \citep{Vri22}, and in this current update we have fit the orbits of \Norbits{} of those targets. Among those orbits, \NorbitsNEW{} are the first such characterizations for those systems, and \NorbitsUPDATED{} are updated from earlier efforts. Many of these results were made possible by the rich data set of incremental results already available in the literature; the positions from HRCam are easily combined with similar astrometry from other instruments, as well as radial velocities.

The \Norbits{} orbits in this current paper are all 30~yr or shorter to contribute to our project exploring orbital periods and eccentricities for systems on those timescales \citep{Vri20}. We will continue to follow the several dozen of our \Ntargets{} targets having no orbits presented here, concentrating primarily on the binaries with orbital periods longer than a decade, where good orbits are lacking. Building this repository of incremental results will also be critical for enhancing the utility of future time-series data from the \Gaia{} mission and the Legacy Survey of Space and Time from Rubin Observatory \citep{LSST}. Our study will ultimately include $\sim$200 orbits spanning multiple observing techniques and orbital periods of 1~d to 30~yr, and the comprehensive evaluation of those results will be presented in a future paper. Only through consistent effort, and the combined data sets from various facilities, can we fully explore the architectures of low mass stellar systems out to the scales of Jupiter and Saturn's orbits in our Solar System.

\acknowledgments
{
This work was made possible by the support of colleagues at the Southern Astrophysical Research (SOAR) Telescope, Cerro Tololo Inter-American Observatory (CTIO), and the SMARTS Consortium. Additionally, the National Science Foundation has been consistently supportive of this effort under grants AST-0507711, AST-0908402, AST-1109445, AST-141206, AST-1715551, and AST-2108373. We also extend our thanks to the Spring 2025 class of AST 200 at Smith College for their eager assistance with Figure~\ref{fig:hrd}. 
Finally, E.H.V.\ particularly appreciates Kim Ward-Duong for her support through the end stages of this project, and Florence W.\ Vrijmoet for her role in the lattermost stages of the manuscript.

This research has made use of several astronomical catalogs and databases, including: the Washington Double Star (WDS) Catalog maintained at the U.S.\ Naval Observatory; the Set of Identifications, Measurements and Bibliography for Astronomical Data (SIMBAD) database, operated at Centre de Donn\'{e}es astronomiques de Strasbourg (CDS), France; and NASA's Astrophysics Data System. This work has also made use of data from the European Space Agency (ESA) mission \Gaia{} (\url{https://www.cosmos.esa.int/gaia}), processed by the \Gaia{} Data Processing and Analysis Consortium (DPAC, \url{https://www.cosmos.esa.int/web/gaia/dpac/consortium}). Funding for the DPAC has been provided by national institutions, in particular the institutions participating in the \Gaia{} Multilateral Agreement.

}

\facilities{CTIO:0.9m, SOAR}

\appendix

To assist with locating our sample's M dwarfs in other catalogs, in Table~\ref{tab:lookup} we have listed our systems' coordinates and alternative names. Column~3 of that table lists each system's ``RECONS Name,'' which are the primary names used in this paper as well as others in the \textit{Solar Neighborhood} series. In column~7 we have also marked whether or not each pair has an orbit presented in this paper.

\startlongtable
\begin{deluxetable}{llllllc}
\tablecaption{
Identifiers used for M dwarf systems throughout this paper and in other commonly used catalogs.
(This table is available in its entirety in machine-readable form.)
\label{tab:lookup}}

\tablehead{
\colhead{RA       } & \colhead{Decl.    } & \colhead{RECONS} & \colhead{WDS} & \colhead{Discov.} & \colhead{SIMBAD} & \colhead{Orbit in  } \\[-1em] 
\colhead{(J2000.0)} & \colhead{(J2000.0)} & \colhead{Name  } & \colhead{   } & \colhead{       } & \colhead{Name  } & \colhead{this paper} \\[-1em] 
\colhead{(1)      } & \colhead{(2)      } & \colhead{(3)   } & \colhead{(4)} & \colhead{(5)    } & \colhead{(6)   } & \colhead{(7)       } 
}
\startdata
00 06 39.24 & $-$07 05 35.9 & 2MA 0006-0705 AB   & 00067$-$0706 & JNN 11        & PM J00066-0705            & ...          \\
00 08 53.92 & +20 50 25.6   & G 131-26 AB        & 00089$+$2050 & BEU 1         & G 131-26                  & $\checkmark$ \\
00 09 45.04 & $-$42 01 39.3 & LEHPM 1-255 AB     & 00098$-$4202 & TSN 119       & LP 988-129                & $\checkmark$ \\
00 15 27.99 & $-$16 08 01.8 & GJ 1005 AB         & 00155$-$1608 & HEI 299       & G 158-50                  & ...          \\
00 15 58.07 & $-$16 36 57.6 & 2MA 0015-1636 AB   & 00160$-$1637 & BWL 2         & BPS CS 31060-0015         & $\checkmark$ \\
00 16 01.97 & $-$48 15 39.1 & L 290-72 AB        & 00160$-$4816 & TOK 808       & CD-48 33                  & $\checkmark$ \\
00 16 14.63 & +19 51 37.5   & GJ 1006 AC         & 00162$+$1952 & TSN 120 Aa,Ab & G 32-6                    & ...          \\
00 24 44.19 & $-$27 08 24.2 & GJ 2005 AB         & 00247$-$2653 & LEI 1   AB    & LP 881-64                 & ...          \\
00 24 44.10 & $-$27 08 24.0 & GJ 2005 BC         & 00247$-$2653 & LEI 1   BC    & LP 881-64                 & ...          \\
00 25 04.31 & $-$36 46 17.9 & LTT 220 AB         & 00251$-$3646 & BRG 2         & G 267-100                 & ...          \\
00 27 55.99 & +22 19 32.8   & LP 349-25 AB       & 00279$+$2220 & FRV 1         & LP 349-25                 & $\checkmark$ \\
00 32 53.14 & $-$04 34 07.0 & GIC 50 AB          & 00329$-$0434 & JNN 12  AB    & GR* 50                    & ...          \\
00 32 53.14 & $-$04 34 07.0 & GIC 50 AC          & 00329$-$0434 & JNN 12  AC    & GR* 50                    & ...          \\
00 48 13.33 & $-$05 08 07.7 & LTT 453            & 00482$-$0508 & ...           & LP 646-17                 & ...          \\
\enddata
 
\tablecomments{
For column~4, we list the identifier we anticipate would be used for each pair in the WDS catalog \citep{WDS}, either currently or when they are eventually resolved. Although many of these systems already have entries in that catalog, several do not yet have entries because they are newly resolved here at SOAR or are not yet resolved by any effort. Pairs only have a discoverer code in column~5 if they are already listed in the WDS catalog, so those newly resolved in this paper do not yet have that identifier. Pairs that are subsets of higher-order multiples have their component identifiers appended to their discoverer code.
}
\end{deluxetable}

\end{document}